\documentclass{elsart}

\usepackage{amsfonts}
\usepackage{amsmath}
\usepackage{amssymb}
\usepackage{enumerate}
\usepackage{epsfig}
\usepackage{epsf}
\usepackage{array}

\journal{Annals of Physics}

\newcommand{\diff}{\d}
\newcommand{\eexp}{\e}
\newcommand{\starcomma}{\stackrel{*}{,}}
\newcommand{\Beta}{\mathcal{B}}
\newcommand{\con}{\!\circ\!}
\newcommand{\rda}{{ \Big\rangle\!\!\Big\rangle}}
\newcommand{\lda}{ \Big\langle\!\! \Big\langle}
\newcommand{\srda}{{ \rangle\!\rangle}}
\newcommand{\slda}{\langle\!\langle}
\renewcommand{\vec}[1]{{\boldsymbol{#1}}}
\renewcommand{\Im}{\mathop{\rm Im}}
\renewcommand{\Re}{\mathop{\rm Re}}
\newcommand{\ep}{\epsilon}
\newcommand{\cL}{\mathcal{L}}

\newcommand{\beq}{\begin{equation}}
\newcommand{\eeq}{\end{equation}}
\newcommand{\be}{\begin{equation}}
\newcommand{\ee}{\end{equation}}
\newcommand{\beqa}{\begin{eqnarray}}
\newcommand{\eeqa}{\end{eqnarray}}
\newcommand{\bea}{\begin{eqnarray}}
\newcommand{\eea}{\end{eqnarray}}

\renewcommand{\r}{\boldsymbol{r}}
\renewcommand{\P}{\boldsymbol{P}}

\newcommand{\Div}{\mathop{\mathrm{div}}\nolimits}

\newcommand{\req}[1]{Eq.~(\ref{#1})}
\newcommand{\reqs}[1]{Eqs.~(\ref{#1})}
\newcommand{\rref}[1]{(\ref{#1})}

\begin{document}

\begin{frontmatter}

\title{
Metal-insulator transition in a weakly interacting many-electron
system with localized single-particle states
}
\author[Princeton]{D.~M.~Basko},
\ead{dbasko@princeton.edu}
\author[Columbia]{I.~L.~Aleiner}, and
\author[Princeton,Columbia,NEC]{B.~L.~Altshuler}
\address[Princeton]{Department of Physics, Princeton University,
Princeton, NJ 08544, USA}
\address[Columbia]{Physics Department, Columbia University, New York, NY
10027, USA}
\address[NEC]{NEC-Laboratories America, 
4~Independence Way, Princeton, NJ 085540, USA}

\begin{abstract}
We consider low-temperature behavior of weakly interacting electrons in
disordered conductors in the regime when all
single-particle eigenstates are localized by the quenched disorder. We
prove that in the absence of coupling of the electrons to any external
bath dc electrical conductivity exactly vanishes as long as the
temperatute~$T$ does not exceed some finite value~$T_c$. At the same
time, it can be also proven that at high enough~$T$ the conductivity
is finite. These two statements imply that the system undergoes a
finite temperature Metal to Insulator transition, which can be viewed
as Anderson-like localization of many-body wave functions in the Fock
space. Metallic and insulating states are not different from 
each other by any spatial or discrete symmetries.

We formulate the effective Hamiltonian description of the system at
low energies (of the order of the level spacing in the single-particle
localization volume). In the metallic phase quantum Boltzmann equation
is valid, allowing to find the kinetic coefficients.
In the insulating phase, $T<T_c$, we use Feynmann diagram technique to
determine the probability distribution function for quantum-mechanical
transition rates. The probability of an escape rate from a given
quantum state to be finite turns out to vanish in every order of the
perturbation theory in electron-electron interaction.
Thus, electron-electron interaction alone is
unable to cause the relaxation and establish the thermal equilibrium.

As soon as some weak coupling to a bath is turned on, conductivity
becomes finite even in the insulating phase. Moreover, in the vicinity
of the transition temperature it is much larger than phonon-induced
hopping conductivity of non-interacting electrons. The reason for this
enhancement is that the stability of the insulating state is gradually
decreasing as the transition point is approached. As a result, a
single phonon can cause a whole cascade of electronic hops.

\end{abstract}


\end{frontmatter}

\tableofcontents

\section{Introduction}

Transport properties of conducting materials 
at low temperature $T$ are determined by an interplay between the 
interaction of the itinerant electrons with each other and the quenched 
disorder which creates a random potential acting on these electrons. In the 
absence of the electron-electron interaction all physics is dominated by 
the phenomenon of Anderson localization~\cite{Anderson58} -- {\em e.g.}, dc
electrical conductivity $\sigma$ can be qualitatively different depending 
on whether one-particle wave functions of the electrons are localized or 
not. In the latter case $\sigma (T)$ has a finite zero-temperature limit, 
while in the former case $\sigma (T)$ vanishes when $T \to 0$. 
Therefore, Anderson localization of electronic states leads to the Metal to 
Insulator Transition at zero temperature. 

When speaking about zero temperature, we need to consider only 
localization of the electronic states close to their Fermi level. The 
conductivity becomes finite at any finite temperature provided that
extended states exist somewhere above the Fermi level. It is commonly 
accepted now that localized and extended states in a random potential can 
not be mixed in the one-electron spectrum and thus this spectrum in a very 
general case is a combination of bands of extended states and bands 
of localized states. A border between a localized and an extended band is 
called mobility edge. If the Fermi level is located inside a localized 
band and inelastic scattering of the electrons are completely absent, the 
conductivity should follow Arrhenius law $\sigma(T)\propto\exp(-E_c/T)$, 
where $E_c$ is the distance from the Fermi level to the closest mobility 
edge.  Another common belief following from the scaling theory of 
Anderson localization~\cite{Thouless,Abrahams}, is that in low
dimensionality $d$, namely at $d=1,2$
all states are localized in an arbitraryly small disorder, while for free 
electrons (no periodic potential) $E_c>0$ is finite at $d=3$
(for $d=1$ this statement was proven rigorously both for
thin~\cite{Gertsen,Berezinskii} and thick~\cite{Efetov,Dorokhov}
disordered wires). It means that 
without inelastic processes $\sigma_{d=1,2}(T)=0$, while for 
$\sigma_{d=3}(T)$ one should expect the Arrhenius law. Note that for 
electrons in a crystal within a given conduction band the latter
conclusion could become incorrect -- strong enough disorder can localize 
the whole band.

Situation becomes more complicated when the inelastic processes are taken 
into account. In particular, electron-phonon interaction leads to the 
mechanism of conductivity known as hopping
conductivity~\cite{Mott,Shklovskii} -- 
with an assistance of phonons electrons hop between the localized states 
without being activated above the mobility edge. As a result, $\sigma (T)$ 
turns out to be finite (although it can be very small) at any finite $T$ even 
when all one-electron states are localized. 

Can interaction between electrons play the same role and cause the hopping 
conductivity?
This question was discussed in literature for a long
time~\cite{Fleishman,Shahbazyan,Kozub,Chalker,Giamarchi}
and no definite conclusion was achieved. The problem is 
that although the electric noise exists inside the material with a
finite ac conductivity\footnote{
In this paper we mostly focus on dc conductivity. As to ac
conductivity, it never vanishes, because at any frequency density of
resonant pairs of states is finite.}
the ``photons'' in contrast with phonons become localized 
together with electrons.

In this paper we demonstrate that electron-electron interaction alone 
cannot cause finite conductivity even when temperature is finite, but 
small enough. In the absence of phonons and extended one-electron states a 
system of interacting electrons has exactly zero conductivity below some 
temperature $T_c$. This critical temperature is infinite if the 
distance between the electrons exceeds the localization lengths of all 
electronic states. In the opposite case $T_c$ turns out to be finite and 
depends on the typical number of electrons within the localization volume 
as well as on the strengths of the electron-electron interaction. We also 
argue that at high temperatures $T > T_c$ the conductivity $\sigma (T)$ is 
finite. It means that at $T = T_c$ the system of interacting electrons 
subject to a random potential undergoes a genuine phase transition that 
manifests itself by the emerging of a finite conductivity!

This transition can be thought of as many-body localization -- it applies 
to many-body eigenstates of the whole system. This localization occurs not 
in the real space, but rather in the Fock space. This fact does not affect 
the validity of the concept of mobility edge.
In fact, the existence of the "metallic" state at $T > T_c$ implies that 
the many-body states with energies $\mathcal{E}$   above
$\mathcal{E}_c$ are  extended. One can estimate the difference between
$\mathcal{E}_c$ and the energy of the many-body ground state
$\mathcal{E}_0$ as  $\mathcal{E}_c-\mathcal{E}_0 \sim 
T\mathcal{N}(T)$, where $\mathcal{N}(T)$ is the total number of one-particle 
states in the energy strip of the width $T$. Note that the existence of 
the extended many-body states above the mobility edge does not contradict 
the fact that below  $T_c$ there is no conductivity -- in contrast with the 
case of one-particle localization there is no Arrhenius regime since 
$\mathcal{E}_c-\mathcal{E}_0$ turns out to be proportional to the
volume of the system,  {\em i.e.}, is macroscopically large (see
Sec.~\ref{sec:MBloc} below for more details).

In order to avoid possible misunderstanding we would like to emphasize at 
the very beginning that throughout this paper we will be focused only on 
the inelastic collisions between the electrons, {\em i.e.}, on creation or 
annihilation of {\em real} electron-hole pairs. There are other effects of 
electron-electron interactions - they can be understood as renormalization 
of the one-particle random potential by the interaction. This 
renormalization is temperature dependent and thus leads to a number of 
interesting effects, for example, the interaction corrections to the 
density of states and conductivity in disordered metals~\cite{AA}. On the 
insulating side of the one-particle localization transition similar 
effects cause the well-known Coulomb gap~\cite{Shklovskii} which suppresses
hopping 
conductivity. On the other hand, this is still a time-independent 
correction to the time-independent random potential. As such, it can maybe 
shift the position of the many-body Metal to Insulator transition,
{\em i.e.}, 
renormalize $T_c$, but is unable to destabilize the insulating or metallic 
phases. From now on we will simply neglect all elastic (Hartree-Fock) 
effects and concentrate on the {\em real} inelastic electron-electron 
collisions.

Localization of the many-body states in the Fock space has been discussed 
in Ref.~\cite{AGKL}
for the case of zero-dimensional systems with finite, 
although maybe large, number of electrons. Authors of
Ref.~\cite{AGKL} used an 
approximate mapping of the Hamiltonian of a metallic grain with large 
Thouless conductance $g$ and moderate interaction between the electrons to
the one-particle Hamiltonian on a lattice with the topology of the 
Cayley tree and an on-site disorder. The latter problem has an exact 
solution~\cite{AbouChacra}
that contains the localization transition. In terms of 
interaction electrons this transition means that below certain energy of 
the excitation the one-particle excitation states are quite close to some 
exact many-body excitations. As to the one-particle excitations with 
energies higher than the critical one, its wave function can be viewed as 
a wave packet, which involves a large number of the many-body eigenstates.
Being a property of  a finite system this transition could be nothing but 
a crossover, which becomes more pronounced with increasing of $g$. 

For an infinite system with $d>0$ the presence of spatial degrees of
freedom makes the situation more complex. In this case,
Cayley tree approximation
turns out to be insufficient. Nevertheless, a consistent
analysis of a model with weak and short range interaction to all
orders of perturbation theory enabled us 
to analyze the many-body localization transition and to demonstrate that 
both the metallic state at high temperatures and the insulating state at 
low temperatures are stable and survive all higher loop corrections to the 
locator expansion. This allows us to claim that the existence of the 
transition is proven on the physical level of rigor. 

It should be noted that such an insulating state that is characterized by 
exactly zero conductivity is quite different from all other known types of 
Metal to Insulator transitions. For example, Mott insulator is believed to 
have finite, though exponentially small conductivity at finite 
temperatures. 

The remainder of the paper is organized as follows. 
Sec.~\ref{sec:qualitative} contains the discussion of the problem on
the qualitative level: we define the many-body localization, show its
macroscopic implications, and discuss the relation to the Anderson
model on a certain graph.
In Sec.~\ref{sec:model} we specify the model many-body system to be
studied throughout the paper.
In Sec.~\ref{sec:Formalism} we show how to treat this model in the
framework of non-equilibrium Keldysh formalism, introduce the
self-consistent Born approximation (SCBA), and derive the general
equations. Using these equations, we demonstrate the existence of the
metallic state at high temperatures and study its properties in
Sec.~\ref{sec:metal}. Sec.~\ref{sec:insulator} is devoted to the proof
of the existence of the insulating phase at low temperatures; we
evaluate the transition temperature as the limit of stability of the
insulating phase. Sec.~\ref{sec:validity} is dedicated to
justification of SCBA; we demonstrate that corrections to SCBA are
indeed small. Finally, in Sec.~\ref{sec:conclusions} we summarize the
results and present an outlook of the future developments.

\section{Qualitative discussion}\label{sec:qualitative}

\subsection{Macroscopic manifestations of the many-body localization
 transition}

\subsubsection{Single-particle localization}

Let us begin with the brief review of the basic notion
emerged in a study of the properties of the
one electron wave functions in a disordered potential in $d$ dimensions.  
Depending on the strength of the disorder potential, a wave function
$\phi_\alpha(\vec{r})$ of an eigenstate~$\alpha$ corresponding to a
given energy can be classified as {\em localized} or {\em extended}:
\be
|\phi_\alpha(\vec{r})|^2\propto
\left\{
\begin{matrix}
\frac{1}{\zeta_{loc}^d}
\exp\left(-\frac{|\vec{r}-\vec\rho_\alpha|}{\zeta_{loc}}\right);
& {\rm localized;}\\
\frac{1}{\mathcal{V}}; & {\rm extended.}
\end{matrix}\right.
\label{eq:2.1}
\ee
where $\zeta_{loc}$ is the localization length which depends on the
eigenenergy $\xi_\alpha$, and $\mathcal{V}$ is the volume of the system. 
Exponentially localized states have a maximum amplitude at some point
$\vec\rho_\alpha$ in the system and a wave function envelope
which falls off exponentially, whereas 
the extended states spread relatively uniformly over the whole volume
of the system. It is believed that the coexistence of the localized
and extended states in the same energy range is not possible, and the
spectrum splits into bands of localized and extended states. The
energies separating such bands are known as {\em mobility edges}.
For example, for free electrons in a disorder potential in the
dimensionality three and higher, only one mobility edge $\mathcal{E}_1$
exists, so that 
\bea
\xi_\alpha < \mathcal{E}_1: &\quad {\rm localized};
\nonumber
\\
\xi_\alpha > \mathcal{E}_1: &\quad  {\rm extended}.
\label{eq:2.2}
\eea 

The statement about the asymptotic behaviour of the single-particle
wavefunctions \rref{eq:2.1} can be translated into the property of the 
matrix elements of a certain local operator $\hat{A}(\vec R)$
(it might be the local mass or current density operator, {\it etc}.):
\be
\begin{split}
A_{\alpha\beta}(\vec R)=\int\diff^d\vec r\,
\phi_\alpha^*(\vec r)\hat{A}(\vec R)\phi_\beta(\vec r).
\end{split}
\label{eq:2.3}
\ee
Then,
\be
\begin{split}
&
\cL^{{A}}_{\alpha\beta}\left(\vec r\right)=
\int \diff^d\vec R\, 
A_{\alpha\beta}(\vec R )A_{\beta\alpha}(\vec R + \vec{r})
\propto
\left\{
\begin{matrix}
\leq \eexp^{-\frac{|\vec{r}|}{\zeta_{loc}}};
& {\rm localized;}\\
\mathcal{F}\left(
\frac{|\vec{r}|}{L_{\omega_{\alpha\beta}}}
\right); & {\rm extended,}
\end{matrix}
\right.
\end{split}
\label{eq:2.4}
\ee
where the linear scale $L_\omega$ is controlled by the transmitted
energy $\omega_{\alpha\beta}=\xi_\alpha-\xi_\beta$ 
only, and $L_\omega \to \infty$ for $\omega\to 0$. No summation over
the repeating indices is implied in \req{eq:2.4}. Energies
$\xi_\alpha$ and $\xi_\beta$ are assumed to be sufficiently close to
each other, so that the localization length is approximately the same
for the two states.

The notion of the matrix elements \rref{eq:2.4} is intimately
related to the observable quantities. As an example, consider the
Kubo formula for the density-density response function
\be
\Pi\left(\omega;\vec{r}\right)
=\frac{1}{\mathcal{V}}
\sum_{\alpha\beta}\frac{\left(f_\alpha-f_\beta
\right)\cL^\varrho_{\alpha\beta}(\vec r)}
{\omega-\xi_\alpha+\xi_\beta+i0},
\label{eq:2.5}
\ee
where the overlap $\cL^\varrho_{\alpha\beta}(\vec r)$ is given by
\req{eq:2.4} built on the operators of the local density,
$\hat{\varrho}(\vec R)$, such that
$\hat{\varrho}(\vec R)|\vec r\rangle=\delta(\vec R-\vec r)|\vec r\rangle$.
In the thermal equilibrium for the fermionic particles,
the occupation numbers $f_\alpha$ are
given by the Fermi distribution.

In a conducting system, the low-frequency asymptotic behavior of
the density-density response function is always determined by
diffusion as the total number of particle is the only conserved
quantity in the system:
\be
\Pi\left(\omega;\vec{r}\right)\propto \int \diff^d\vec Q
\frac{\eexp^{i\vec{Q}\cdot\vec{r}}}{-i\omega+D\vec{Q}^2},
\label{eq:2.6}
\ee
where $D$ is the diffusion coefficient. 
First, let assume that all the states are localized.
Then, comparing \req{eq:2.5} with 
\req{eq:2.4}, we see that the asymptotic behaviour of the
wave function precludes the diffusion propagation \rref{eq:2.6}.
Therefore,
the diffusion coefficient vanishes for any temperature~$T$.
The same is true for the dc conducivity, as it is related to the
diffusion coefficient via Einstein relation, $\sigma\propto D$.
Physical meaning of vanishing $D$ and $\sigma$ is
the impossiblity for
an excitation caused by a local external perturbation to
propagate in a localized system, and uniformly span all the
phase space allowed by the conservation of energy.

If a finite mobility edge~\rref{eq:2.2} exists and the Fermi level~$\ep_F$
lies in the energy region of localized states, the conducitivity
is determined by the exponentially small occupation number of the
delocalized states 
\be
\sigma(T)\propto{e}^{-(\mathcal{E}_1-\ep_F)/T}.
\label{Arrhenius}
\ee
In this paper we will be interested in transport properties of the
systems where {\em all} single-particle states are localized, and thus
without many-body effects $\sigma=0$ at any temperature.
It is well established now that the mobility edge does not exist
for one- and two-dimensional systems for any disorder, and all
single-particle states
are indeed localized. Such a situation can also arise in an arbitrary
high dimensionality as well. Indeed, consider as an example Anderson
model with one state per lattice site~\cite{Anderson58}. It is well
known that all eigenstates become localized as soon as the on-site
disorder exceeds a critical value.

\subsubsection{Many-body localization}\label{sec:MBloc}

Let us now turn to the discussion of the many-body localization.
From now on, we assume that all one-particle state are localized in the
sense of the previous subsection. Can the interaction cause finite
conductivity?
Consider a many-body eigenstate $|\Psi_k\rangle$ of the interacting
system, with the corresponding eigenenergy~$E_k$. Our purpose is to
generalize the notion of localization to such many-body states.

In the coordinate representation, the many-body wave function
$\Psi_k\left(\left\{\vec\r_j\right\}_{j=1}^N\right)$ depends on the
coordinates of all $N$ particles in the system. The single-particle
states forming this many-body state are located everywhere in the
volume~$\mathcal{V}$. Thus, no definition analogous to
\req{eq:2.1} can be constructed. On the other hand, the
relation~\rref{eq:2.4} can be used, if one takes a local
additive one-particle operator:
\be
\hat{\mathbb A}(\vec{R})=\sum_{\alpha\beta}A_{\alpha\beta}(\vec{R})
\hat{c}^\dagger_\alpha\hat{c}_\beta
\equiv \int \diff^d\vec r \hat{c}^\dagger(\vec r) 
\hat{A}(\vec{R})
\hat{c}(\vec r),
\label{eq:2.3prime}
\ee
where we introduced the
fermionic creation and annihilation operators
$\hat{c}^\dagger_\alpha$, $\hat{c}^\dagger(\vec r)$, $\hat{c}_\alpha$,
$\hat{c}(\vec r)$ in the basis of the one-particle eigenstates and in
the coordinate representation, respectively.

We consider the matrix elements of the local operator between
two exact many-body eigenstates: 
\be
{\mathbb A}_{kk^\prime}(\vec R)
=\left\langle \Psi_k\left|\hat{\mathbb A}(\vec{R}) \right| \Psi_{k^\prime}\right\rangle.
\label{eq:2.3dprime}
\ee
Then we can define localized states by a relation, analogous to \req{eq:2.4}:
\be
\begin{split}
&
\cL^{\mathbb{A}}_{kk'}\left(\vec r\right)=
\int \diff^d\vec R\,
\mathbb{A}_{kk'}(\vec R )\,\mathbb{A}_{k'k}(\vec R + \vec{r})
\propto
\left\{
\begin{matrix}
\leq \eexp^{-\frac{|\vec{r}|}{\zeta(E_k)}};
& {\rm localized;}\\
\mathcal{F}\left(
\frac{|\vec{r}|}{L_{\omega_{kk'}}}
\right); & {\rm extended.}
\end{matrix}
\right.
\end{split}
\label{eq:2.4prime}
\ee
Again, $\omega_{kk'}=E_k-E_{k'}$ and
the energies $E_k$ and $E_{k'}$ are taken sufficiently close to
each other, so that the difference between $\zeta(E_k)$ and
$\zeta(E_{k'})$ need not be taken into account.\footnote{
Let us emphasize that the criterion (\ref{eq:2.3dprime}) can be extended
to the arbitrary number of electron-hole excitations,
$\hat{A}(\vec{R})\rightarrow
\hat{A}_1(\vec{R})\hat{A}_2(\vec{R})\dots\hat{A}_n(\vec{R})$,
corresponding to the local heating of the system. The asymptotic
behaviour will still be determined by Eq.~(\ref{eq:2.4prime}).
}

Equation \rref{eq:2.4prime} is the definition of the localized
many-body states. To elucidate its meaning, let calculate
the matrix element \rref{eq:2.3dprime} between two Hartree-Fock states
\be
\left| \Psi_{k}^{HF}\right\rangle
=\prod_{\alpha}
\left(1-f_\alpha^k+
f_\alpha^k
c^\dagger_\alpha\right)
\left|{\rm vacuum}\right
\rangle,
\label{eq:2.7}
\ee
where the state $|\Psi_k^{HF}\rangle$ is completely characterized by the
set of occupation numbers $f^k_\alpha=0,1$, corresponding to empty and
filled single-particle states, respectively.
Substituting \req{eq:2.7} into \req{eq:2.3prime} we find for $k\neq k^\prime$
\be
{\mathbb A}_{kk^\prime}=
\sum_{\alpha\beta} A_{\alpha\beta}
\left[f_\alpha^k(1-f_\alpha^{k'})
f_\beta^{k'}(1-f_\beta^{k})
\right]
\prod_{\gamma\neq\alpha,\beta}\delta_{f^k_\gamma,f^{k'}_\gamma},
\label{eq:2.8}
\ee
where we omitted the argument $\vec R$ in both sides of the
equation. The states $k$ and $k'$ connected by the operator
$\hat{\mathbb{A}}(\vec{R})$ are obtained from each other by creation of
a single electron-hole pair, so that only one term in the sum over
$\alpha$ and $\beta$ is different from zero.
We note that the matrix elements $A_{\alpha\beta}(\vec{R})$
are exponentially suppressed unless both states $\alpha$ and $\beta$
are located near the point~$\vec{R}$. Thus, the distance between the
electron and the hole cannot exceed the single-particle localization
length~$\zeta_{loc}$, and the number of states $k'$ which can be
connected to the given state $k$ by the local perturbation is
effectively finite, even though
the total number of the Slater determinat states (\ref{eq:2.7}) 
scales exponentially with the size of the system.

We substitute \req{eq:2.8} into the localization criterion
\rref{eq:2.4prime} and obtain
\be
\begin{split}
&\cL^{\mathbb{A}}_{kk'}\left(\vec r\right)
=
\sum_{\alpha\beta}
\cL^{{A}}_{\alpha\beta}\left(\vec r\right)
\left[f_\alpha^k(1-f_\alpha^{k'})
f_\beta^{k'}(1-f_\beta^{k})
\right]
\prod_{\gamma\neq\alpha,\beta}\delta_{f^k_\gamma,f^{k'}_\gamma}.
\label{eq:2.9}
\end{split}
\ee
Because of the $\delta$-symbols,
$|E_k-E_{k^\prime}|=|\sum_\gamma \xi_\gamma
(f_\gamma^k-f_\gamma^{k^\prime})|= |\xi_\alpha-\xi_\beta|$, where
$\alpha$~and~$\beta$ are the only two states contributing to the sum
for given $k,k'$. If $\xi_\alpha,\xi_\beta$ are not too far from the Fermi
level, so that $\zeta_{loc}(\xi_\alpha)\approx\zeta_{loc}(\xi_\beta)$, we
find that $\cL^{\mathbb{A}}_{kk'}\left(\vec r\right)$ has the same
spatial dependence as the kernel \rref{eq:2.4} for non-interacting
system. Thus, according to definition \rref{eq:2.4prime} any
Hartree-Fock state \rref{eq:2.7} is localized with the localization
length $\zeta(E_k) = \zeta_{loc}$.

\begin{subequations}
Let us apply the same ideas to the Kubo formula 
for the many-body
density-density response function
\bea
 &&\Pi\left(\omega;\vec{r}\right)
 =\sum_k P_k \Pi_{k}\left(\omega,\vec r\right)
 =\sum_{kk'} \left[P_k -P_{k'}\right]
 \Pi_{kk'}\left(\omega,\vec r\right)
 \label{eq:2.10a}\\
 &&
 \Pi_k\left(\omega,\vec r\right)=
 \sum_{k'} \left[\Pi_{kk'}\left(\omega,\vec r\right)
- \Pi_{k'k}\left(\omega,\vec r\right)
\right]
;
\\
 &&
 {\rm Im}\Pi_{kk'}\left(\omega,\vec r\right)=
 \frac{\pi}{\mathcal{V}}\cL^\varrho_{kk'}(\vec{r})
\delta\left(\omega+E_k-E_{k'}\right)
 ,
\label{eq:2.10c}
\eea
where $P_k$ is the probability for the system to be in the eigenstate
$k$ and  $\Pi_k\left(\omega,\vec r\right)$ is the linear response of
the system in this particular eigenstate. The overlap function
$\cL^\varrho_{kk'}(\vec{r})$ is given by \req{eq:2.4prime} with the
operator of local density $\hat\varrho (\vec R)=\hat{c}^\dagger(\vec
R)\hat{c} (\vec R)$.
\label{eq:2.10}
\end{subequations}

Formulas \rref{eq:2.10} are valid for any many-body eigenstates
$\left|\Psi_k\right\rangle$. For the Hartree-Fock state
\rref{eq:2.7},
we can substitute \req{eq:2.9} into \reqs{eq:2.10} and obtain
\req{eq:2.5} with
\be
f_\alpha=\sum_k P_k f_\alpha^k.
\label{eq:2.11}
\ee
For the initial Gibbs distribution
\be
P_k=\frac{\exp\left(-\frac{E_k}{T}\right)}{\sum_k\exp\left(-\frac{E_k}{T}\right)},
\label{eq:2.12}
\ee
and the Hartree-Fock spectrum, $E_k=\sum_\alpha f_\alpha^k \xi_\alpha$,
formula \rref{eq:2.11} gives the Fermi distribution function.

According to \reqs{eq:2.9} and \rref{eq:2.10c}, the spatial dependence
of the correlator $\Pi(\omega,\vec r)$ is still given by \req{eq:2.4}
for the localized single-particle wave functions. Therefore, both the
conductivity
and the diffusion coefficient vanish for the Hartree-Fock states
constructed in the basis of exact one-particle wave functions of
disordered systems. The same statement can be made about certain wave
functions which are formally very different from a single Slater
determinant. Namely, consider the creation of the electron-hole
pair on top of some eigenstate $\left|\Psi_k\right\rangle$
and expand the result in terms of other eigenstates
\begin{subequations}
\label{eq:2.13}
\be
\hat{c}^\dagger_\alpha\hat{c}_\beta \left|\Psi_k\right\rangle
 = \sum_{k'}C^{kk'}_{\alpha\beta} \left|\Psi_{k'}\right\rangle;
\quad
\sum_{k'}\left|C^{kk'}_{\alpha\beta}\right|^2=1.
\label{eq:2.13a}
\ee
For the Hartree-Fock state \rref{eq:2.7}, only one term contributes to
the sum, see \req{eq:2.9}. It is possible to check that the state
would remain localized in a sense of \req{eq:2.4prime},
if the number of terms contributing to the sum in \req{eq:2.13a}
is large but finite, {\em i.e.}
\be
\lim_\mathcal{V\to \infty}
\left[\sum_{k'}\left|C^{kk'}_{\alpha\beta}\right|^4\right]^{-1} <\infty.
\label{eq:2.13b}
\ee
In complete analogy with the non-interacting
system, this corresponds to the {\em insulating} or
{\em localized many-body} state; excitation can not propagate over 
all states allowed by the energy conservation.

Conductivity can be different from zero only if the wave functions
of the excitations can be broken onto
the  {\em infinite} number of eigenstates
\be
\lim_\mathcal{V\to \infty}
\left[\sum_{k'}\left|C^{kk'}_{\alpha\beta}\right|^4\right]^{-1} = \infty,
\label{eq:2.13c}
\ee
which would correspond to the {\em metallic} 
or {\em extended many-body} state.

Developed metallic state corresponds to the case when
electron-electron interaction mixes the excited state with
all the eigenstates in the system with close enough energy:
\be
|C^{kk'}_{\alpha\beta}|^2 \propto ``\delta(E_k
+\omega_{\alpha\beta}-E_{k'})\mbox{''},
\label{eq:2.13d}
\ee
\end{subequations}
where $\delta$-function should be understood in the thermodynamic
sense: its width, although sufficiently large to include many states,
vanishes in the limit $\mathcal{V}\to\infty$.
Only in this regime, which may also be called {\em ergodic many-body}
state, the
electron-electron interaction can bring the 
system from the intitial Hartree-Fock state
to the equilibrium correspoding to spanning all the states
permitted by the energy conservation.
In this case, the averaging over the exact many-body
eigenfunction is equivalent to averaging over the microcanonical
distribution, and temperature~$T$ can be defined as a usual Lagrange
multiplier. It is related
to~$E_k$ by the thermodynamic relation:
\begin{equation}
E_k-E_{0}=\int\limits_0^T C_V(T_1)\,\diff{T}_1,
\label{Tdef=}
\end{equation}
where $E_{0}$ is the ground state energy, and
$C_V(T)\propto\mathcal{V}$ is the specific heat. 

The main result obtained in the present paper is the proof of
existence of the extensive {\em many-body mobility edge}
$\mathcal{E}_c\propto\mathcal{V}$. This proof is based on two
statements: (i)~states with sufficiently large energies $E_k-E_0$
are extended; (ii)~states with sufficiently small energies $E_k-E_0$
are localized.

The first statement follows from the validity of high-temperature
expansion for the quantum corrections to conductivity~\cite{AA}.
The quantum corrections to
conductivity are divergent for $d=1,2$; for non-interacting systems
these weak localization corrections have no cutoff other than the
size of the system. In interacting systems two kinds of new phenomena
appear: (a)~interference due to the scattering off the self-consistent
Hartree-Fock potential (see Ref.~\cite{AAG}), and (b)~inelastic
electron scattering. Effects of type~(a) are regularized by the
temperature itself and do not produce any consequences for the present
paper. At the same time, inelastic electron scattering leads to
regularization of the weak localization corrections due to
appearence of the inelastic rate and the corresponding length:
\be
\frac{1}{\tau_\phi}\simeq\frac{\lambda^2T}{g(L_\phi)},\quad
L_\phi\simeq\sqrt{D\tau_\phi},
\label{tauphi=}
\ee
where $\lambda\lesssim{1}$ is the dimensionless interaction constant,
$D$~is the diffusion coefficient, and $g(L)$ is the dimensionless
conductance of the $d$-dimensional cube of linear size~$L$. All the
interference corrections are finite if
\be
\frac{1}{\tau_\phi}\gtrsim\delta_\zeta.
\label{tauphi<}
\ee
Using $g(\zeta_{loc})\simeq{1}$, we rewrite Eqs.~(\ref{tauphi=})
and~(\ref{tauphi<}) as
\be
T\gtrsim{T}^{(el)}\simeq\frac{\delta_\zeta}{\lambda^2}.
\label{mcondition0}
\ee
Inequality~(\ref{mcondition0}) is the condition of applicability of
the expansion from the metallic state. In fact, it is also valid for
higher-dimensional systems with the finite bandwidth~$E_b$ and all
single-particle eigenstates localized, if $E_b\gg T^{(el)}$.

At $T<T_{el}$ the perturbation theory breaks down.
It may indicate either (i) a simple insufficiency of the
perturbation theory to describe the metallic state, or (ii) existence
of the  {\em many-body mobility edge} $\mathcal{E}_c$.
It important to emphasize that $\mathcal{E}_c$ is an extensive quantity
$\mathcal{E}_c \propto \mathcal{V}$.

In case (i), states $\Psi_{k}$ are extended for all~$E_k$, so that the
conductivity would remain finite (no matter how small) down to zero
temperature.
In case (ii) the many-body eigenstates with  $E_k-E_{0}\leq\mathcal{E}_c$
are localized, see \req{eq:2.13b}.

In case (ii), the partial conductivity of one state, $\sigma_k$,
defined analogously to $\Pi_k$ from  \reqs{eq:2.10},
is zero for  $E_k-E_{0} \leq \mathcal{E}_c$.

\begin{figure}
\includegraphics[width=0.7\textwidth]{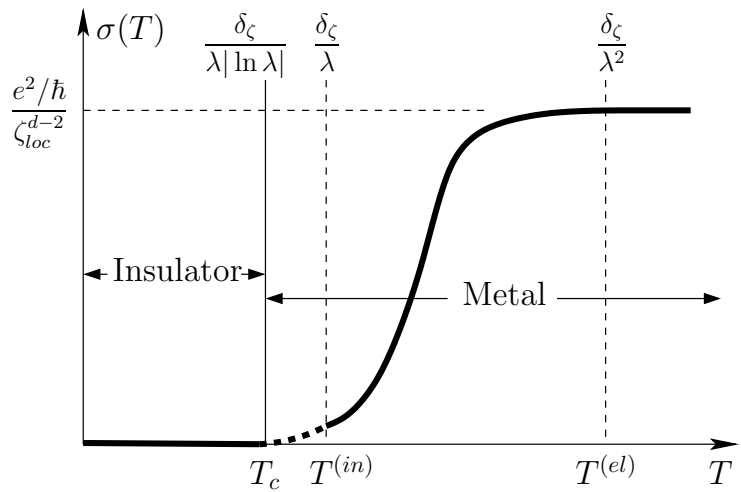}
\caption{Schematic temperature dependence of the dc conductivity
  $\sigma(T)$. Below the point of the many-body metal-insulator
  transition, $T<T_c$, $\sigma(T)=0$, as shown in
  Sec.~\ref{sec:insulator}. Temperature interval
  $T>T^{(in)}>T_c$ corresponds to the developed metallic phase, where
  Eq.~(\ref{eq:2.13d}) is valid. In this regime for the model
  described in Sec.~\ref{sec:model} $\sigma(T)$ is
  given analytically by
  Eqs.~(\ref{observables})--(\ref{results}) and plotted on
  Fig.~\ref{Plot1d}. At $T>T^{(el)}$ the high-temperature metallic
  perturbation theory of Ref.~\cite{AA} is valid.
}\label{fig:summary}
\end{figure}

Results of Ref.~\cite{AGKL} strongly indicate that the case
(ii) is realized. We will review, extend and refine the arguments of
Ref.~\cite{AGKL} in the next subsection; here, we simply proceed
with the discussion of the macroscopic manifestations of this
scenario. Let us assume that the equilibrium occupation is
given by the Gibbs distribution \rref{eq:2.12}. One could think that
it would still imply the Arrhenius law (\ref{Arrhenius}) for the
conductivity. However, this is not the case for
the many-body mobility threshold. In fact, in the limit $\mathcal{V}
\to \infty$
\begin{subequations}
\be
\sigma(T)=0;\quad T < T_c,
\label{transition1}
\ee
where the critical temperature is determined by Eq.~(\ref{Tdef=}):
\be
\label{transition2}
\int_0^{T_c} \diff{T}_1\, C_V(T_1)=\mathcal{E}_c.
\ee
\label{transition}
\end{subequations}
The schematic temperature dependence of the conductivity is
summarized on Fig.~\ref{fig:summary}.
Therefore, the temperature dependence of the dissipative 
coefficient in the system shows the singularity typical
for a phase transition.

To prove \reqs{transition}  we use the Gibbs distribution
and find
\[
\sigma(T)=\sum_k P_k \sigma(E_k)
=\frac{\int_0^\infty\diff E \eexp^{S(E)-E/T}\sigma(E)}
{\int_0^\infty \diff E \eexp^{S(E)-E/T}},
\]
where the entropy $S(E)$ is proportional to volume,
and $E$ is counted from the ground state. The integral
is calculated in the saddle point or in
the steepest decent approximations, exact for $\mathcal{V}\to\infty$. 
The saddle point $E(T)$ is
given by 
\[
\left.\frac{dS}{dE}\right|_{E=E(T)}=\frac{1}{T}\,.
\]
Taking into account $\sigma(E)=0$ for $E < \mathcal{E}_c$ we find
\[
\begin{split}
&\sigma(T)=\sigma\left[E(T)\right], \quad E(T)>\mathcal{E}_c;
\\ 
& \sigma(T) \propto \exp\left(-\frac{\mathcal{E}_c-E(T)}{T}\right);
\quad E(T)< \mathcal{E}_c
\end{split}
\]
As both energies entering the exponetial are extensive,
$E(T),\mathcal{E}_c \propto \mathcal{V}$, we obtain  \reqs{transition}.

As we already mentioned, vanishing of the dissipative conductivity at
$T<T_c$
means freezing of all relaxation processes. In particular
the microcanonical distribution could never be established for
the closed system. In this respect, the dynamics of the system
resembles the glassy state~\cite{Glass}.


To establish the thermal equilibrium in
such  insulating state requires finite
coupling of the system with the external
reservoir ({\em i.e.}, phonons). The presence of the finite
electron-phonon interaction  (as phonons are usually
delocalized), smears out the transition, and $\sigma(T) >0$ for any
temperature. Nevertheless, for the weak electron-phonon interaction,
 the phenomenon of the many-body
metal-insulator transition  \rref{transition} manifests itself as 
a very sharp crossover from phonon induced hopping at $T<T_c$ to the
conductivity independent of the electron-phonon coupling at  $T>T_c$.




\subsection{Microscopic mechanism of the many-body localization transition}

As discussed in the previous subsection, the existence of extended
many-body states at high energies is an established fact~\cite{AA}.
Here we focus on the existence of localized states at low energies,
and discuss the correspondence between a
many-electron interacting system and the Anderson model on
a certain graph. As we have already mentioned, it is
convenient to analyze many-body localization in terms of
single-particle excitations $\hat{c}^{\dagger}_\alpha|\Psi_k\rangle$
above a certain eigenstate of the interacting system, namely, how
these excitations spread over other many-bogy eigenstates
(consideration of the electron-hole excitation is performed in
a same fashion and does not bring anything qualitatively different).
This discussion has a close relation with that of Ref.~\cite{AGKL}. 

Let the system initially be in the eigenstate $|\Psi_k\rangle$.
At time $t=0$ an extra electron is created in the single-particle state
$\alpha$.
The many-body Schr\"odinger equation describing the subsequent time
evolution of such state $|\tilde{\Psi}_{k\alpha}(t)\rangle$ is
\begin{equation}
\left[i\partial_t-\hat{H}\right]|\tilde{\Psi}_{k\alpha}(t)\rangle=
\delta(t)\,\hat{c}^{\dagger}_\alpha|\Psi_k\rangle,
\label{Schrt=}
\end{equation}
where the right-hand side determines the initial condition at
$t=0$, and the Hamiltonian, written in the basis
of the exact single electron wavefunctions, is given by
\be
\hat{H}=\sum_{\alpha}
\xi_{\alpha}\,\hat{c}^{\dagger}_{\alpha}\hat{c}_{\alpha}+
\frac{1}{2}\sum_{\alpha\beta\gamma\delta}
V_{\alpha\beta\gamma\delta}\,
\hat{c}_{\alpha}^{\dagger}\hat{c}_{\beta}^{\dagger}
\hat{c}_{\gamma}\hat{c}_{\delta}\,.
\label{H=}
\ee
The interaction part of the Hamiltonian contains only non-diagonal
terms, $\alpha\neq \gamma,\delta;\ \beta\neq \gamma,\delta$,
and assumed to be antisymmetrized,
$V_{\alpha\beta\gamma\delta}=-V_{\beta\alpha\gamma\delta}
= -V_{\alpha\beta\delta\gamma}$. The diagonal matrix elements are
already included into the definition
of the Hartree-Fock spectrum $\{\xi_\alpha\}$.

We make the Fourier transform of Eq.~(\ref{Schrt=}):
\begin{equation}
\left(\ep+E_k-\hat{H}\right)|\tilde{\Psi}_{k\alpha}(\ep)\rangle=
\hat{c}^{\dagger}_\alpha|\Psi_k\rangle
\label{Schrep=}
\end{equation}
(here we count the energy $\ep$ from that of the reference state $E_k$),
and solve this equation for $|\tilde{\Psi}_{k\alpha}(\ep)\rangle$ by iterations:
\begin{equation}
|\tilde{\Psi}_{k\alpha}(\ep)\rangle=\frac{1}{\ep-\xi_\alpha}
\left(\left|\psi_{k\alpha}^{(0)}(\ep)\right\rangle
+\left|\psi_{k\alpha}^{(1)}(\ep)\right\rangle+\ldots\right).
\label{psiexpansion}
\end{equation}
The zeroth order in the electron-electron interaction is just
the one-particle excitation itself:
\begin{equation}
\left|\psi_{k\alpha}^{(0)}\right\rangle=
\hat{c}^{\dagger}_\alpha|\Psi_k\rangle.
\end{equation}
The first order corresponds to a three-particle excitation above
$|\Psi_k\rangle$:
\begin{equation}\begin{split}
\left|\psi_{k\alpha}^{(1)}\right\rangle&=
\sum_{\beta,\gamma,\delta}
\frac{V_{\delta\gamma\beta\alpha}}
{\ep-\Xi_{\gamma\delta}^\beta}\,
\hat{c}^{\dagger}_\delta\hat{c}^{\dagger}_\gamma\hat{c}_\beta|\Psi_k\rangle,
\label{Psi1=}
\end{split}\end{equation}
where we introduced a shorthand notation
\begin{equation}
\Xi_{\gamma\delta}^\beta=\xi_\gamma+\xi_\delta-\xi_\beta
\end{equation}
for the energy of the three-particle excitation.

As usual in the theory of metal-insulator transition~\cite{Anderson58},
one starts from estimating the probability for the 
sum \rref{Psi1=} not to be small. 
We notice that the geometric distance between particles is of the
order of the one-particle localization length $\zeta_{loc}$,
as the interaction is short-ranged.
Thus, the denominators in \req{Psi1=}
are the random quantities with the maximal value of the order
of the level spacing in the localization length $\delta_\zeta$.
Assuming weak interaction,
$\lambda={\rm max |V_{\alpha\beta\gamma\delta}|}/\delta_\zeta \ll 1$,
we see that the value of sum is, in fact, determined by
the smallest denominator. 

Let us assume that the typical number of terms ({\em i.e.}, terms not
suppressed by the matrix elements) is $K \gg 1$.
As the denominators are distrubited approximately uniformly
for $-\delta_\zeta < \ep-\Xi_{\gamma\delta}^\beta < \delta_\zeta$,
the smallest denominator is of the order of $\delta_\zeta/K$.
Therefore, if $K\lambda \ll 1$, the probability to find
the large mixing is small, whereas for 
\be
K\lambda \gtrsim 1
\label{Klambda}
\ee
it becomes of the order of unity. This estimate
up to a factor of the order of $|\ln \lambda|$ is the basis for finding
the position of the transition~\cite{Anderson58}.

The remaining  non-trivial problems are: (i) to find the connectivity
$K$, and (ii) to check that higher-order terms 
indeed match the locator structure of Anderson, {\em i.e.}, the
the number $\mathcal{N}_n$ of  the $n$th order terms  scales as 
\be
\mathcal{N}_n \simeq K^n,
\label{calN=}
\ee
where the prefactor can be any algebraic function of $n$.
Condition \rref{calN=} is very important:  if   $\mathcal{N}_n
\gg K^n$, {\em e.g.},  $\mathcal{N}_n \simeq n!$, then the probability
 to find small enough denominators in high orders of the
 perturbation theory $\mathcal{N}_n\lambda^n\simeq 1$ 
would be always of the order
of unity no matter how small the interaction constant $\lambda$ is;
the system in this case would always remain metallic.
If, oppositely, the number of terms in high orders is small,
{\em e.g.},   $\mathcal{N}_n \simeq K^n/n!$, the higher-order terms
will not contain the small denominators, even though the lowest
orders of the perturbation theory do. In this case, the system 
would always remain insulating. In both cases, the conlusions drawn
from the lowest order perturbation theory~\cite{AGKL} would be misleading.

To estimate $K$ from \req{Psi1=}, one has to account for
 (i) the
structure of the eigenfunction $\left|\Psi_k\right\rangle$, {\em i.e.},
which  indices $\alpha$ of fermionic operators ${\hat c}_\alpha$ are
allowed not to produce zero result after action on the wave function;
(ii) the structure of the matrix elements $V_{\alpha\beta\gamma\delta}$.

The energy of a typical Hartree-Fock state is almost uniformly
distributed   among the localization volumes. The fact that the
eigenstate $\left|\Psi_k\right\rangle$ is a linear combination
of an infinite number of  Hartree-Fock states does not affect this
assumption. Therefore, the state $\left|\Psi_k\right\rangle$ is
contributed by electron-like and hole-like excitations with the energy
smaller than $T$, where $T$ is of the order of temerature given by
\req{Tdef=}. Considering the action of the fermionic operators on such
a state we estimate
\be
-T < \xi_{\gamma},\xi_\delta; \quad
\xi_\beta < T.
\label{Trestriction}
\ee

Second restriction is based on the structure of the matrix elements.
As we already mentioned, they restrict the final states to be at
the distance not exceeding the localization length $\zeta_{loc}$
from each other. In addition to this spatial restriction,
the matrix elements decrease rapidly when the energy difference,
say $\xi_\alpha-\xi_\gamma$, exceeds the level spacing $\delta_\zeta$,
{\em i.e.}, $V_{\alpha\beta\gamma\delta}\simeq \lambda\delta_\zeta$
only for
\be\begin{split}
&|\xi_\alpha-\xi_\delta|,|\xi_\beta-\xi_\gamma|\lesssim\delta_\zeta
\quad{\rm or}\quad
|\xi_\alpha-\xi_\gamma|,|\xi_\beta-\xi_\delta|\lesssim \delta_\zeta,
\label{Mrestriction=}
\end{split}\ee
and vanish rapidly otherwise.

Counting the number of states ${\beta,\gamma,\delta}$ within one
localization volume, such that they produce the value of the denominator
in \req{Psi1=} smaller than $\delta_\zeta$.
and satisfy both restrictions \rref{Trestriction} -- \rref{Mrestriction=},
we obtain
for $T \gg \delta_\zeta$\footnote{The estimate of Ref.~\cite{Mirlin97}
for the quantum dot at $T=0$ 
is $K\simeq \ep^2/\delta^2$, where $\delta$ is the level
spacing in the quantm dot.
Our estimate is different because the energy restriction 
on the matrix elements
\rref{Mrestriction=} is absent in a quantum dot.
}
\begin{equation}
K\simeq\frac{T}{\delta_\zeta}.
\label{connectivity=}
\end{equation}
Comparing \req{connectivity=} with the estimate 
\rref{Klambda} we conclude that at $T\gtrsim \frac{\delta_\zeta}{\lambda}$
the first correction to the wavefunction \rref{Psi1=} contains at least one
term of the order of unity, which may signify the transition
in the system.

\begin{figure}
\includegraphics[width=0.7\textwidth]{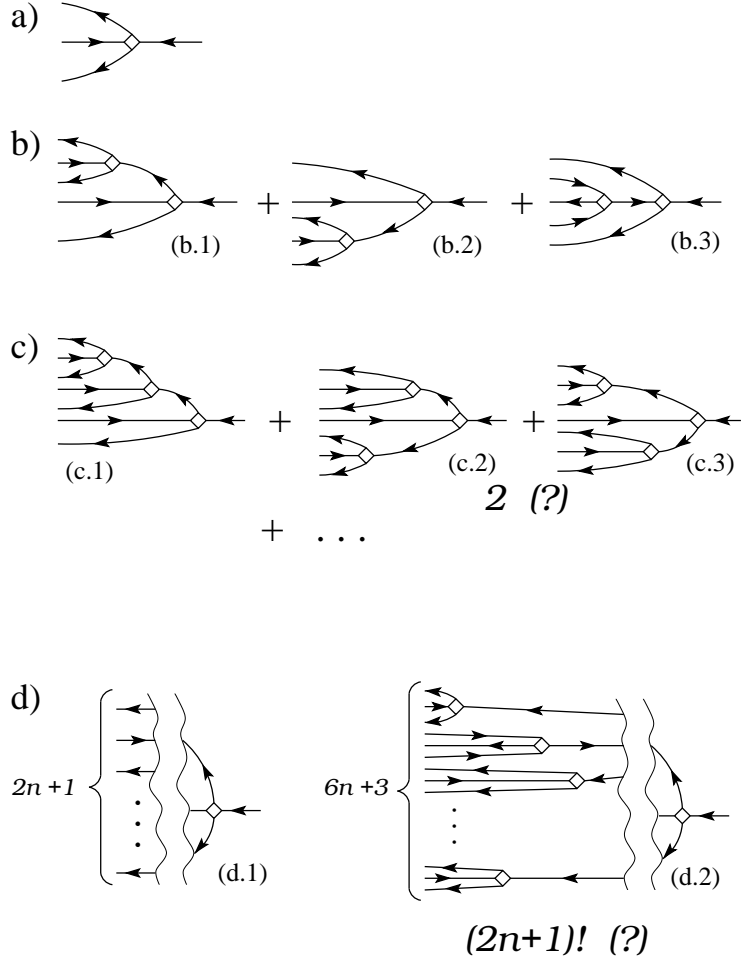}
\caption{Pictorial representation of the paths corresponding
to the first (a), second (b), third (c),
and higher (d) orders of the perturbation theory for the mixing of
one-particle excitations with many-particle states. 
Process (d.2) is obtained from the process (d.1) by the
application of the interaction operators once to each of (2n+1)
final particles of the process (d.1). The sequence of such applications 
can be ordered in $(2n+1)!$ ways, each producing formally different path.}
\label{fig:factorial}
\end{figure}

To check that this conclusion is not an artefact of the lowest order
perturbation theory, one has to analyze further terms in the
expansion~\rref{psiexpansion}.
The second order correction is given by
\begin{equation}\begin{split}
\left|\psi_{k\alpha}^{(2)}\right\rangle&=
\sum_{\alpha_1\beta_1}
\sum_{\beta,\gamma,\delta}
\frac{V_{\alpha_1\beta_1\gamma\delta}}
{\ep-\Xi_{\alpha_1\beta_1}^\beta}\,
\frac{V_{\delta\gamma\beta\alpha}}
{\ep-\Xi_{\gamma\delta}^\beta}\,
\hat{c}^{\dagger}_{\alpha_1}\hat{c}^{\dagger}_{\beta_1}
\hat{c}_\beta|\Psi_k\rangle\\
&+\sum_{\alpha_1,\beta_1,\gamma_1}
\sum_{\beta,\gamma,\delta}
\frac{2V_{\alpha_1\beta_1\gamma_1\delta}}
{\ep-\Xi_{\gamma\alpha_1\beta_1}^{\beta\gamma_1}}\,
\frac{V_{\delta\gamma\beta\alpha}}
{\ep-\Xi_{\gamma\delta}^\beta}\,
\hat{c}^{\dagger}_{\alpha_1}\hat{c}^{\dagger}_{\beta_1}
\hat{c}^\dagger_\gamma\hat{c}_\beta\hat{c}_{\gamma_1}|\Psi_k\rangle
\\
&+\sum_{\alpha_1,\gamma_1,\delta_1}
\sum_{\beta,\gamma,\delta}
\frac{V_{\alpha_1\beta\gamma_1\delta_1}}
{\ep-\Xi_{\gamma\delta\alpha_1}^{\gamma_1\delta_1}}\,
\frac{V_{\delta\gamma\beta\alpha}}
{\ep-\Xi_{\gamma\delta}^\beta}\,
\hat{c}^{\dagger}_{\alpha_1}\hat{c}^{\dagger}_{\delta}
\hat{c}^\dagger_\gamma\hat{c}_{\beta_1}\hat{c}_{\gamma_1}|\Psi_k\rangle,
\label{Psi2=}
\end{split}\end{equation}
where the $(2n+1)$-particle energies are defined as
\begin{equation}
\Xi_{\alpha_1\dots\alpha_{n+1}}^{\beta_1\dots\beta_n}=
\sum_{i=1}^{n+1}\xi_{\alpha_i}-\sum_{i=1}^{n}\xi_{\beta_i}.
\label{newXi}
\end{equation}
The first term in Eq.~(\ref{Psi2=}) is again a three-particle
excitation. We can estimate its typical value as $\lambda^2T$,
which is smaller than the first-order contribution~(\ref{Psi1=}) by a
factor $\lambda\ll{1}$; similar type terms arise also from
the second and the third lines for $\hat{c}^\dagger,\hat{c}$
with coinciding indices.  The actual meaning of
this terms is the renormalization of the value of the two-particle
interaction strength.  They are analyzed with more rigor in
Sec.~\ref{sec:renint}, here we simply neglect them as they do not
affect the statement about the transition.

Once again, we estimate the number of the relevant terms in 
the multiple sum in the last two lines \req{Psi2=}.
They are shown pictorially on Fig.~\ref{fig:factorial}b.
The structure of the state $\left|\Psi_k\right\rangle$ gives
the restriction similar to \req{Trestriction} on
the electron (entering \req{newXi} with plus sign) and the hole  
(entering \req{newXi} with minus sign) 
energies respectively. The matrix element
$V_{\alpha\beta\gamma\delta}$  restricts
the energies by \req{Mrestriction=}. Combining those two restrictions
and requiring  each denominator in \req{Psi2=} to be smaller than
the level spacing $\delta_\zeta$, one estimates
\be
\mathcal{N}_2 \simeq K^2,
\label{K2}
\ee 
where $K$ is given by \req{Klambda}.
Aparently, \req{K2} agrees with the requirement~\rref{calN=}. However
some questions may arise already on this stage.

One notices from \req{Psi2=} that there are terms which
can be obtained from each other by the simple permutation of indices
of the creation or annihilation operators. The states obtained by
such permutations from each other are in fact identical, so one may
be tempted to amend \req{K2} as~\cite{Mirlin97}
\be
\mathcal{N}_n \stackrel{?}{\simeq} \frac{K^n}{n!(n+1)!}, 
\label{K?}
\ee
where the combinatorial factors describe the indistinguishability
among $n+1$ electrons and $n$ holes [$n=2$ for \req{Psi2=}].

However, in the theory of Anderson transition~\cite{Anderson58},
the relevant parameter is not the number of available final states,
but the number of statistically independent paths leading to
these states from the initial state ({\em e.g.}, for the
Anderson transition in three dimensions the number of states grows
as $n^3$ whereas the number of paths grows exponentially with $n$).
By construction,
the summation in \req{Psi2=} is performed over the single-particle
indices and not over the final states. 
Moreover, the denominators in each terms involve different
combinations of levels and thus produce statisitically independent
contributions.
This corresponds to the
summation over paths number of which cancels the factorials in
\req{K?}.

Having demonstrated the absence of the factorial factors 
suppressing the number of terms $\mathcal{N}_n$, one may
worry about the opposite extreme: factorial growth
of~$\mathcal{N}_n$. The way to obtain this factorial is shown on
Fig.~\ref{fig:factorial}d. The transition
between the two states with $(2n+1)$ excitations and $(6n+3)$
excitations may be realized in $(2n+1)!$ ways. These ways
are different by the order in which interaction operator
acts on the particles, thereby producing different intermediate
states. If these paths were statistically independent,
it would mean 
\be
\mathcal{N}_{3n+1} \stackrel{?}{\simeq} (2n+1)! K^{2n+1}\mathcal{N}_{n}, 
\label{K??}
\ee
so that no transition would occur and the system would always 
be delocalized.

However, such paths are, in fact, correlated and the sum of the
corresponding amplitudes produces always the result of the order
of an amplitude of a single path; the terms essentially cancel each other. 
To demonstrate this cancellation we consider the third order of
the perturbation theory. The terms schematically shown on
Fig.~\ref{fig:factorial}c correspond to
\begin{equation}
\begin{split}
\delta\left|\psi_{k\alpha}^{(3)}\right\rangle  = &
\sum_{\alpha_2,\beta_2,\gamma_2}
\sum_{\alpha_1,\beta_1,\delta_1}\sum_{\beta,\gamma,\delta}
\frac{2V_{\alpha_2\beta_2\gamma_2\delta}}
{\ep-\Xi_{\alpha_1\beta_1\alpha_2\beta_2}^{\beta\delta_1\gamma_2}}
\left(\frac{V_{\alpha_1\beta_1\gamma\delta_1}}
{\ep-\Xi_{\delta\alpha_1\beta_1}^{\beta\delta_1}}
+\frac{V_{\alpha_1\beta_1\gamma\delta_1}}
{\ep-\Xi_{\gamma\alpha_2\beta_2}^{\beta\gamma_2}}\right)
\\ &\times
\frac{V_{\delta\gamma\beta\alpha}}
{\ep-\Xi_{\gamma\delta}^\beta}\,
\hat{c}^{\dagger}_{\alpha_2}\hat{c}^{\dagger}_{\beta_2\textmd{}}
\hat{c}^{\dagger}_{\alpha_1}\hat{c}^{\dagger}_{\beta_1}
\hat{c}_\beta\hat{c}_{\delta_1}\hat{c}_{\gamma_2}|\Psi_k\rangle.
\end{split}
\label{Psi3=}
\end{equation}
Two terms in the second factor correspond to  two ways to obtain the
final seven-particle state. One immediatley notices that
the matrix elements in the two terms are the same. Moreover, the
denominators in two terms can be combined as 
\be
\frac{1}{\ep-\Xi_{\delta\alpha_1\beta_1}^{\beta\delta_1}}
+\frac{1}{\ep-\Xi_{\gamma\alpha_2\beta_2}^{\beta\gamma_2}}=
\frac{\ep-\Xi_{\gamma\delta}^\beta
+\ep-\Xi_{\alpha_1\beta_1\alpha_2\beta_2}^{\beta\delta_1\gamma_2}}
{\left(\ep-\Xi_{\delta\alpha_1\beta_1}^{\beta\delta_1}\right)
\left(\ep-\Xi_{\gamma\alpha_2\beta_2}^{\beta\gamma_2}\right)}.
\label{Psi3=prime}
\ee
Generally, one seeks to maximize the transition amplitude by choosing
the smallest possible denominators for each of the three factors in
Eq.~(\ref{Psi3=}) independently. Each denominator being of the order of
$\delta_\zeta/K$, the whole expression would be proportional to~$K^3$.
However, one can see immediately from
\req{Psi3=prime} that small denominators of the first and the
third factors of \req{Psi3=} appear in the numerator of the second factor.
Thus the amplitudes shown Fig.~\ref{fig:factorial} (c.2)--(c.3) are not
indepenedent, and in fact, cancel each other, producing a contribution
proportional to~$K^2$ instead of~$K^3$. One can follow
similar cancellations in any order of the perturbation theory.
Therefore, the estimate \rref{K??} is not valid and scaling
\rref{calN=} remains intact. Therefore,
one can use the Anderson result~\cite{Anderson58} for the critical
connecitivity and obtain the estimate for the transition temperature
$T_c$
\be
K \simeq \frac{T_c}{\delta_\zeta}
\simeq 
\frac{1}{\lambda\left|\ln{\lambda}\right|};
\label{Testim}
\ee
the many-body mobility edge $\mathcal{E}_c$ is then found from
\req{transition2}.
The qualitative estimate~\rref{Testim} is in agreement with the
result of a quantitative calculation performed in subsequent sections
for a specific model [Eq.~(\ref{Tc4prime})].

The arguments given in this section show that, even though the
many-body problem~\rref{H=} exhibits a close analogy with the problem
of Anderson localization~\cite{Anderson58}, an exact mapping to the
Anderson model on some graph, like one suggested in
Ref.~\cite{AGKL}, is problematic. First, the geometric structure of this
graph is unknown. Analogy with the Cayley tree, popular due to the
exact solvability of the corresponding Anderson
model~\cite{AbouChacra},
is not applicable, strictly
speaking, because in the many-body problem two states can be connected
in more than one way, which would be prohibited for the Cayley tree.
Second, the many-body problem does not allow for simple counting of
statistically independent paths, which was the
main idea of the Anderson's solution~\cite{Anderson58}, as transition
amplitudes, corresponding to seemingly different paths, exhibit
striking correlations.
Therefore, the appropriate starting point is the formalism
where the cancellation of the factorial terms is taken
into account automatically. This formalism is in fact none
but the well-known diagrammatic technique for the many-body
system~\cite{AGD}, as one diagram includes the sum of all the processes
obtained by trivial permutations. The correlations between
different diagrams are much weaker and can be treated perturbatively.
The subsequent sections are devoted to the statistical analysis for
the many-body Green functions in a basis of the localized
one-particle states. The transition resembling the Anderson's
transition on the Cayley tree~\cite{AbouChacra,Efetov87} will be indeed
demonstrated,
despite of the subtleties discussed above.

\section{Choice of the model}\label{sec:model}

The purpose of this section is to introduce the simplest
model describing the metal-insulator transition for interacting
electrons as a coarse-grained version of the initial Hamiltonian of
interacting electrons in the disorder potential.

The Hamiltonian $\hat{H}$ of electrons placed in a disorder potential
$U(\r)$ and interacting with each other via a two-body interaction
potential
$V(\vec{r},\vec{r}^\prime)=V(\vec{r}^\prime,\vec{r})$
can be written in the basis
of the {\em exact} single-electron wavefunctions\footnote{For
  simplicity,
we consider the electrons as the spinless
fermions. 
There is no reason to believe that the straightforward
inclusion of the spin degrees of freedom 
affects qualitatively the final conclusions.}
as
\begin{subequations}
{\setlength{\arraycolsep}{0pt}
\label{eq:1.1}
\bea
&&\hat{H}=\hat{H}_0+\hat{V}_{int};
\label{eq:1.1a}
\\
&&\hat{H}_0=\sum_{\alpha}
\xi_{\alpha}\,\hat{c}^{\dagger}_{\alpha}
\hat{c}_{\alpha};\\
&&\hat{V}_{int}=
\frac{1}{2}\sum_{\alpha\beta\gamma\delta}
V_{\alpha\beta\gamma\delta}
\hat{c}_{\alpha}^{\dagger}\hat{c}_{\beta}^{\dagger}
\hat{c}_{\gamma}\hat{c}_{\delta};
\\
&&V_{\alpha\beta\gamma\delta}
=\frac{1}{2}\int
V(\vec{r},\vec{r}')
\varrho_{\alpha\delta}(\vec{r})
\varrho_{\beta\gamma}(\vec{r}')
\diff^d\vec{r}\diff^d\vec{r}';
\label{eq:1.1int}
\\
&&\varrho_{\alpha\delta}(\vec{r})= \phi^*_{\alpha}(\vec{r})
\phi_{\delta}(\vec{r}),
\label{eq:1.1rho}
\eea
where  $\hat{c}^{\dagger}_{\alpha}$~and~$\hat{c}_{\alpha}$ are the
fermion creation and annihilation operators:
\[
\left\{\hat{c}^{\dagger}_{\alpha}; \hat{c}_{\beta}\right\}
=\delta_{\alpha\beta},\quad
\left\{\hat{c}_{\alpha}; \hat{c}_{\beta}\right\}=
\left\{\hat{c}^\dagger_{\alpha}; \hat{c}^\dagger_{\beta}\right\}
=0,
\]
and $\{\dots;\dots\}$ stand for the anticommutator.
}

Index $\alpha$ labels the one-particle state and $\xi_\alpha$
is the corresponding eigenvalue:
 \be
 \left[-\frac{\vec{\nabla}^2}{2m} + U(\r) -\ep_F\right]\phi_\alpha(\r) =
 \xi_\alpha\phi_\alpha(\r).
 \label{eq:1.1e}
 \ee
\end{subequations}

To characterize the single-particle spectrum,
we introduce average density of states (DoS) per unit volume
 \be
 \nu = \frac{1}{\mathcal{V}}\sum_\alpha \langle \delta(\xi_\alpha) \rangle
 \label{eq:1.1f}
 \ee
 where $\langle\dots \rangle$ denote the averaging over the disorder
realization and $\mathcal{V}$ is the volume of $d$-dimensional system.

Both one-particle energies and interaction matrix
elements are random quantities which are
functionals of the random  potential $U(\vec{r})$.
The usual program~\cite{AA} of averaging over the disorder realizations
faces difficulty when the relevant spatial scale becomes
comparable with the one-particle localization length
$\zeta_{loc}$. However, we would like to discuss
the physics associated with length scale much larger
than $\zeta_{loc}$.
To perform this task, we adopt the reduced
statistical model for the matrix elements and eigenfunctions
which, however,  keeps all essential physics intact.
We will not discuss how the parameters
of this low-energy effective model are connected to the properties
of the system in the high-temperature regime~\cite{AA}.

To write down this model, we use the following
 properties of the localized non-interacting
system~\cite{wavefunctions}:
\begin{enumerate}[{(}i{)}]
\item

 The typical wavefunctions have  exponential envelopes,
\be
- \ln |\phi_\alpha(\r)| =
 \frac{\left|\r-\vec{\rho}_\alpha\right|}{\zeta_{loc}},
\label{zetaloc}
\ee
where $\zeta_{loc}$ is
the localization length, and $\vec{\rho}_\alpha$ characterizes the
position
of the ``center of mass'' of the wavefunction.\footnote{Everywhere
we assume that at the distances
smaller than the localization length $\zeta_{loc}$ the wave functions
are well described by the semiclassical approximation, {\em i.e.},
$(m\ep_F)^{1/2}\zeta_{loc} \gg 1$. We will also neglect
the weak energy dependence of the localization length
as well as of the average DoS in \req{eq:1.1f}.}

\item

Levels $\xi_\alpha$, $\xi_\beta$ repel each other
for $|\xi_\alpha-\xi_\beta| \lesssim \delta_\zeta \exp\left[-
\frac{\left|\vec{\rho}_\beta - \vec{\rho}_\alpha\right|}{\zeta_{loc}}
\right]$, and are almost independent otherwise. Here we introduce
the main energy scale of the problem:
\be
\delta_\zeta \equiv \frac{1}{\nu\zeta_{loc}^d},
\label{deltaloc}
\ee
which has the meaning of one-particle level spacing on the
localization length ($d$ is the dimensionality of the system);

\item

The overlap between wavefunctions decays exponentially
with the distance, see \req{zetaloc}, whereas the overlap between
the wavefunctions with the centers of mass at the distance
much smaller than the localization  length strongly depends on
the corresponding one-particle energies $\xi_{\alpha}$.
In particular, for the estimate of the contribution to the interaction
matrix elements \rref{eq:1.1int} from the distances smaller than the
localization length one can approximate functions $\varrho(\vec{r})$
from \req{eq:1.1rho} as  Gaussian variables with the correlation
function\footnote{We will neglect the correlation of the wavefunctions
stemming from the Cooperon contributions, {\em i.e.}, assume the
unitary ensemble}
\be
\begin{split}
&\langle
\varrho_{\alpha\beta}(\vec{r}_1)\varrho_{\alpha\beta}^*(\vec{r}_2)
\rangle \simeq
\frac{
  \delta_{\zeta}
{\Pi}_d\left(\frac{\left|\r_1-\r_2\right|}{L_{\omega_{\alpha\beta}}}\right).
}{\omega_{\alpha\beta}\zeta_{loc}^d L_{\omega_{\alpha\beta}}^d}
\end{split}
\label{diffuson}
\ee 
where $\omega_{\alpha\beta}\equiv |\xi_\alpha-\xi_\beta|$,
diffusion length is given by $L_{\omega}=\left(D/\omega\right)^{1/2}$,
and $D$ is the classical diffusion coefficient. Dimensionless diffuson
${\Pi}_d(x)$ is given by
\[
{\Pi}_d(x) \equiv {\rm Re}
\int \frac{\diff^d\vec{Q}}{(2\pi)^d} \frac{\eexp^{iQ_xx}}{i+{Q}^2},
\]
and it decays exponentially at $x \gg 1$. Equation (\ref{diffuson}) is
valid provided $|\r_{1,2} - \vec{\rho}_{\alpha,\beta}| \ll \zeta_{loc}$
so that the results obtained for the metallic states are 
applicable\footnote{For $d\geq 3$ Anderson model with
all the one-particle states localized  the wavefunctions
at distances smaller than $\zeta_{loc}$ are critical rather than
metallic. It does not affect the final form of the effective model
proposed in this section.}.
It is important to emphasize that \req{diffuson} implies the relation
between the spatial distance and the energy transfer and does not
impose any restrictions on the values of energy themselves.
\end{enumerate}

{\setlength\arraycolsep{-7pt}
\begin{subequations}
\label{Heff}
These facts suggest the following coarse-grained version of the
Hamiltonian \rref{eq:1.1}. We
discretize the space into a $d$-dimensional cubic lattice
with the lattice constant $\zeta_{loc}$, the coordinate
of each site will be labeled as $\vec{\rho}$.
We will call the unit cell of this lattice a {\em localization cell}.
Each localization cell, $\vec{\rho}$,
 contains large number of levels, $N\to \infty$, 
labeled
by integer $l$,
\be
-\frac{N\delta_\zeta}{2}< \xi_{l}(\vec{\rho}) < \frac{N\delta_\zeta}{2};
\quad 1\leq l \leq N.
\label{delta}
\ee
Two levels $\xi_l(\vec{\rho}_1)$, $\xi_m(\vec{\rho}_2)$ are independent
for  $\vec{\rho}_1\neq \vec{\rho}_2$ and
repel each other otherwise. To characterize this repulsion,
let us consider the probability, $P(n,E)$, to find $n$ levels
in the energy interval of the width $E \gg \delta_\zeta$.
We will write this probability in the form
\be
P(n,E)=
\left[\frac{\eexp^{-E/\delta_\zeta}}{n!}
\left(\frac{E}{\delta_\zeta}\right)^n\right]
\exp\left[-\mathcal{P}\left(\frac{n\delta_\zeta}{E}\right)\right].
\label{levelrepulsion}
\ee
The first factor in this expression characterizes the Poisson
distribution of the independent random levels. The last exponential
describes the level repulsion. The precise functional form of this
repulsion is not important for us,  we require only
\[
\lim_{x\to \infty}x^{-1}\mathcal{P}(x) =\infty,
\]
which is a natural assumption for any repelling levels
with the scale of the repulsion determined by $\delta_\zeta$.

As follows from previous discussion of the structure of the wave functions,
the interaction matrix elements are largest for the states
belonging to one localization cell, 
and decay exponentially with the distance. Thus,
we will take into account only interaction 
within one localization cell. 

The interaction within one cell, however, can not cause
the delocalization in space and, in fact, does not cause
any qualitative effects at all. Therefore, we will take into account
the single-electron hopping from one localization cell $\vec{\rho}_1$ to
its nearest neighbour $\vec{\rho}_2$.
Being small,
this hopping does not change the localization properties
of the single electron wave function, however, taken together
with the electron-electron interaction would eventually lead to the
transiton.

The resulting Hamiltonian, thus, takes the form
\bea
&&{\hat{H}=\hat{H}_0+\hat{V}_{int}};
\label{Heffb}\\
&&\hat{H}_0=\sum_{\vec{\rho},l}\,
\hat{c}^{\dagger}_{l}(\vec{\rho})
\left[\xi_{l}(\vec{\rho})
\hat{c}_{l}(\vec{\rho})
 +I\delta_\xi
\sum_{\vec{a},m}
\hat{c}_{m}(\vec{\rho}+\vec{a})\right].
\\
&&\hat{V}_{int}=\frac{1}{2}
\sum_{\substack{l_1l_2j_1j_2;\vec{\rho}}}
V_{l_1l_2}^{j_1j_2}(\vec{\rho})
\hat{c}_{l_1}^{\dagger}(\vec{\rho})
\hat{c}_{l_2}^{\dagger}(\vec{\rho})
\hat{c}_{j_2}(\vec{\rho})
\hat{c}_{j_1}(\vec{\rho}),
\eea
where 
$\left\{\hat{c}^{\dagger}_{i}(\vec{\rho}_1); \hat{c}_{j}(\vec{\rho}_2)\right\}
=\delta_{ij}\delta_{\vec{\rho}_1\vec{\rho}_2}$,
$\left\{\hat{c}_{i}(\vec{\rho}_1); \hat{c}_{j}(\vec{\rho}_2)\right\}=
\left\{\hat{c}^\dagger_{i}(\vec{\rho}_1);
\hat{c}^\dagger_{j}(\vec{\rho}_2)\right\}
=0$.

Here $\vec{a}$ are the vectors connecting the
cell~$\vec\rho$ to its nearest neighbors.
Dimensionless hopping parameter, $I$, such that
\be
I \ll \frac{1}{2d\ln 2d},
\label{Ismall}
\ee
is introduced to control further perturbative expansion.
We will chose $I > 0$: this choice does not affect
any conclusions, as $I$ will connect terms with
the random signs.

The antisymmetrized coefficients $V_{l_1l_2}^{j_1j_2}(\vec{\rho})=
V_{l_2l_1}^{j_2j_1}(\vec{\rho})=-V_{l_1l_2}^{j_2j_1}(\vec{\rho})$
are random  numbers. Because the physical processes
discussed in this paper are associated with the
counting of the resonant denominators, the particular choice of the
statisitical distribution of $V_{l_1l_2}^{j_1j_2}(\vec{\rho})$
is not really important. For the calculational convenience
we choose the binary distribution
\be
\begin{split}
 V_{l_1l_2}^{j_1j_2}
= \frac{\lambda \delta_{\zeta}\sigma_{l_1}^{j_1}
\sigma_{l_2}^{j_2}
}{2}
\Upsilon\left(\frac{\omega_{j_1l_1}}{\delta_\zeta}\right)
\Upsilon\left(\frac{\omega_{j_2l_2}}{\delta_\zeta}\right)
-\left(l_1\leftrightarrow l_2\right),
\end{split}
\label{Heffintdist}
\ee
where $\lambda \simeq I \ll 1$ is a dimensionless parameter allowing to
control the perturbative expansion, $\omega_{lj}=\xi_l-\xi_j$
and we omitted argument $\vec\rho$ on both sides of the
equation. We chose $\lambda>0$ without loss of the generality,
as all the effects which will be considered are not sensitive to
the sign of $\lambda$.
Function $\Upsilon(x)$
is introduced to describe
the interaction  decaying
rapidly with the distance between
the levels in the energy space, see \req{diffuson}.
As the smallest linear scale in the reduced model is $\zeta_{loc}$
the maximal {\em energy transfer} which is permissible to consider 
in the model is of the order of 
$\delta_\zeta$. Thus, we
choose\footnote{It can be shown that taking into account
the algebraic decay of $\Upsilon(x),\ x\gg 1$ would require
the consideration of the spatial correlation of the wave functions,
so the approximation for the interaction matrix elements to be
independent of each other and of the hopping would be
false even on a qualitative level.}
\be
\begin{split}
& \Upsilon(x)=\theta\left(\frac{M}{2}-|x|\right);
\quad  1 \ll M
\lesssim \frac{1}{\sqrt{\lambda}}.
\end{split}
\label{Upsilon}
\ee
Actual value of the parameter $M$ is not well defined and we will
consider it as the initial data for the effective Hamiltonian.
The significance of the upper bound to $M$ for the consistency of the
theory will be clear later, see discussion after \req{mcondition2} as
well as Sec.~\ref{sec:renint}.

The signs for different
wave functions are not correlated, so
\be
\begin{split}
&\left[\sigma_{l}^{j}(\vec{\rho})\right]^2=1;\quad
\langle
\sigma_{l}^{j}(\vec{\rho})
\sigma_{l^\prime}^{j^\prime}
(\vec{\rho}^\prime)
\rangle
 = \delta_{\vec{\rho},\vec{\rho}^\prime}
\delta_{l l^\prime}\delta_{j j^\prime}
.
\end{split}
\label{sigmas}
\ee

Equations \rref{Heff} constitute a complete formulation of the
reduced model for the interacting electrons in system with
localized one-particle states. This model will be analyzed in
the subsequent sections to show the stability of both
high-temperature phase (metal) and low-temperature phase (insulator).
\end{subequations}}

\section{Formalism.}
\label{sec:Formalism}

The purpose of this section is to describe the machinery
which enables us to put the previous qualitative arguments
into the context of the usual many-body theory of non-equlibrium
systems. We will start with the outline of the Keldysh formalism
for the exact (non-averaged) Green functions
corresponding to Hamiltonian \rref{Heff} in Sec.~\ref{sec:Keldysh}
and formulate which
quantity describes the metal-insulator transition in Sec.~\ref{sec:problem}.
Next, in Sec.~\ref{sec:SCBA} we will describe our main working
approximation which
corresponds to the summation of all the rainbow diagrams (SCBA).
The justification of the validity of this approximation for
the description of the transition is postponed
until Sec.~\ref{sec:validity}.

\subsection{Time evolution equations and basic definitions.}
\label{sec:Keldysh}

We intend to describe both metallic and insulating regimes.
In the latter regime
relaxation dynamics is absent, there is no mechanism to establish
the thermal equilibrium, and the temperature itself is not
defined. Therefore, the only appropriate formal
framework is the non-equilibrium
(Keldysh) formalism~\cite{Keldysh}.
We define the corresponding Green functions as
\be
\begin{split}
&\mathcal{G}^R_l(t_1, t_2;\vec{\rho}) =
-i\theta(t_1-t_2)\lda \left\{\hat{c}_l (t_1,\vec{\rho} ) ;
\hat{c}_l^
\dagger(t_2,\vec{\rho} )\right\} \rda,
 \\
 &\mathcal{G}^A_l(t_1, t_2;\vec{\rho}) =
 i\theta(t_2-t_1)\lda
 \left\{
  \hat{c}_l (t_1,\vec{\rho}) ;\hat{c}_l^\dagger(t_2,\vec{\rho})\right\}
 \rda,
\\
 &\mathcal{G}^K_l(t_1, t_2;\vec{\rho})  = -i
\lda\left[
 \hat{c}_l (t_1,\vec{\rho}); \hat{c}_l^\dagger(t_2,\vec{\rho})
 \right]
\rda,
\end{split}
\label{GFs}
\ee
where $\theta(t)$ is the Heaviside step function,
and the fermionic operators are written in the Heisenberg representation.
Quantum mechanical averaging $\slda\dots\srda$ is
performed over an arbitrary density matrix to be found from the
solution of the kinetic equation.
To avoid misunderstanding, we emphasize that no averaging over the disorder
realization is assumed in \req{GFs}.

We parametrize the Keldysh Green function as
\be
\mathcal{G}^K_l(\vec{\rho})=\mathcal{G}^R_l(\vec{\rho})\con n_l(\vec{\rho})-
 n_l(\vec{\rho})\con \mathcal{G}^A_l(\vec{\rho})
\label{n}
\ee
where we omitted the time arguments for brevity and introduced
the short-hand notation
\be
\mathcal{C}\con \mathcal{D} \equiv \int \diff{t}_3\,
\mathcal{C}(t_1,t_3)\,\mathcal{D}(t_3,t_2).
\label{notation}
\ee
for arbitrary functions $\mathcal{C},\ \mathcal{D}$.

In the thermodynamic equilibrium
\be
n_l(\vec{\rho};\ep,t)= 1-2f_F(\ep),
\label{equilibrium}
\ee
where $f_F(\ep)$ is the Fermi distribution function with
arbitray temperature and chemical potential,
and the time Wigner transform is defined as usual:
\be
\mathcal{D}\left(t+\frac{\tau}{2},t-\frac{\tau}{2}\right)
=\int \frac{\diff\ep}{2\pi}\,\eexp^{-i\ep\tau}
\mathcal{D}\left(\ep, t\right).
\label{WT}
\ee
In the absence of interaction $n_l(\vec{\rho};\ep=\xi_l(\vec{\rho},t))$
characterizes the occupation of the level
$n_l(\vec{\rho};\xi_l(\vec{\rho}), t)=1 (-1)$ for an empty (filled)
level $(\vec{\rho},l)$.

\begin{figure}
\includegraphics[width=0.9\textwidth]{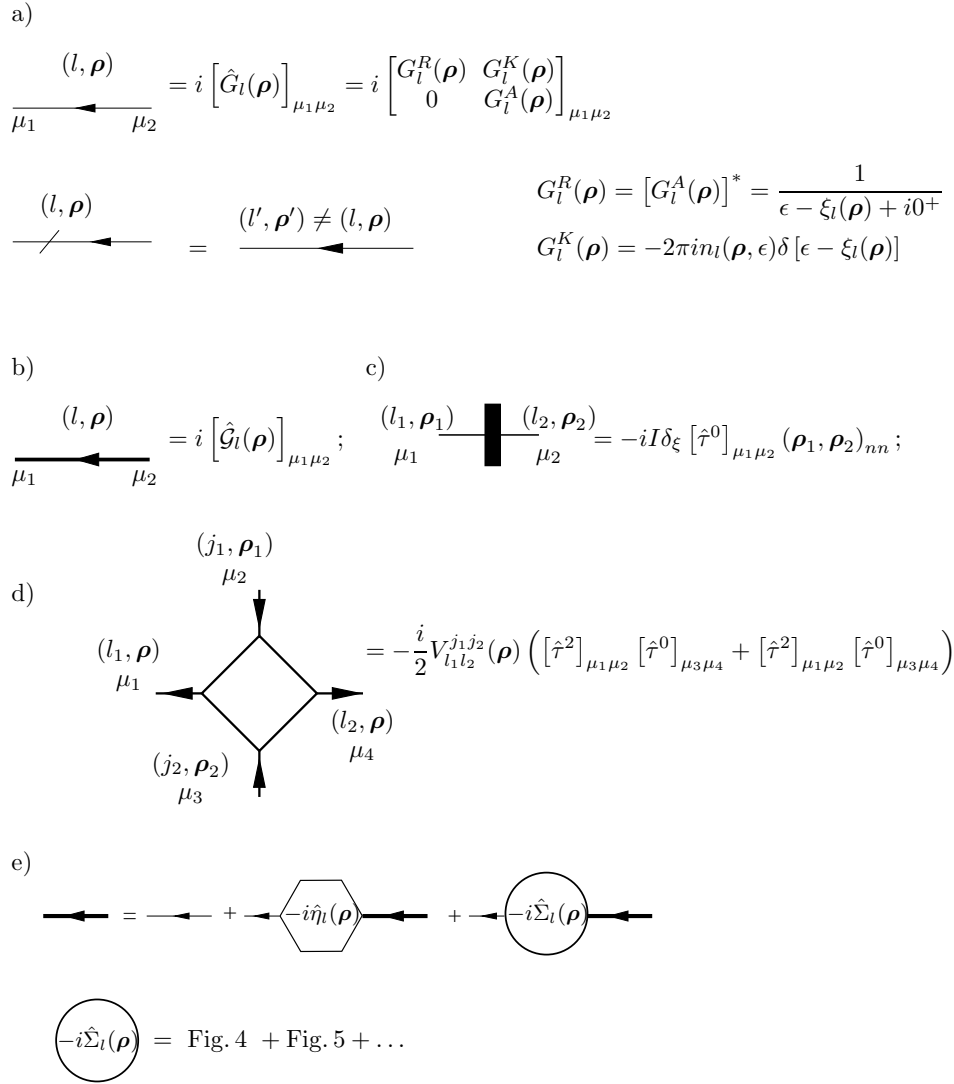}
\caption{(a-d) Basic elements of the diagrammatic technique.
The Keldysh space is labeled by $\mu_i$, and the Pauli matrices
in Keldysh space are defined in \req{GFeqn}.
Line crossing the Green function excludes the orbital $(l,\vec\rho)$
from the summation.
(e) Representation of \reqs{GFeqn} -- \rref{sigma1}.
}
\label{fig:diag1}
\end{figure}

In what follows we will use standard diagrammatic technique for
the perturbative expansion for the model~\rref{Heff}.
The basic elements of this technique 
are defined in Fig.~\ref{fig:diag1}.

The Green functions \rref{GFs} corresponding
to the Hamiltonian \rref{Heff} satisfy the equations
(we omitted time arguments for brevity)
\be\begin{split}
&\left[i\partial_{t_1}
-\xi_l(\vec{\rho})
\right]\hat{\mathcal{G}}_l(\vec{\rho})
=\hat{\tau}_0\delta(t_1-t_2)
 +\hat{\boldsymbol{\Sigma}}_l(\vec{\rho})\con
\hat{\mathcal{G}}_l(\vec{\rho});
\\
&\left[-i\partial_{t_2}
-\xi_l(\vec{\rho})
\right]\hat{\mathcal{G}}_l(\vec{\rho})
=\hat{\tau}_0\delta(t_1-t_2) + 
\hat{\mathcal{G}}_l(\vec{\rho})\con \hat{\boldsymbol{\Sigma}}_l(\vec{\rho});
\\ & \\
&\hat{\mathcal{G}}=\begin{bmatrix}
\mathcal{G}_l^R(\vec{\rho}) & \mathcal{G}_l^K(\vec{\rho})\\
0 & \mathcal{G}_l^A(\vec{\rho})
\end{bmatrix}_K\!\!;
\;
\hat{\boldsymbol{\Sigma}}=
\begin{bmatrix}
{\boldsymbol{\Sigma}}_l^R(\vec{\rho}) & {\boldsymbol{\Sigma}}_l^K(\vec{\rho})\\
0 & {\boldsymbol{\Sigma}}_l^A(\vec{\rho})
\end{bmatrix}_K\!\!\!;
\\ &
\hat{\tau}^0=\begin{bmatrix}
1 & 0\\
0 & 1
\end{bmatrix}_K\!\!;\quad
\hat{\tau}^2=\begin{bmatrix}
0 & 1\\
1 & 0
\end{bmatrix}_K\!\!.
\label{GFeqn}
\end{split}\ee

\begin{figure}
\includegraphics[width=0.9\textwidth]{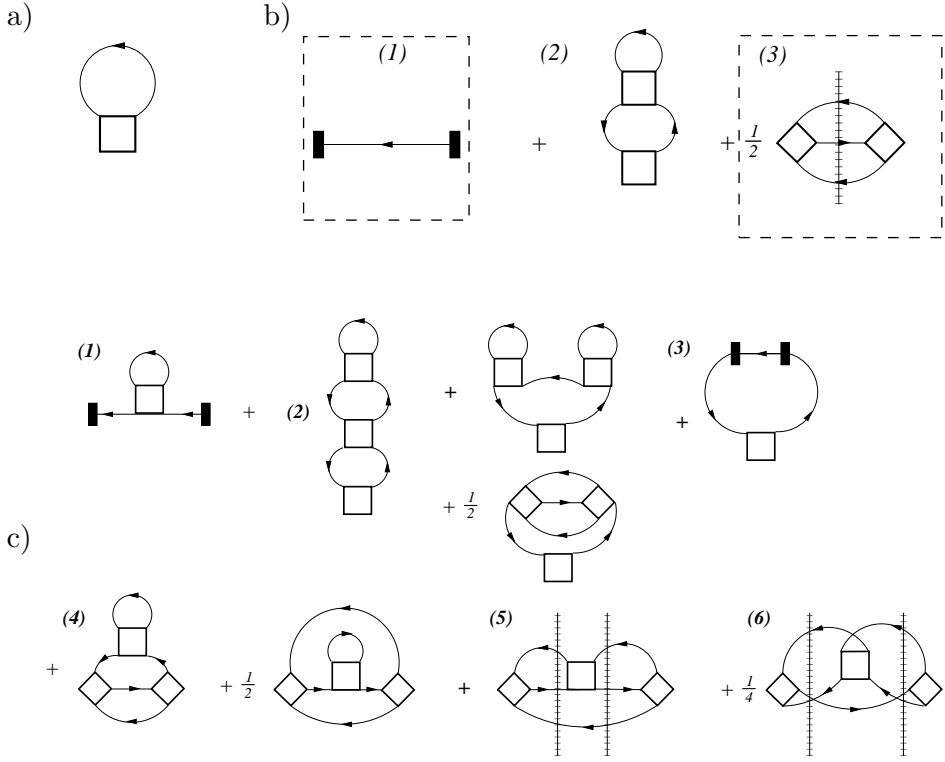}
\caption{
First (a), second (b) and third (c)
orders contributions to the
self-energy $\hat{\Sigma}_l(\vec{\rho})$.
The diagrams which will be taken into account in the self-consistent
Born approximation are highlighted by the dashed frames.
Cross-sections of the diagram which produce the imaginary part of
the self-energy are denoted by the dotted lines.
}
\label{fig:diag2}
\end{figure}

\begin{figure*}
\includegraphics[width=\textwidth]{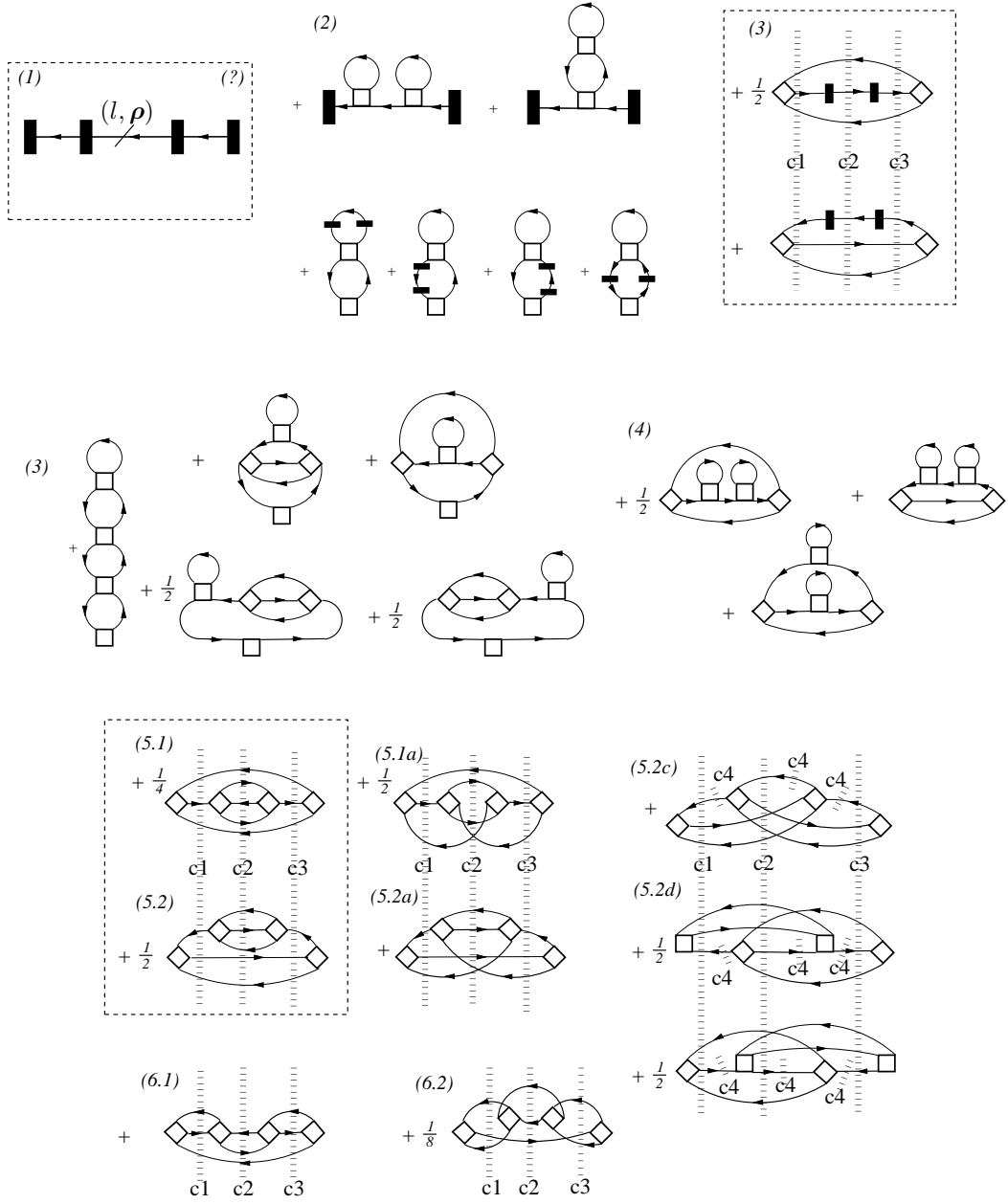}
\caption{Fourth-order contributions to the
self-energy $\hat{\Sigma}_l(\vec{\rho})$.
The diagrams which will be taken into account in the self-consistent
Born approximation are highlighted by the dashed frames.
Cross-sections of the diagram which produce the imaginary part of
the self-energy are denoted by the dotted lines and denoted by
c1 -- c4. The crosssection c4 implies cutting
three lines  at the same time.
}
\label{fig:diag3}
\end{figure*}

The self-energy $\hat{\boldsymbol{\Sigma}}_l(\vec{\rho})$
is given by
\be
 \hat{\boldsymbol{\Sigma}}_l(\vec{\rho})=
\hat{\eta} +  \hat{{\Sigma}}_l(\vec{\rho}),
\label{sigma1}
\ee
where $\hat{\eta}$ originates from the coupling of the sysstem
to an external bath with regular continuous
spectrum (it will be discussed in more details later).
This coupling has to be kept small but finite and can be put
to zero only in the end of the
calculation. Self-enery $\hat{{\Sigma}}_l(\vec{\rho})$ originates
from the electron-electron interaction and hopping and represents the
sum of all diagrams which cannot be separated by cutting
one-electron line $(l,\vec{\rho})$, see
Figs.~\ref{fig:diag2},~\ref{fig:diag3}.

Substitution of \req{n} into the Keldysh component of \req{GFeqn}
yields
\be
\begin{split}
&\left(\partial_{t_1}
+\partial_{t_2}\right)
n_l(\vec{\rho})
=
-i{\boldsymbol{\Sigma}}_l^R(\vec{\rho})\, \con \, n_l(\vec{\rho})
+ i n_l(\vec{\rho}) \,\con\,
{\boldsymbol{\Sigma}}_l^A(\vec{\rho})
 +  i{\boldsymbol{\Sigma}}_l^K(\vec{\rho}).
\end{split}
\label{neqn1}
\ee

On the next step we restrict ourselves to the consideration
of very slow dynamics. 
In this case one can perform the time Wigner transform
\rref{WT} in \req{neqn1} and obtain the quantum Boltzmann
equation
\be
\begin{split}
&
\partial_t n_l(\ep; \vec{\rho},t)
+  \left[\delta\xi_l(\ep; \vec{\rho};\{n\})\starcomma n_l(\vec{\rho},\ep,t)
\right]
 = {\rm St}_{l}\left(\ep;\vec{\rho}; \{n\}, \{\partial_t n\} \right)
+  {\rm St}_{l}^{bath},
\end{split}
\label{neqn2}
\ee
where the collision integrals are defined as
\be
\begin{split}
 &
 {\rm St}_l(\ep;\vec{\rho}; \{n\}, \{\partial_t n\} )
 = -2 {\Gamma}_l(\ep; \vec{\rho}; \{n\},  \{\partial_t n\})
n_l(\vec{\rho},\ep,t)
  + i{{\Sigma}}_l^K(\vec{\rho};\ep; \{n\}, \{\partial_t n\} );
\\ & \\
&
 {\rm St}_l^{bath}= -2\Im\eta^A n_l(\vec{\rho},\ep,t) + i\eta^K.
\end{split}
\label{stoss}
\ee
Hereinafter, the time Poisson brackets are defined as
\be
\left[\mathcal{C}\starcomma \mathcal{D}\right]\equiv
{\partial_t \mathcal{C}}\,{\partial_\ep \mathcal{D}}
- {\partial_\ep \mathcal{C}}\,{\partial_t \mathcal{D}}.
\label{PB}
\ee
for arbitrary functions $\mathcal{C}(\ep,t)$, $\mathcal{D}(\ep,t)$.
The entries in \req{neqn2} are defined by
\be
\begin{split}
{\Sigma}^{R}=\delta\xi - i{\Gamma};
\quad  {\Sigma}^{A}=\delta\xi + i{\Gamma};
\end{split}
\label{Gamma}
\ee
where we suppressed all arguments which are the same as in
\req{stoss}.
Using \reqs{Gamma} and the analytic properties of the retarded and
advanced Green functions one obtains
{\setlength{\arraycolsep=0pt}
\bea
&&\mathcal{G}_l^R(\vec{\rho};\ep,t)=
\left[\mathcal{G}_l^A(\vec{\rho};\ep,t)\right]^*
=\int \frac{\diff\ep^\prime {A}_l (\vec{\rho};\ep^\prime,t)}
{\ep-\ep^\prime +i0^+},
\label{WTGRA}\\
&&{A}_l(\vec{\rho};\ep,t)=
\frac{{\pi}^{-1}\boldsymbol{\Gamma}_l(\ep; \vec{\rho}; \{n\},  \{\partial_t n\})}
{\left[\ep-\xi_l(\vec{\rho}) -
\delta\xi(\ep; \vec{\rho}; \{n\}) \right]^2+
\left[\boldsymbol{\Gamma}_l(\ep; \vec{\rho}; \{n\},  
\{\partial_t n\})\right]^2 },
\nonumber
\eea
and $\boldsymbol{\Gamma}_l={\Gamma}_l+{\rm Im}\eta^A_l$.
Notation $\mathcal{F}(\{n\})$ means that  $\mathcal{F}$ is a functional
depending upon all the functions $n_l(\ep,\vec{\rho})$ but local in time.
The latter functions enter the expressions through the
quasistationary version of \req{n}:
\be
\mathcal{G}^K_l(\vec{\rho};\ep,t)=
-2\pi i n_l(\vec{\rho};\ep,t) {A}_l (\vec{\rho};\ep^\prime,t)
\label{WTGK}
- i \int\diff\ep^\prime
\left[{A}_l (\vec{\rho};\ep^\prime,t)\starcomma n_l(\vec{\rho};\ep,t)\right]
P\frac{1}{\ep-\ep^\prime},
\ee
where $P$~denotes the principal value.
}

What remains now, is to specify the thermal bath. As
we already mentioned, the particular form of this choice
is not important. We will require only that it preserves the
number of particles
\[
\int\diff\ep\,{A}_l (\vec{\rho};\ep,t)\,{\rm St}^{bath}(\ep,t)=0,
\]
it is local, and
the collision integral ${\rm St}^{bath}$ is nullified by the
equlibrium distribution function \rref{equilibrium}.
It is easy to check that the choice
\be
\begin{split}
2 \Im\eta^A&=\int\diff\omega\, \omega\, b(\omega)\,
{A}_l (\vec{\rho};\ep-\omega,t)
\left[\coth \frac{\omega}{2T} + n_l(\vec{\rho};\ep-\omega,t)\right];\\
- i\eta^K &= \int\diff\omega\, \omega\, b(\omega)\,
{A}_l (\vec{\rho};\ep-\omega,t)
\label{eq:bath}
\left[\coth \frac{\omega}{2T}\, n_l(\vec{\rho};\ep-\omega,t) + 1 \right],
\end{split}
\ee
where $b(\omega)=b(-\omega)>0$, saisfies both those
requirements\footnote{
The subsequent formulas for
non-zero coupling $\hat{\eta}$ actually may be used to
describe 
the effect of the short-rangle electron-electron interaction
on the phonon-assisted nearest neighbors hopping.}.

\subsection{Formulation of the problem.}
\label{sec:problem}

\begin{figure}
\includegraphics[width=0.7\textwidth]{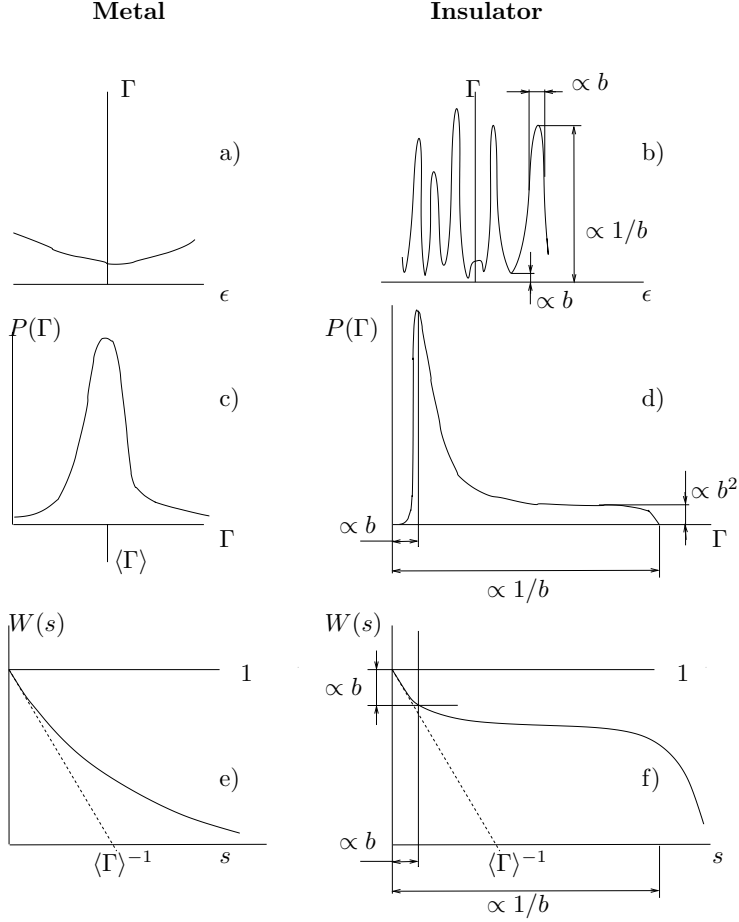}
\caption{
Schematic energy dependence of the
quasiparticle decay rate $\Gamma(\ep)$ for
(a) metallic and (b) insulating phases.
The corresponding distribution functions
are sketched on panels (c),~(d).
The characteristic functions are plotted on panels
(e),~(f).
}
\label{figP}
\end{figure}

Having introduced the definitions, we are ready to
reformulate the criterion distinguishing insulating and metallic
states.\footnote{Subsequent
discussion is a straightforward
generalization of the Anderson's argument~\cite{Anderson58} to a
many-body system.} The
left-hand side of the Boltzmann equation \rref{neqn2} describes the
evolution of the occupation of the levels in a
self-consistent field created by all the other electrons. This
evolution is deterministic and time reversible.
It is the right-hand side of the kinetic equation (collision integral)
that makes the time evolution probabilistic and specifies the
direction of the time arrow. Thus, the energy dependence of
the decay rate $2\Gamma_l(\ep)$ determines whether or
not the irreversible
evolution occurs in the system. If coupling with
the environment, $b(\omega)$ from \req{eq:bath}, is finite,
$\Gamma_l(\ep)$ is positive for any energies. However, if
this coupling tends to zero
(but not faster 
than the exponential function of the volume of the system $\mathcal{V}$,
$b > \exp\left[-\mathcal{V}/\mathcal{V}_0\right]$), two situations
are possible, see Fig.~\ref{figP}a,b.
\begin{enumerate}[{(}i{)}]
\item Despite  $b(\omega)$ tending to zero the number of
the intermediate states via which the excitation can decay
goes to infinity. This results in $\Gamma_l(\ep)$ being
a smooth function of energy even at $b(\omega)\to 0$.
This situation corresponds to the applicability of the
Fermi golden rule, the thermal equilibrium
within the system established at times independent
of the external bath, so it is natural to classify this regime
as {\em metallic}.

\item  The number of
the intermediate states via which the excitation can decay
remains finite and independent
on $b(\omega)$ as $b(\omega)\to 0$.
This results in $\Gamma_l(\ep)$ to be
a sequence of resonant peaks positioned
at the energies of the exact excitations of
the many-body system. In this case $n(\ep)$
will remain extremely singular function whose relaxation
rate is determined by $b(\omega)$.
At $b(\omega)\to 0$ the thermal equilibrium
can  never be reached and
this regime is {\em insulating}.
\end{enumerate}

Since these two behaviors are qualitatively different,
there could be no smooth crossover between them, and
only {\em phase transition} is possible. Therefore, to
show the existence of the transition, it is
sufficient to formulate the conditions at which
either metallic (i) or insulating (ii) regimes are stable.
It will be done in Secs.~\ref{sec:metal} and \ref{sec:insulator}.
The investigation of the behavior of the kinetic coefficients
near the phase transition point itself will be the subject of a
separate paper.

Next question one has to ask is how to distinguish between
metallic and insulating phases within a statistical framework.
It is clear that the positions of the
peaks in $\Gamma(\ep)$ for the insulating regime
deviate randomly with the variation of random
energies $\xi_\alpha$ from \req{Heff}. Therefore, the averaged
value of the decay rate $\langle\Gamma(\ep)\rangle$ is qualitatively
similar in both phases and can not be used for the distinction.
The magnitude of the fluctuations
$\langle \left[\delta \Gamma(\ep)\right]^2\rangle$
is, however, qualitatively different
in both cases. In the metallic case the averaging is performed
with respect to the smooth positive functions, whereas in the
insulating regime the fluctuations are determined by the squares
of the separated delta-peaks and thus diverge as the width of the
delta peaks goes to zero. Thus, we have the criterion
\be
\lim_{b(\omega)\to 0}\lim_{\mathcal{V}\to \infty}
\frac{\langle
\left[\delta \Gamma(\ep)\right]^2\rangle}
{\langle
 \Gamma(\ep)\rangle^2} =
\left\{
\begin{matrix}
{\rm finite}; & {\rm metal}\\
\infty; & {\rm insulator.}
\end{matrix}
\right..
\label{criterion}
\ee
Another way to address the same problem
is to investigate the distribution function
$P(\Gamma)$, see Fig.~\ref{figP}(c-d).
One finds by simple inspection of  Fig.~\ref{figP}(a-b)
\be
\lim_{b(\omega)\to 0}\lim_{\mathcal{V}\to \infty}
P(\Gamma > 0 )
 =
\left\{
\begin{matrix}
> 0; & {\rm metal}\\
0; & {\rm insulator.\textmd{}}
\end{matrix}
\right.
\label{criterion2}
\ee
For us,
it will be more technically convenient to perform the 
Laplace transform and calculate the characteristic function 
\be
W(s)=\left\langle \exp\left[-s\Gamma(\ep)\right]\right\rangle,
\label{Wgeneral} 
\ee 
where the precise definition of the averaging procedure is given in 
\textmd{}Sec.~\ref{sec:canonical}.
The criterion~\rref{criterion2} [see also Fig.~\ref{figP}(e-f)]
of the insulating phase translates into
\be
\lim_{b\to 0}\lim_{\mathcal{V}\to \infty}W(s) = 1, \label{linearconditiongen}
\ee 
for any fixed $s>0$.

Closing this subsection, we emphasize that the transition
occurs as a function of temperature (which corresponds to the 
extensive energy
of the many-body state) and not as a function of $\ep$ which
charaterizes the energy of the one-particle excitation on
top of this many-body state. The latter energy is not an extensive
quantity and can not be a characteristic of any phase transition,
in contrast with conclusions of Ref.~\cite{AGKL} for a
finite-size system.


\subsection{Self-Consistent Born Approximations (SCBA and ImSCBA).}
\label{sec:SCBA}

In this subsection we introduce our main approximation
for the summation of the infinite series of perturbative
expansion. We will discuss motivation for this approximation
here in quite loose terms and justify it further in
Sec.~\ref{sec:validity}.

\begin{figure}
\includegraphics[width=0.6\textwidth]{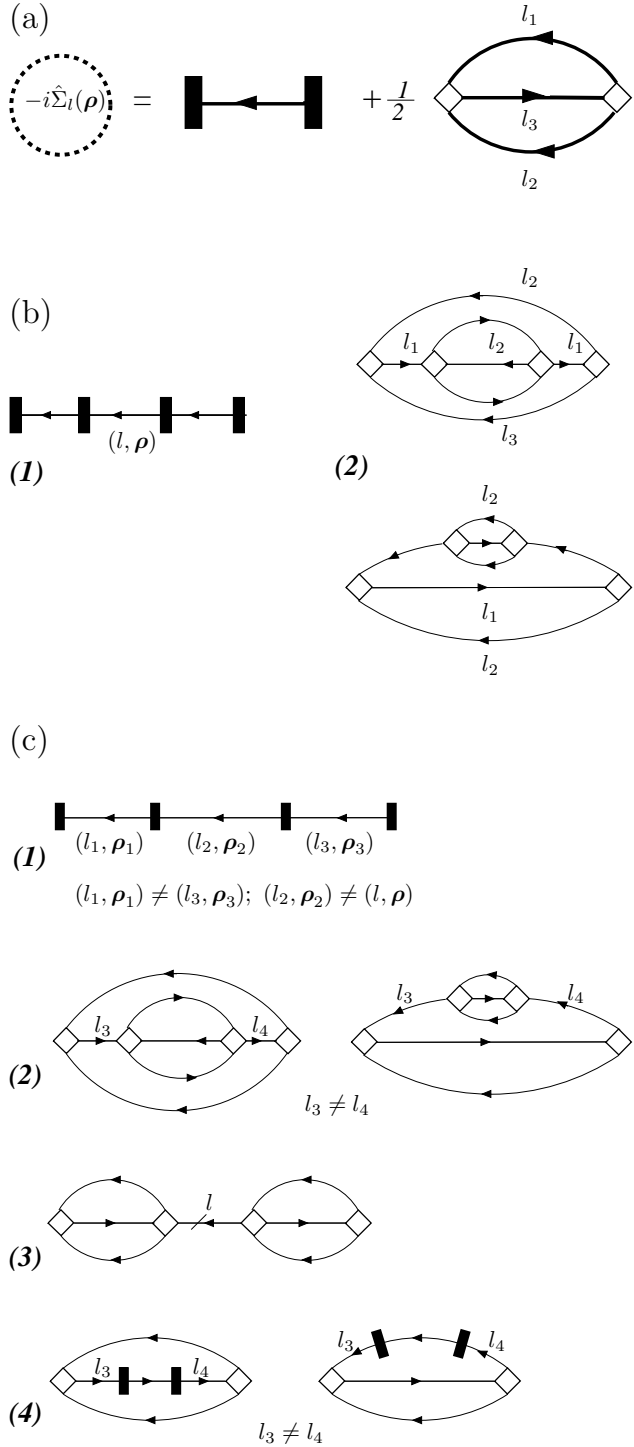}
\caption{a) Self-consistent Born approximation. The thick
fermionic lines are defined in Fig.~\ref{fig:diag2}a.
b) Spurious contributions of the fourth
order generated by iteration of SCBA:
b.1 contradicts the definition of the self-energy;
b.2 contains the intermediate particles in the same states;
such contributions are cancelled by diagrams (5.1a) and
(5.2a)--(5.2c) of Fig.~\ref{fig:diag2}.
c)  Fourth order tunneling and interaction
contributions not taken into account by
SCBA.  These contributions have a random sign.}
\label{figSCBA}
\end{figure}

Contributions to the self-energy shown in
Figs.~\ref{fig:diag2},\ref{fig:diag3} bear different
physical significance. For instance, the Hartree-Fock diagrams
(b), (d2) of Fig.~\ref{fig:diag2}, and diagrams (3)~of
Fig.~\ref{fig:diag3}
characterize the self-consistent one-particle
spectrum. As we explained before, the structure of this spectrum is
not relevant for the transition. On the contrary, diagrams (c1) and
(c3) of Fig.~\ref{fig:diag2} can lead to irreversible processes,
and the appearance of the
imaginary part of those diagrams signals the metal-insulator
transition. Thus, those diagrams will be taken into account.

The third order diagrams,  Fig.~\ref{fig:diag2}d, describe
the effect of the change of the self-consistent potential
on the tunneling process (d1); effect of the tunneling on the
self-consitent potential (d3); and the
effect of the self-consistent potential on the spectrum
of the decay channel.
  These diagrams will be neglected for
the reasons explained above.

Fourth order diagrams  Fig.~\ref{fig:diag3}
describe further potentially irreversible
process: tunneling out of the localization cell~(1);
three-particle production with the consequent tunneling~(3);
five-particle production~(5). Those contributions
must be taken into account. On the other hand, diagrams
(2)~and~(4) once again represent the self-consistent potential
affecting the lower order processes (c1), (c3) of
Fig~\ref{fig:diag2}. Diagrams (6.1), (6.2) of  Fig.~\ref{fig:diag3}
describe the interaction of the quasiparticles created by the
lower-order process (c3) of Fig.~\ref{fig:diag2}.
Such effect has the same physical significance
and the relative value as the effect of the self-consistent potential
and will be neglected. The effect of those diagrams will be estimated
in Sec.~\ref{sec:validity}. Finally, diagrams (5.1a), and (5.2a)--(5.2c)
are exchange counterparts of the main diagrams (5.1)-(5.2)
[we will elaborate it further in Sec.~\ref{sec:validity}]. Their
role is to cancel out contributions in  (5.1),~(5.2)  which are
forbidden by Pauli principle (more than one particle in one
intermediate state) and also to give the random sign
interference corrections.
The latter corrections will be neglected here and estimated
further in    Sec.~\ref{sec:validity}.

The above discussion of the lowest orders of the perturbation theory
suggests the following prescription for the summation of the
leading series: (i) we take into account only the even orders of the
perturbation theory; (ii) we require that each contribution
maximize the sum of the tunneling events and the extra quasiparticle
production; (iii) we neglect contributions with random sign.
By inspection, one can see that such series with the correct
combinatorial coefficients is generated by the self-consistent
Born approximation (SCBA) shown on Fig.~\ref{figSCBA}a.

Iterations of the SCBA equations also produce spurious
contributions shown on Fig.~\ref{figSCBA}b. For the SCBA scheme
to be valid, we will make sure that those spurious contributions are
always smaller than those responsible for the final result.

We calculate the self-energy (Fig.~\ref{figSCBA}a) according
to the rules of Fig.~\ref{fig:diag1}. We neglect the
shifts $\delta\xi_l(\vec{\rho})$ in accord with our previous
discussion, {\em i.e.}, we  make the additional approximation
to the usual SCBA scheme
\be
\Sigma^{(SCBA)}_l(\vec\rho;\ep) \to i \Im\Sigma^{(SCBA)}_l(\vec\rho;\ep)
\label{ImSCBA}
\ee
We  will refer to approximation \rref{ImSCBA} as
Im-Self-Consistent-Born-Approxima\-tion (ImSCBA).
In the two subsequent sections, we will work 
only within this approximation.

Using spectral representation \rref{WTGRA}
and \rref{WTGK} and performing the integration over
the intermediate energies, we obtain
\begin{subequations}
\label{SCBA}
\be
\begin{split}
&\Gamma_l(\ep) =\Gamma_l^{(el)}(\ep)+\Gamma_l^{(in)}(\ep)
+\Gamma_l^{(bath)}(\ep);
\\
&\Gamma_l^{(el)}(\ep,\vec{\rho})
=\pi I^2\delta_{\zeta}^2\sum_{l_1,\vec{a}}
A_{l_1}\left(\ep,\vec{\rho}+\vec{a}\right);
\\
&\Gamma_l^{(in)}(\ep)=
\pi \lambda^2\delta_\zeta^2
\sum_{l_1,l_2,l_3}
Y_{l_1,l_2}^{l_3,l}
\int{\diff\ep_1}{\diff\ep_2}{\diff\ep_3}\,
A_{\l_1}(\ep_1)\,A_{l_2}(\ep_2)\,A_{l_3}(\ep_3)\\
&\qquad\qquad\times\delta(\ep-\ep_1-\ep_2+\ep_3)\,
F_{l_1, l_2; l_3}^{\Rightarrow}(\ep_1,\ep_2;\ep_3);
\\
&
\Gamma_l^{(bath)}(\ep)=\frac{1}{2}\int\diff\omega\, \omega\, b(\omega)
{A}_l (\ep-\omega)
\left[\coth \frac{\omega}{2T} + n_l(\ep-\omega)\right]
\\
&{A}_l(\ep)=
\frac{{\pi}^{-1}{\Gamma}_l(\ep)}
{\left[\ep-\xi_l\right]^2+
\left[{\Gamma}_l(\ep)\right]^2 }\\
&
Y_{l_1,l_2}^{l_3,l}\equiv
\frac{1}{2}\left[
\Upsilon\left(\frac{\xi_{l_2}-\xi_{l}}{\delta_\zeta}\right)
\Upsilon\left(\frac{\xi_{l_1}-\xi_{l_3}}{\delta_\zeta}\right)
-
\Upsilon\left(\frac{\xi_{l_1}-\xi_{l}}{\delta_\zeta}\right)
\Upsilon\left(\frac{\xi_{l_2}-\xi_{l_3}}{\delta_\zeta}\right)
\right]^2
\\
&
F^{\Rightarrow}_{l_1, l_2;l_3}(\ep_1,\ep_2;\ep_3)=
\frac{1}{4} \Big\{1+ n_{l_1}(\ep_1) n_{l_2}(\ep_2)
-n_{l_3}(\ep_3)
\left[n_{l_1}(\ep_1) + n_{l_2 }(\ep_2)\right]
\Big\};
\end{split}
\label{eqSCBA1}
\ee
where $\eta$ is defined in \req{eq:bath},
and we utilized the notation used in \reqs{Heff} for the nearest
neighbours.
Everywhere, the coordinate $\vec{\rho}$ and time $t$ are assumed
to be same in all terms in the equations unless it is specified
explicitly otherwise.

Equations \rref{eqSCBA1} form a closed set for
finding the decay rate for fixed occupation numbers
$n_l(\ep,\vec{\rho})$. In the delocalized regime the
time evolution of those occupation numbers is governed
by the ImSCBA version of the kinetic eqution \rref{neqn2}.
Calculating the Keldysh component of Fig.~\ref{figSCBA}a and using
\reqs{stoss} and \rref{eq:bath}, we find
\be
\begin{split}
&\partial_t n_l(\ep)
={\rm St}_{l}^{(el)} + {\rm St}_{l}^{(in)}
+  {\rm St}_{l}^{(bath)};
\\
&{\rm St}_{l}^{(el)} =
2\pi I^2\delta_{\zeta}^2\sum_{l_1,\vec{a}}
A_{l_1}(\ep,\vec{\rho}+\vec{a})
\left[n_{l_1}(\ep,\vec{\rho}+\vec{a}) - n_l(\ep,\vec{\rho})\right];
\\
&{\rm St}_{l}^{(in)}=
2\pi
\lambda^2\delta_\zeta^2
\sum_{l_1,l_2,l_3}Y_{l_1,l_2}^{l_3,l}
\int{\diff\ep_1}{\diff\ep_2}{\diff\ep_3}\,
A_{\l_1}(\ep_1)\,A_{l_2}(\ep_2)\,A_{l_3}(\ep_3)\,
\delta(\ep-\ep_1-\ep_2+\ep_3)
\\
&
\qquad\qquad\times
\left[- n_l(\ep)
F_{l_1, l_2;l_3}^\Rightarrow (\ep_1,\ep_2;\ep_3)
+ F_{l_1, l_2;l_3}^{\Leftarrow}(\ep_1,\ep_2;\ep_3)
\right];
\\
& F_{l_1, l_2;l_3}^{\Leftarrow}(\ep_1,\ep_2;\ep_3)=
\frac{1}{4} \Big\{-n_{l_3}(\ep_3)
\left[1+ n_{l_1}(\ep_1) n_{l_2}(\ep_2)\right]
+
\left[n_{l_1}(\ep_1) + n_{l_2}(\ep_2)\right]
\Big\};
\\
&{\rm St}_{l}^{(bath)}=\int\diff\omega\, \omega\, b(\omega)\,
{A}_l (\ep-\omega)\\
&\qquad\qquad\times\left\{
\coth \frac{\omega}{2T_b}\left[
 n_l(\ep-\omega)-n_l(\ep)\right] + 1
-  n_l(\ep-\omega)n_l(\ep)
\right\}.
\end{split}
\label{eqSCBA2}
\ee
\end{subequations}
Equation \rref{eqSCBA2} is the usual quantum Boltzmann equation written
in terms of the exact (not averaged) single electron levels
$\xi_l(\vec{\rho})$. As any Boltzmann equation, it must respect the
fundamental symmetries of the system: conservation
of the number of particles for any collision; conservation of
electron energy for the processes involving the electrons
only; and the conservation of the number of particles
for a given energy for elastic collisions. It is straightforward
to check that the collision integral indeed possesses the
desired properties:
\be
\begin{split}
&\sum_{l,\rho}A_l(\ep,\vec{\rho})\,{\rm St}_{l}^{(el)}(\ep,\vec{\rho})=0;
\\
&\sum_{l}
\int\diff\ep\, A_l(\ep,\vec{\rho})\,{\rm St}_{l}^{(in)}(\ep,\vec{\rho})=0;
\\
&
\sum_{l}\int\diff\ep\, \ep\,
 A_l(\ep,\vec{\rho})\,{\rm St}_{l}^{(in)}(\ep,\vec{\rho})=0;
\\
& \sum_{l}\int \diff\ep\,
 A_l(\ep,\vec{\rho})\,{\rm St}_{l}^{(bath)}(\ep,\vec{\rho})=0.
\end{split}
\label{conservation}
\ee

The properties~(\ref{conservation}) of the collision integrals
enable one to write the continuity equations
for  the particle and energy densities $\mathcal{N}$ and $\mathcal{E}$, and
introduce the corresponding currents $\vec{J},\ \vec{J}_\mathcal{E}$:
\begin{subequations}
\be
\begin{split}
&\partial_t\mathcal{N}(\vec\rho)+\Div\vec{J}(\vec\rho)=0;
\\
&
\partial_t\mathcal{E}(\vec\rho)+\Div\vec{J}_\mathcal{E}(\vec\rho)=-Q_{bath}(\vec\rho);
\end{split}
\ee
where $Q_{bath}$ is the thermal flow to the thermal bath; it will
not be important for the further cosideration. The
lattice version of the divergence of the currents is defined as
\[
\Div\vec{J}(\vec\rho)=\frac{1}{\zeta_{loc}}\sum_{k=1}^d
\left[J^{k}\!\left(\vec\rho+\frac{\vec{a}^{(k)}}{2}\right)-
J^{k}\!\left(\vec\rho-\frac{\vec{a}^{(k)}}{2}\right)\right],
\]
and it becomes the usual divergence in the continuum limit.
Index $k$ labels the direction in the cartesian coordianat system in
$d$ dimensions. Vector $\vec{a}^{(k)}$ is the lattice vector
along the $k$th direction.

The densities are defined on the sites $\vec{\rho}$
\be
\begin{split}
&
\begin{bmatrix}
\mathcal{N}\\
\mathcal{E}
\end{bmatrix}(\vec\rho)\
=
\frac{1}{\zeta_{loc}^d}
\int \diff\ep\begin{bmatrix}
1
\\
\ep
\end{bmatrix}
 \sum_l \,A_l (\ep,\vec\rho)\,
\frac{1-n_l(\ep,\vec\rho)}{2},
\end{split}
\label{dens}
\ee
whereas the currents are defined on the links 
$\vec\rho^{(k)}=\vec{\rho}+\vec{a}^{(k)}/2$:
\be
\begin{split}
\begin{bmatrix}
J^{k}
\\
J^{k}_\mathcal{E}
\end{bmatrix}\left(\vec\rho^{(k)}\right)
=&
\int {\diff\ep} \begin{bmatrix}
1
\\
\ep
\end{bmatrix}\sum_{l,l_1} \,
A_l(\ep,\vec\rho)\,A_{l_1}\!\left(\ep,\vec\rho+\vec{a}^{(k)}\right)
\\ & \times
\frac{\pi I^2\delta_\zeta^2}{\zeta_{loc}^{d-1}}
\left[n_{l_1}\!\left(\ep,\vec\rho+\vec{a}^{(k)}\right)
-n_l(\ep,\vec\rho)\right]
,
\end{split}
\label{cur}
\ee
\label{curdens}
\end{subequations}

Finally, ${\rm St}_{l}^{(el)}$ is nulled by any function
$n_l(\ep;\vec{\rho})=n(\ep)$, electron-electron inelastic collision
integral ${\rm St}_{l}^{(in)}$ is
nulled by any Fermi finction $n_l(\ep;\vec{\rho})
=\tanh\frac{\ep-\mu(\vec{\rho})}{2T(\vec{\rho})}$;
and the  ${\rm St}_{l}^{(bath)}$ vanishes for
$n_l(\ep;\vec{\rho})
=\tanh\frac{\ep-\mu(\vec{\rho})}{2T_b}$.

The closed system of \reqs{SCBA} is a drastic (though parametrically
justifiable) simplification in comparison with the original
problem. However, it is still a non-linear system which depends on an
infinite number of random energies $\xi_l(\vec{\rho})$. Substantial
progress can be achieved within the statistical analysis. This
analysis is a subject of two following sections.

\section{Stability and properties of the metallic phase.}
\label{sec:metal}

\subsection{Condition for stability}
\label{sec:metalA}

\begin{figure}
\includegraphics[width=0.6\textwidth]{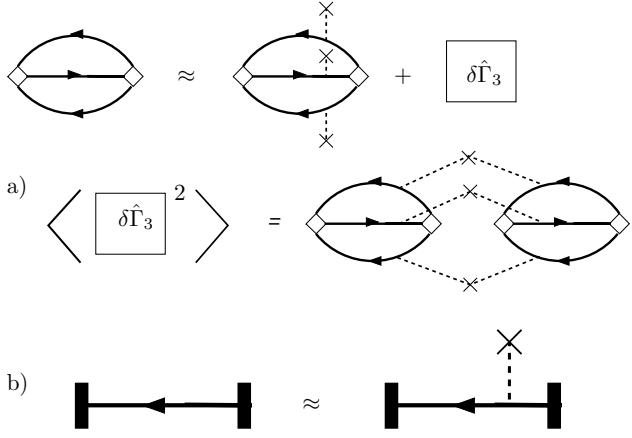}
\caption{a) Simplification of
the self-consistent Born approximation
at $T > T_{in}$.
b) The self-averaging of the hopping term
at  $T > T_{el}$. Cross with the dashed line(s)
stands for the averaging of the corresponding
Green function (product of the Green functions)
over the realization of random energies $\xi_l(\vec{\rho})$,
see \req{delta}.
}
\label{figSCBAm}
\end{figure}

The hallmark of the developed metallic phase is self-averaging of the
kinetic coefficients, see Sec.~\ref{sec:problem}.
To establish a sufficient condition for the
existence and stability of this phase,
we assume that the inelastic
decay rate is indeed self-averaging and then justify
this assumption by explicit calculation of its mesoscopic
fluctuations, see Fig.~\ref{figSCBAm}.
As a result, we will see that this condition
reduces to a certain integral inequality for the distribution function,
see \req{mcondition1}. We will also see that even when this condition
is satisfied, the elastic rate still may not be self-averaging.
This, however, will not violate the criterion~\rref{criterion}, as
the fluctuations of the elastic rate remain finite.

First, we take into account only the inelastic rate
in the $A_l$ of \req{eqSCBA1}. We assume
and justify {\em a posteriori} that the main contribution
originates from
\be
|l-l_i|,|l_i-l_j| \gg 1,\ i,j=1,2,3.
\label{1cond}
\ee
Statistical averaging
over the distribution of the levels $\xi_l$ can be
performed independently,
\[
\Big\langle\sum_l(\ldots)\Big\rangle =
 \delta_{\zeta}^{-1}\int\diff\xi_l\, (\ldots)\,.
\]
For the needed products of the spectral densities,
see \req{eqSCBA1}, we find
\be
\begin{split}
&\Big\langle \sum_l A_{l}(\ep)\Big\rangle=\frac{1}{\delta_{\zeta}};\\
&\Big\langle \sum_lA_{l}(\ep_1)
A_l(\ep_2)\Big\rangle=
\frac{1}{\pi\delta_{\zeta}}
\frac{2\Gamma^{(in)}(\ep_1)}{(\ep_1-\ep_2)^2+
\left[2\Gamma^{(in)}(\ep_1)\right]^2},
\end{split}
\label{average}
\ee
where we assumed that $\Gamma(\ep)$ is a smooth function on the
scale of $\Gamma(0)$.

Averaging \req{eqSCBA1} with the help of \req{average}, see also
Fig.~\ref{figSCBAm}a, and assuming
that the distribution functions $n_l(\ep,\vec{\rho})$
do not depend explicitly on
the orbital index $l$,
we find:\footnote{
All the formulas of Sec.~\ref{sec:metal} are written under the
condition $\langle\Gamma_l^{(in)}(\ep)\rangle \ll M\delta_\zeta$.
In this case one can neglect the dependence on the index~$l$ of
$\Gamma_l(\ep)$, $n_l(\ep)$, and the collision integral. Such
dependence is present only for $|\ep-\xi_l| \gtrsim M\delta_\zeta$,
while all physical properties are determined by the region
$|\ep-\xi_l|\sim\Gamma^{(in)}$; in particular, the distribution
function enters the observables only as
$\sum_lA_l(\ep)\,n_l(\ep)$.}
\be
\begin{split}
\langle\Gamma_l^{(in)}(\ep)\rangle=&
\frac{\pi \lambda^2}{\delta_\zeta}
\int{\diff\ep_1}{\diff\omega}\,
F^\Rightarrow(\ep_1+\omega,\ep_1;\ep-\omega)
\\ & \times 
\left[
\Upsilon^4\!\left(\frac{\omega}{\delta_\zeta}\right)
-
\Upsilon^2\!\left(\frac{\omega}{\delta_\zeta}\right)
\Upsilon^2\!\left(\frac{\ep_1+\omega-\ep}{\delta_\zeta}\right)
\right].
\end{split}
\label{gammam1}
\ee
As we will see shortly, the metallic regime is realized
when the characteristic energy scale of the distribution function
is much larger than $M\delta_\zeta$ [\req{Upsilon}] -- typical
transferred energy in \req{gammam1}. Under such conditions, the
second term in the second line of this equation is more restrictive
on the phase volume and that is why it can be neglected:
\be
\begin{split}
&\langle\Gamma^{(in)}_l(\ep)\rangle=\frac{\pi \lambda^2}{\delta_\zeta}
\int{\diff\ep_1}{\diff\omega}\,
\Upsilon^4\!\left(\frac{\omega}{\delta_\zeta}\right)
F^\Rightarrow(\ep_1+\omega,\ep_1;\ep-\omega).
\end{split}
\label{gammam1p}
\ee
Calculating mesoscopic fluctuations of the inelastic rate
shown in Fig.~\ref{figSCBAm}b with the help of \req{average},
and keeping the terms with the largest phase volume we find
\be
\begin{split}
&\left\langle
 \left[\delta   \Gamma^{(in)}_l(\ep)\right]^2
 \right\rangle=
\frac{\pi\lambda^4\delta_\zeta}{2}
\int\diff\ep_1\diff\omega\,
\frac{\Upsilon^8(\omega/\delta_\zeta)\left[
F^\Rightarrow(\ep_1+\omega,\ep_1;\ep-\omega)\right]^2}
{\Gamma^{(in)}(\ep+\omega)+\Gamma^{(in)}(\ep_1)
+\Gamma^{(in)}(\ep_1-\omega)}.
\end{split}
\label{gamma2}
\ee

Our initial assumption that the inelastic rate is self-avergaing is
justified provided that
\be
\left\langle
 \left[\delta   \Gamma^{(in)}_l(\ep)\right]^2
 \right\rangle \lesssim
\left[\left\langle\Gamma^{(in)}_l(\ep)
 \right\rangle\right]^2.
\label{mcondition1}
\ee
According to \reqs{gammam1p} and \req{gamma2},
both sides of this inequality are determined by the
distribution function only, so that \req{mcondition1}
is a sufficient (but not necessary)
condition for the arbitrary non-equilibrium
state  to  be metallic.

For the thermal distribution $n(\ep)=\tanh\frac{\ep}{2T}$, the
explicit expressions  can be obtained.
One finds
\be
\left\langle\Gamma^{(in)}\right\rangle=\pi \lambda^2 M T,
\label{gamma3}
\ee
where $M$ is the coefficient defined in \req{Upsilon},
and
\be
\left\langle\left(\delta\Gamma^{(in)}\right)^2\right\rangle=
\frac{\pi \lambda^4M\delta_\zeta^2T}{36\langle\Gamma^{(in)}\rangle}.
\label{gamma2p}
\ee
The condition \rref{mcondition1}, thus, reduces to
the lower bound for the temperature
\be
T \gtrsim  T_{in}\equiv
\frac{\delta_\zeta}{6\pi\lambda M}
.
\label{mcondition2}
\ee

To complete our discussion of the properties of
inelastic rate, we justify our assumptions.
To check the condition \rref{1cond}.
we anlalyze the structure of the energy integrals in
\reqs{gammam1}--\rref{gamma2} and find
$|l-l_1|, |l_2-l_3| \simeq M \gg 1$,  $|l-l_3| \simeq T/\delta_\zeta
\gg M$, which is consistent with \req{1cond}. Deriving
\req{gamma3} we assumed $T \gg M\delta_\zeta$. It is
consitent with \req{mcondition2} provided that the condition
\rref{Upsilon} is fullfilled.

Let us turn now to the properties of the elastic decay rate,
$\Gamma^{(el)}$, of \req{eqSCBA1}
\be
\begin{split}
&\Gamma_l^{(el)}(\ep,\vec{\rho})
=\pi\delta_{\zeta}^2I^2\sum_{\vec{a}}
A(\ep,\vec{\rho}+\vec{a})
;\\
&A(\ep,\vec{\rho})
=\frac{1}{\pi}\sum_{l_1}
\frac{\Gamma^{(in)}(\ep)}
{\left[\ep-\xi_{l_1}(\vec{\rho})\right]^2
+ \left[\Gamma^{(in)}(\ep)\right]^2}.
\end{split}
\label{gel1}
\ee
Let us note that only the inelastic width enters the
right-hand sides of these equations. Further iterations
of the elastic processes in expression for $A_l(\vec{\rho}+\vec{a})$,
see \req{eqSCBA1}, generate either terms small as
$I^2(\Gamma^{(in)}(\ep)/\delta_\zeta) \ll 1$ (originating from
off-resonant levels), or the divergent
term corresponding to the elastic return on the same level.
The latter contribution, however, corresponds to the spurious
diagram Fig.~\ref{figSCBA}b.1 and must be discarded.

Average and fluctuations of the elastic rate \rref{gel1} is
calculated with the help of the \req{average} and we find
\be
\begin{split}
&\left\langle\Gamma^{(el)}\right\rangle=
\left(2d\right) \pi I^2\delta_{\zeta};\quad
\left\langle
\left[\Gamma^{(el)}(\ep)\right]^2\right\rangle=
\frac{d \pi I^4\delta_{\zeta}^3}{\Gamma^{(in)}(\ep)},
\end{split}
\label{gelaverage}
\ee
where $2d$ is the number of the nearest neighbors.

\begin{figure}
\includegraphics[width=0.8\textwidth]{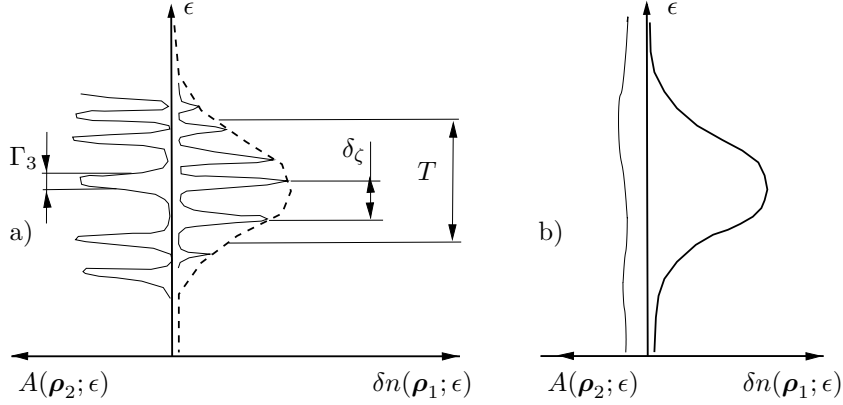}
\caption{
Sketches of the shapes of the
non-equilibrium influx function $\delta n(\vec{\rho}_1,\ep)$
and the spectral density  of
the other neighbours $A(\vec{\rho}_2;\ep)\equiv \sum_l
A_l(\vec{\rho}_2;\ep)$ of the other neighbour;
for a) low-temperature metal $T_{in}< T < T_{el}$;
b) $T > T_{el}$. See text for the further explanation.
}
\label{DoSm}
\end{figure}

Using \req{gamma3}, we find
\be
\begin{split}
&\frac{\langle[\Gamma^{(el)}]^2\rangle}
{\langle\Gamma^{(el)}\rangle^2}
= \frac{4T_{el}}{T},\quad
T_{el}=\frac{\delta_\zeta}{16\pi^2dM\lambda^2}
\simeq \frac{T_{in}}{\lambda},
\end{split}
\label{gelratio}
\ee
where the numerical factor is chosen for the convenience in
the further formulas,
At $T\gg{T}_{el}$ the level discreteness plays no role, so that both
elastic and inelastic decay rates are self-averaging,
see also Fig.~\ref{DoSm}. At
$T_{in}\ll{T}\ll{T}_{el}$ only $\Gamma^{(in)}$ is self-averaging,
while the fluctuations of~$\Gamma^{(el)}$ are large
compared to the average. However,
they are finite, so that the system is in the metallic state
according to criterion~\rref{criterion}.

\subsection{Kinetic equation and transport coefficients.}
\label{sec:metalB}

Having studied the statistical distribution of the decay rates,
we are ready to apply the same ideas to the kinetic equation:
\begin{equation}
\label{kinurav}
\begin{split}
&\partial_t n(\ep,\vec\rho) =
2\pi I^2 \delta_\zeta^2
\sum_{\vec{a}}A(\ep,\vec\rho+\vec{a})\Big[
 n(\ep,\vec\rho+\vec{a})
-
n(\ep,\vec{\rho})
\Big]
\\
&\quad+\frac{\pi \lambda^2}{2 }
\int \diff\omega\, \diff\ep_1 \Upsilon^4\!\left(\frac{\omega}{\delta_\zeta}\right)
A(\ep+\omega,\vec\rho)
\\
&\quad\times
\Big\{\left[n(\ep+\omega,\vec\rho)
-n(\ep,\vec\rho)\right]
\left[1-n(\ep_1,\vec\rho)\,n(\ep_1-\omega,\vec\rho)\right]
\\
&\quad\
+\left[1-n(\ep,\vec\rho)\,n(\ep+\omega,\vec\rho)\right]
\left[n(\ep_1-\omega,\vec\rho)-n(\ep_1,\vec\rho)\right]\Big\}.
\end{split}
\end{equation}


Equation \rref{kinurav} enables us to
find the kinetic coefficients in the system.
We look for the distribution function in the form
\begin{equation}\label{ntrial}
n(\ep,\vec{\rho},t)=
\tanh\frac{\ep}{2T}+\Phi(\ep,\vec{\rho},t)
+\varphi(\ep,\vec{\rho},t),
\end{equation}
where the function $\Phi(\ep,\vec{\rho},t)$ describes the shape
of the distribution function on the energy scale
$\ep\gtrsim{M\delta_\zeta}$, whereas the function $\varphi(\ep,\vec{\rho})$
encodes the structure on the scale $\ep\sim\Gamma^{(in)},\delta_\zeta$.
Namely, we impose the condition
\begin{equation}
\label{orthog}
\langle \varphi (\ep,\vec{\rho})\rangle_\ep
\equiv \int\limits^{\ep+{\Delta}}_{\ep-{\Delta}}
\frac{\diff\ep_1}{2\Delta}\, \varphi (\ep_1,\vec{\rho})=0,
\end{equation}
where  $\Delta$ is an  energy interval
$\delta_\zeta \ll \Delta \lesssim M\delta_\zeta$.
We substitute \req{ntrial} into \reqs{kinurav}
and linearize with respect to $\Phi$ and $\varphi$.
For the smooth part of the distribution function we find
\begin{subequations}
\begin{equation}
\label{smooth}
\begin{split}
&\partial_t \Phi(\ep,\vec\rho) =
2\pi I^2 \delta_\zeta
\sum_{\vec{a}}\Big[
 \Phi(\ep,\vec\rho+\vec{a})
-
\Phi(\ep,\vec{\rho})
\Big]
\\
&\quad+
2\pi I^2 \delta_\zeta^2
\sum_{\vec{a}}\left\langle A(\ep,\vec\rho+\vec{a})\Big[
 \varphi(\ep,\vec\rho+\vec{a})
-
\varphi(\ep,\vec{\rho})
\Big]\right\rangle_\ep
\\
&\quad
+ \widehat {\rm St}_{\Phi} \Phi(\ep,\vec{\rho})+
\left\langle
\widehat {\rm St}_{\varphi} \varphi(\ep,\vec{\rho})\right\rangle_\ep.
\end{split}
\end{equation}
Deriving \req{smooth} we used the fact that $\langle A(\ep)\rangle_\ep=
1/\delta_\zeta$ is a self-averaging quantity.

The smooth part, $\Phi$, contains,  in particular, density and energy density
which propagate diffusively through the system.
In contrast,
the oscillatory contribution decays due to the inelastic processes.
Thus, function $\varphi$ can be considered in the stationary limit
\begin{equation}
\label{sharp}
\begin{split}
0 =&2\pi I^2 \delta_\zeta^2
\sum_{\vec{a}}\left\|A(\ep,\vec\rho+\vec{a})\right\|_{osc}
\Big[
 \Phi(\ep,\vec\rho+\vec{a})
-
\Phi(\ep,\vec{\rho})
\Big]
\\
&
+
2\pi I^2 \delta_\zeta^2
\sum_{\vec{a}}\left\|A(\ep,\vec\rho+\vec{a})
\Big[
 \varphi(\ep,\vec\rho+\vec{a})
-
\varphi(\ep,\vec{\rho})
\Big]\right\|_{osc}
\\
&
+\left\|
\widehat {\rm St}_{\varphi} \varphi(\ep,\vec{\rho})\right\|_{osc}.
\end{split}
\end{equation}
where we introduced the notation for the oscillatory
part of the expression
\[
\left\|\dots\right\|_{osc} \equiv \dots- \langle\dots\rangle_{\ep}.
\]
The smooth  part of the linearized collision integral
$\widehat{\rm St}_\varphi\varphi$
is given by
\be
\begin{split}
\label{phismooth}
\left\langle\widehat{\rm St}_\varphi \varphi(\ep)\right\rangle_\ep
&=2 \pi\lambda^2T
\int\, \diff\omega \Upsilon^4\!\left(\frac{\omega}{\delta_\zeta}\right)
\left\langle A(\ep+\omega)\left[
\varphi(\ep+\omega)-\varphi(\ep)\right]\right\rangle_\ep
\\
&
\approx
2 \pi \lambda^2T
\int\, \diff\omega \Upsilon^4\!\left(\frac{\omega}{\delta_\zeta}\right)
\left\langle A(\ep+\omega)
\varphi(\ep+\omega)\right\rangle_\ep
\\ &
\approx
 2\Gamma^{(in)}\delta_\zeta\left\langle A(\ep)
\varphi(\ep)\right\rangle_\ep.
\end{split}
\ee
Here we used the fact that $\left\langle A(\ep+\omega)
\varphi(\ep)\right\rangle_\ep = \left\langle A(\ep+\omega)\right\rangle_\ep
\left\langle\varphi(\ep)\right\rangle_\ep =0$, for $|\omega| \gtrsim
\delta_\zeta$ and the contribution to the integral is determined by
$|\omega| \simeq M\delta_\zeta$. We also
used the expression for
the inelastic rate \rref{gamma3}. By the same token,
we find the oscillatory part as
\be
\begin{split}
\label{phisharp}
\left\|\hat{\rm St}_\varphi \varphi(\ep)\right\|_{osc}
&=\frac{2 \pi\lambda^2T}{ \delta_\zeta}
\int\, \diff\omega \Upsilon^4\!\left(\frac{\omega}{\delta_\zeta}\right)
\left[\varphi(\ep+\omega)-\varphi(\ep)\right]
\approx -2\Gamma^{(in)} \varphi(\ep).
\end{split}
\ee
We remind the reader that the relationship $T \gg T_{in}\gg M\delta_\zeta$ is
widely used, see also \req{mcondition2}.
Finally, the shape of the smooth distriution function is
stabilized by the linearized collision integral
\begin{equation}
\begin{split}\label{hatStPhi=}
\widehat{\mathrm{St}}_\Phi\Phi(\ep)=&
\frac{{2}\pi\lambda^2}{\delta_\zeta} \int\diff\omega\,
\Upsilon^4\!\left(\frac\omega{\delta_{\zeta}}\right)
\left\{\frac\omega{2}\,\coth\frac\omega{2T}
\left[\Phi(\ep+\omega)-\Phi(\ep)\right]\right.\\
&+\frac\omega{2}
\left[\Phi(\ep+\omega)\tanh\frac\ep{2T}+
\Phi(\ep)\tanh\frac{\ep+\omega}{2T}\right]\\
&+\frac{1}{4}
\left(\tanh\frac{\ep-\omega}{2T}-\tanh\frac{\ep}{2T}\right)
\\&\quad\times\left.
\int\diff\ep_1
\left[\Phi(\ep_1+\omega)\tanh\frac{\ep_1}{2T}+
\Phi(\ep_1)\tanh\frac{\ep_1+\omega}{2T}\right] \right\}.
\end{split}\end{equation}

This linear operator has two zero modes
\be
\Phi_\mu(\ep)=-\partial_\ep \tanh\frac{\ep}{2T};
\quad \Phi_T(\ep)=\partial_T \tanh\frac{\ep}{2T},
\label{zeromodes}
\ee
which reflect the conservation of number of particles and energy
by the inelastic processes.

Because of the condition $T\gg M\delta_\zeta$, \req{hatStPhi=}
can be rewritten in the energy diffusion approximation
\be
\begin{split}
&\widehat{\mathrm{St}}_\Phi\Phi(\ep)=
{D}^{(\ep)}
\Bigg\{\partial_\ep^2\Phi(\ep)
+\frac{\partial_\ep}{2}
\left[\Phi(\ep)\tanh\frac{\ep}{2T}\right]
\\&\qquad\qquad
-\frac{\partial_\ep^2}{4}\tanh\frac{\ep}{2T}
\int \diff\ep_1 \Phi(\ep_1)\tanh\frac{\ep_1}{2T}
\Bigg\};
\\ &
{D}^{(\ep)}=\frac{\pi M^3\delta_\zeta^2 \lambda^2T}{12}.
\end{split}
\label{DEp}
\ee

\label{smoothsharp}
\end{subequations}

The simple form of the collision integral \rref{phisharp} enables
us to solve \req{sharp}. We adopt the following relation
\be
\Gamma^{(in)} \gg I\delta_\zeta,
\label{relation}
\ee
which is automatically fullfilled\footnote{
The regime $\Gamma^{(in)} \ll
I\delta_\zeta$, which may occur for somewhat artificial for our model choice
$I\gg \lambda$, corresponds to coherent oscillations of the
population between two resonant levels. In this regime one should
modify Eqs.~(\ref{solsharp}) and~(\ref{finalsmooth}) as
\begin{eqnarray*}
&&\varphi(\ep,\vec{\rho})=
\sum_{\vec{a}}\Big[ \Phi(\ep,\vec\rho+\vec{a})-\Phi(\ep,\vec{\rho})\Big]
\\&&\qquad\times
\left\|\frac{2\pi I^2 \delta_\zeta^2A(\ep,\vec\rho+\vec{a})}
{2\Gamma^{(in)}+2\pi I^2 \delta_\zeta^2
[A(\ep,\vec\rho+\vec{a})+A(\ep,\vec\rho)]}\right\|_{osc},\\
&&\partial_t \Phi(\ep,\vec\rho)
-\widehat {\rm St}_{\Phi}\Phi(\ep,\vec{\rho})
=\sum_{\vec{a}}\Big[\Phi(\ep,\vec\rho+\vec{a})
-\Phi(\ep,\vec{\rho})\Big]
\\&&\qquad \times
\left\langle\frac{2\pi I^2 \delta_\zeta^2 A(\ep,\vec\rho)\,
2\Gamma^{(in)} A(\ep,\vec\rho+\vec{a})}
{2\Gamma^{(in)}+2\pi I^2 \delta_\zeta^2
[A(\ep,\vec\rho+\vec{a})+A(\ep,\vec\rho)]}
\right\rangle_\ep\,.
\end{eqnarray*}
One should also use the exact spectral densities for the resonant
pairs, as obtained from the solution of a $2\times{2}$ problem.
The results for the low-
and high-temperature regimes remain unchanged, but an  intermediate
region appears.
This regime is analogous to that studied in 
Ref.~\cite{GMR} for the conductivity due to the scattering
of localized electrons on phonons.
}
for $T \gg{T}^{(in)}$, see \req{mcondition2}, and $I\simeq \lambda$.
Then $\Gamma^{(in)} \gg I^2 \delta_\zeta^2  A(\ep)$ and \req{sharp}
yields
\be
\begin{split}
\varphi(\ep,\vec{\rho})&=
\frac{\pi I^2 \delta_\zeta^2}{\Gamma^{(in)}}
\sum_{\vec{a}}\left\|A(\ep,\vec\rho+\vec{a})\right\|_{osc}
\\
&
\times\Big[
 \Phi(\ep,\vec\rho+\vec{a})
-
\Phi(\ep,\vec{\rho})
\Big]
\left[1+\mathcal{O}\left(\frac{I^2\delta_\zeta^2}{
\left[\Gamma^{(in)}\right]^2}\right)\right].
\end{split}
\label{solsharp}
\ee
Substituting \req{solsharp} into
\reqs{phismooth} and \rref{smooth}, and keeping once again only
the terms leading in $({I\delta_\zeta}/{\Gamma^{(in)}})^2$, we
find the equation for the smooth part of the distribution function
\be
\begin{split}
&\partial_t \Phi(\vec\rho) =\widehat {\rm St}_{\Phi} \Phi(\vec{\rho})
+
2\pi I^2 \delta_\zeta^3
\sum_{\vec{a}}
\left\langle A(\ep,\vec\rho)
A(\ep,\vec\rho+\vec{a})
\right\rangle_\ep
\Big[
 \Phi(\vec\rho+\vec{a})
-
\Phi(\vec{\rho})
\Big],
\end{split}
\label{finalsmooth}
\ee
where we omitted the energy argument of the distribution function.

Equation \rref{finalsmooth} deserves some discussion. The first
term on the right-hand side determines dynamics of the population
at different energies~$\ep$ on the site~$\vec\rho$. The second
term describes transport between neighboring sites. The inspection
of this term shows that for $\Gamma^{(in)} \ll \delta_\zeta$ this
term is contributed mostly by rare overlaps of the peaks in the
spectral densities on the neighboring sites, see Fig.~\ref{DoSm}.
We will call such rare events {\em ``pin-holes''}.
The typical energy separation between the pin-holes is
$\delta_\zeta^2/\Gamma^{(in)}$. Thus, for a given energy this term
may not be self-averaging, and the calculation of the observable
transport coefficients requires further statistical analysis.
We note that the situation when transport is dominated by rare
configurations is quite common in systems with strongly fluctuating
local transmission, see Ref.~\cite{Raikh}.

To perform this analysis, we  assume that the energy
relaxation rate within each site is larger than the rate of the
tunnelling into the neighbors (this 
assumption is
justified for our model in Appendix~\ref{subslow}). 
In this case the solution is restricted to the zero modes
of the inelastic collision integral \rref{zeromodes}
\be
\Phi(\ep,\vec\rho;t)= \delta\mu(\vec\rho,t)\,\Phi_\mu(\ep) +
\delta{T}(\vec\rho,t)\,\Phi_T(\ep). \label{sol1}
\ee
Substituting \req{sol1} into
\req{finalsmooth}, performing the corresponding energy integration
to utilize the properties $\int \diff\ep\, \widehat {\rm St}_{\Phi}
\Phi(\ep,\vec{\rho})=0$ and $\int \diff\ep\, \ep\, \widehat {\rm St}_{\Phi}
\Phi(\ep,\vec{\rho})=0$, and neglecting thermopower and Peltier
coefficients (which have random signs), we find
\begin{subequations}
\be
\begin{split}
&\frac{e^2}{\delta_\zeta}
\frac{\partial \mu(\vec\rho)}{\partial t}
=\sum_{\vec{a}}G\!\left(\vec\rho +\frac{\vec{a}}{2}\right)
\left[\mu(\vec\rho+\vec{a})-\mu(\vec\rho)\right];\\
&c_V\,\frac{\partial T(\vec\rho)}{\partial t}=
\sum_{\vec{a}}\mathcal{K}\!\left(\vec\rho +\frac{\vec{a}}{2}\right)
\left[T(\vec\rho+\vec{a})-T(\vec\rho)\right];
\label{cv}
\end{split}
\ee
where $c_V =\pi^2\nu T/3$ is the specific heat per localization
cell. The electrical conductance~$G$ and the thermal
conductance~$\mathcal{K}$ are defined for each link $\vec\rho+\vec{a}/2$
connecting two sites $\vec\rho$ and $\vec\rho+\vec{a}$.
They are given by
\be
\begin{split}
&G\!\left(\vec\rho +\frac{\vec{a}}{2}\right)
=\frac{2\pi e^2 I^2}{\hbar}\,
\Beta_\sigma\!\left(\vec\rho +\frac{\vec{a}}{2}\right);
\qquad
\mathcal{K}\!\left(\vec\rho +\frac{\vec{a}}{2}\right)
=\frac{2\pi^3 e^2I^2T}{3\hbar}\,
\Beta_\kappa\!\left(\vec\rho +\frac{\vec{a}}{2}\right),
\end{split}
\label{GK}
\ee
where $\Beta_{\sigma,\kappa}$ are
dimensionless random quantities determined
by the overlaps of the densities of states:
\be
\begin{split}
\Beta_{\sigma,\kappa}\!\left(\vec\rho+\frac{\vec{a}}{2}\right)&=
\delta_{\zeta}^2\int \frac{\diff\ep}{2T}\,
\beta_{\sigma,\kappa}\!\left( \frac{\ep}{2T}\right)
A(\ep,\vec\rho)\,A(\ep,\vec\rho')
\\
&=
\sum_{l,l'}
\frac{\delta_\zeta}{2T}\,\beta_{\sigma,\kappa}\!
\left( \frac{\xi_{l}(\vec{\rho})}{2T}\right)
\frac{2\delta_\zeta\Gamma^{(in)}/\pi }
{
\left[\xi_{l}(\vec{\rho})-
\xi_{l'}(\vec{\rho}')\right]^2
+ \left[2\Gamma^{(in)}\right]^2},
\end{split}
\label{smallg}
\ee
where $\vec{\rho}'\equiv \vec\rho+\vec{a}$. We used the explicit
expression \req{gel1} for $A(\ep,\vec\rho)$ and the condition
$\Gamma^{(in)}\ll T$.
Dimensionless functions $\beta_{\sigma,\kappa}(x)$ such that
$\int\beta_{\sigma,\kappa}(x)\diff{x}=1$,
are given by
\begin{equation}\begin{split}
&\beta_\sigma(x)=\frac{1}{2\cosh^2x}\,,\quad
\beta_\kappa(x)=\frac{6x^2}{\pi^2\cosh^2x}\,.
\end{split}
\label{smallbeta}
\end{equation}
\label{resistornetwork}
\end{subequations}

Equations \rref{resistornetwork} are nothing but the description
of a network of random conductors (thermal conductors). It is easy
to check that the average of these quantities over the
realizations of $\xi_l$ gives the temperature-independent result
$\langle \Beta_{\sigma,\kappa}\rangle=1$. However, this result is
meaningless as the observable conductivities are determined by
typical rather than rare events. In fact, observable
conductivity~$\sigma$ and observable thermal conductivity~$\kappa$
are given by
\be
\begin{split}
\sigma=\frac{2\pi e^2 I^2\zeta_{loc}^{2-d}}{\hbar}
&\times
\left\{
\begin{matrix}
\left[\left\langle\frac{1}{\Beta_\sigma}\right\rangle\right]^{-1}; & d=1;
\\
\\
\exp\left[\langle \ln {\Beta_\sigma}\rangle\right] ;& d=2;
\\
\\
\exp\left[\langle
\ln ^{\alpha_d} {\Beta_\sigma}\rangle^{1/\alpha_d}\right]; & d>2;
\end{matrix}
\right.
\\
\\
\kappa=\frac{2\pi^3 e^2 T I^2\zeta_{loc}^{2-d}}{3\hbar}
&\times
\left\{
\begin{matrix}
\left[\left\langle\frac{1}{\Beta_\kappa}\right\rangle\right]^{-1}; & d=1;
\\
\\
\exp\left[\langle \ln {\Beta_\kappa}\rangle\right]; & d=2;
\\
\\
\exp\left[\langle
\ln ^{\alpha_d} {\Beta_\kappa}\rangle^{1/\alpha_d}\right]; & d>2;
\end{matrix}
\right.
\end{split}
\label{observables}
\ee
where $d$ is the dimensionality of the system and $\alpha_d$ are
the numerical coefficients of the order of unity which are not
known analytically for $d>2$.
The formula for $d=1$ is the trivial result for the random
resistors connected in series whereas the result for $d=2$ follows from the
duality arguments~\cite{Dykhne}.

The averages entering \req{observables} can be immediately
calculated if the characteristic functions
$\tilde{P}_{\sigma,\kappa}(s)=
\left\langle \eexp^{-s\Beta_{\sigma,\kappa}}\right\rangle$
of the distributions are
known:\footnote{The average of the logarithm is calculated using the
identity
\[
\int\limits_0^\infty\diff{s}\,
\frac{\eexp^{-as}-\eexp^{-bs}}{s}=\ln\frac{b}{a}\,.
\]
}
\be
\begin{split}
&\left\langle\frac{1}{\Beta_{\sigma,\kappa}}\right\rangle =
\int_0^\infty\diff{s}\,
\tilde{P}_{\sigma,\kappa}(s);
\qquad
\left\langle\ln{\Beta_{\sigma,\kappa}}\right\rangle =
\int_0^\infty\frac{\diff{s}}{s}\left[\eexp^{-s}-
\tilde{P}_{\sigma,\kappa}(s)\right].
\end{split}
\label{fromLaplace}
\ee
The characteristic functions $\tilde{P}_{\sigma,\kappa}(s)$ can be
found from the definition~\rref{smallg}  by a straightforward
calculation given in Appendix~\ref{app1}: 
\be
\begin{split}
&\tilde{P}_{\sigma,\kappa} (s)
=\exp\left\{-\int\limits_{-\infty}^{\infty}\diff{x}
\left[r
  S_2\left(\frac{s\beta_{\sigma,\kappa}(x)}{r}\right)\right]\right\},
\quad
S_2(y)=y
\eexp^{-y}\left[\mathcal{I}_0(y)+\mathcal{I}_1(y)\right],
\end{split}
\label{FromAppendix}
\ee
where $\mathcal{I}_0$~and~$\mathcal{I}_1$
are the modified Bessel functions, and $\beta_{\sigma,\kappa}$ are
given by \req{smallbeta}.
The result is controlled by a
dimensionless parameter
\be
r(T)=\frac{{8}\pi\Gamma^{(in)}{T}}{\delta_\zeta^2}=
\frac{8\pi^2\lambda^2 MT^2}{\delta_\zeta^2}
=\frac{1}{2d} \left(\frac{T^2}{\delta_\zeta T_{el}}\right).
\label{rT}
\ee
The meaning of this parameter is the typical
number of resonances in the energy strip $|\ep|\lesssim{T}$.

\begin{figure}
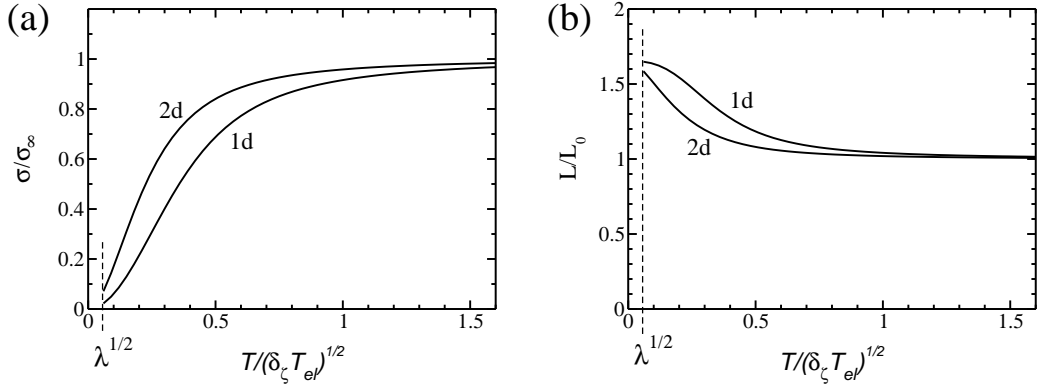

\includegraphics[width=0.47\textwidth]{fig10a}\hspace{0.5cm}
\includegraphics[width=0.47\textwidth]{fig10b}
\caption{
(a) Lower and upper curves represent the conductivity~$\sigma$
in the units of the Drude conductivity~$\sigma_\infty$ for the 1d and
2d cases, respectively [{\em i.e.}, $\langle\Beta_\sigma^{-1}\rangle^{-1}$
and $\exp(\langle\ln\Beta_\sigma\rangle)$] as a function of
$\sqrt{2dr(T)}$ --
temperature in the units of $\sqrt{\delta_\zeta{T}_{el}}$, as given
by Eqs.~(\ref{observables}),~(\ref{fromLaplace}) in the fast energy
relaxation appproximation (this curve does not depend on any parameters).
(b) The relative Lorentz
number $\mathrm{L}/\mathrm{L}_0$ versus temperature in the units of
$\sqrt{\delta_\zeta{T}_{el}}$ for 1d (upper curve) and 2d (lower
curve) cases. Vertical dashed lines indicate the limit
of the applicablity of the theory to the metallic region,
see \req{mcondition2}.                                   
}
\label{Plot1d}
\end{figure}

The result of numerical integration of Eqs.~(\ref{fromLaplace}),
(\ref{FromAppendix}) is plotted on Fig.~\ref{Plot1d}.
In two limiting cases $r\gg{1}$ and $r\ll{1}$ we have
\be\begin{split}
\label{answerLaplace}
&r\gg{1}:\quad\left\{\begin{array}{l}
\tilde{P}_\sigma(s)\approx \eexp^{-s}[1+(1/6)s^2/r],\\
\tilde{P}_\kappa(s)\approx
\eexp^{-s}[1+(7/10-6/\pi^2)s^2/r],
\end{array}\right.\\
&r\ll{1}:\quad\left\{\begin{array}{l}
\tilde{P}_\sigma(s)\approx \eexp^{-\sqrt{\pi{r}s}},\\
\tilde{P}_\kappa(s)\approx
\eexp^{-\sqrt{(192\mathrm{G}^2/\pi^3)rs}},
\end{array}\right.
\end{split}\ee
where $\mathrm{G}\approx{0}.916\cdots$ is the Catalan's constant.
Substituting \req{answerLaplace} into \req{fromLaplace} we
find that for $r\gg{1}$, $P_{\sigma,\kappa}(\Beta)$ are strongly
peaked around~1, so that the temperature indepedent Drude resistivity
is restored:
\be
\begin{split}
&\sigma(T \gg \sqrt{\delta_\zeta T_{el}})
\approx\sigma_\infty
\left(1-\frac{2}{3}\,\frac{\delta_\zeta T_{el}}{T^2}\right);
\\
&
\kappa(T \gg \sqrt{\delta_\zeta T_{el}})
\approx\kappa_\infty(T)
\left[1-\left(\frac{14}5-\frac{24}{\pi^2}\right)
\frac{\delta_\zeta T_{el}}{T^2}\right];
\\
&
\sigma_\infty\equiv\frac{2\pi e^2 I^2\zeta_{loc}^{2-d}}{\hbar}\,,\quad
\kappa_\infty(T)\equiv\frac{2\pi^3 e^2 T I^2\zeta_{loc}^{2-d}}{3\hbar}\,.
\end{split}
\ee
In other words, the effect of localization of one-particle
wave functions on transport is completely removed by the inelastic processes
even in the ``non-ergodic'' regime of $\Gamma^{(in)} \ll
\delta_\zeta$, where the peaks in the one-particle density of states
are still well-resolved.

In the opposite case, $r\ll{1}$ we have for $d=1,2$:
\be\begin{split}
&{\sigma(T \ll \sqrt{\delta_\zeta T_{el}})}
=
\sigma_\infty
\frac{\pi}{4}\left(\frac{T^2}{\delta_\zeta T_{el}}\right),
\quad
\kappa
(T \ll \sqrt{\delta_\zeta T_{el}})
=
\kappa_\infty(T)\,
\frac{48\mathrm{G}^2}{\pi^3}\left(\frac{T^2}{\delta_\zeta T_{el}}\right),\\
\label{results}
\end{split}\ee
(for larger dimensionalities the temperature dependence is the same but
the numerical coefficients could not be found analytically).

These results have the following meaning.
At $T \gg \sqrt{\delta_\zeta{T}_{el}}$
the quantities are self-averaging and the result
is temperature independent.
At $T \ll \sqrt{\delta_\zeta{T}_{el}}$ the electron on a site cannot
explore enough pin-holes to find the rare resonant
one which would determine the average.
As a result, it chooses the best available pin-hole ({\em i.e.}, the
one with the smallest separation between the levels).
The denominator in the \req{smallg} can be estimated
as $|\xi(\vec\rho)-\xi(\vec\rho')| \simeq \delta_\zeta/n$,
where $n \simeq T/\delta_\zeta$ is the number of the levels
available for the electrons. As a result, the largest
term entering into the sums in \req{smallg} is $\simeq \Gamma^{(in)}T
\propto T^2$. It explains the power law dependence at low temperature.

It is instructive to investigate the validity of the Wiedemann-Frantz
law. From \req{results} we find for the Lorentz number
\be
\frac{\mathrm{L}(T)}{\mathrm L_0}\equiv
\frac{3e^2 \kappa (T)}{ \pi^2\sigma(T)T} =
\left\{
\begin{matrix}\displaystyle
1+\left(\frac{24}{\pi^2}-\frac{32}{15}\right)
\frac{\delta_\zeta T_{el}}{T^2}\,,& T\gg \sqrt{\delta_\zeta T_{el}},\\ \\
\displaystyle{\frac{192\mathrm{G}^2}{\pi^4}}\approx 1.65\dots,&T\ll
\sqrt{\delta_\zeta T_{el}}.
\end{matrix}
\right.
\label{WF}
\ee
The origin of the violation of the Wiedemann-Franz is the following.
For the conductivity all the tunnelling pairs with the energies $\simeq T$
are roughly of the same importance. For the thermal conduction, however,
the contributions of the low energy part is less important, and the best
tunnelling pairs are different from those for the conductivity. It
leads to the renormalization of the Lorentz number \rref{WF} by the factor
of the order of unity.

We conclude this section by noticing that the contribution of the
rare pin-holes leads to the deviation from the ``natural''
assumption $\sigma \propto \zeta^2_{loc}\Gamma^{(in)}$, {\em cf.}
Refs.~\cite{GMR,Gornyiv1}. We think,
however, that the relative contribution of such configurations
at $T_{in}<T<T_{el}$ is a model dependent question, and do not pursue
this line further.

\section{Stability of the insulating phase.}
\label{sec:insulator}

Even though the condition \rref{mcondition2} gives the sufficient
condition for the metallic state to be stable, the consideration
of the previous section does not prove the existence or the
stability of the insulating phase, but rather gives the
indications of the breakdown of the calculational scheme at $T
\lesssim T_{in}$. To show the existence of the transition, we
have to prove that perturbation theory from insulating side is
also convergent under certain conditions. This analysis is the
subject of the present section.
 
The notion of the statisitical averaging is somewhat non-standard
in the insulating phase and its general aspects are discussed in
Sec.~\ref{sec:canonical}. To demonstrate 
the stability of the insulating phase we linearize the
nonlinear equation \rref{eqSCBA1} in Sec.~\ref{SCBAlinear} and
obtain an equation with the random coefficients somewhat similar
to that of Ref.~\cite{AbouChacra}. Further consideration will be
based on the statistical analysis of this equation, using
the technique described in Sec.~\ref{sec:Mayer}.
To begin with, we will
investigate the simpler but yet instructive case of the
zero-dimensional system $I=0$ to illustrate the general structure
and make the connections with Ref.~\cite{AGKL}, see
Sec.~\ref{0D}. To complete the study, 
we generalize the  consideration to higher dimensions in
Sec.~\ref{bulk} and calculate the transition temperature.

\subsection{Averaging procedure}
\label{sec:canonical}

Equations \rref{SCBA} form an infinite set of
coupled non-linear equations whose coefficients are functions of
the random energies $\xi_l$ and occupation numbers $n_l(\ep)$. 
Moreover, even though the equilibrium distribution
$n_l(\ep)= \tanh\frac{\ep}{2T}$ nullifies the
collision integral, the time of the relaxation of an arbitrary  
distribution $n_l(\ep)$ to the equilibrium one becomes infinite
for the insulating phase.
Therefore, $n_l(\ep)= \tanh\frac{\ep}{2T}$ has the meaning of the 
occupation number averaged over an infinitely long period of
time (taken to infinity {\em prior} to setting the coupling to
the bath equal to zero) and bears no information about the state
of the system at a given instant of time.

The basis in the space of many-body states of the system is formed
by states represented as Slater determinants built on single-particle
states $(l,\vec\rho)$. Such Slater determinants correspond to the
occupation numbers $n_l(\ep)=\pm{1}$.
Thus, the decay rates $\Gamma_l$ can be considered for each given
set $\{n_l\}$. 
Because the transition may occur only if the number of terms
contributing to $\Gamma_l$ is large, we can also perform the
statistical average with respect to those occupation numbers.
We assume them to be arbitrary with the only constraint being
to fix the global energy of the system (microcanonical ensemble).
Because the number of excitations involved into
the formation of the decay rate is much smaller than
the total number of the excitations in the system (which scales
proportionally to the volume), averaging
over the microcanonical ensemble with the energy $E$ (counted from
the ground state) is equivalent to the averging over the
canonical ensemble with temperature $T$ such that
\be
E=\frac{\pi^2
}{6}\nu \mathcal{V} T^2,
\label{Temp}
\ee
where $\nu$ is the averaged density
of states per unit volume and  $\mathcal{V}$ is the volume of the system.
We assume $T \gg M\delta_\zeta$ and verify later that the transition
occurs only at such temperature.

To find the distribution functions of the local and thus strongly
fluctuating quantities, we perform the ensemble averaging on
the final stage of the calculation.
Thus, the averaging procedure in this section is defined as
\be
\begin{split}
&\Big\langle \dots \Big\rangle
= \lim_{N\rightarrow\infty}
\int\limits_{-N\delta_\zeta/2}^{N\delta_\zeta/2}
\prod_{l=1}^N\prod_{\rho} \frac{\diff\xi_l(\vec\rho)}{N\delta_\zeta}
\sum_{n_l(\vec\rho)=\pm
1}\frac{\exp\frac{n_l(\vec\rho)\xi_l(\vec\rho)}{2T}
}{2\cosh\frac{\xi_l(\vec\rho)}{2T}}\;
 \dots,
\end{split}
\label{averaging} 
\ee 
where the number of levels $N$ and the domain of integration
is in accordance with \req{delta}.
Formula \rref{averaging} does not take into account the level
repulsion, see \req{levelrepulsion}. This repulsion will be
included later when it is necessary. 

Let us notice that the same averaging formula would arise if
one considered the probability of the occupation (and not the
average occupation) of the levels formed
as a result of the arbitrarily weak interaction with equilibrium phonons
kept at temeperature $T$. In this respect temperature $T$ in
\req{averaging}  has a meaning of the experimentally measurable quantity.

Having discussed the issue of how to average, we 
are prepared to calculate the characteristic function \rref{Wgeneral}
to establish the conditions of the stability of the insulating phase.

\subsection{Linearized ImSCBA equations}
\label{SCBAlinear}

To begin the actual analysis of the insulating phase, we notice
that in the absence of the external bath $\Gamma_l^{(bath)}(\ep)$,
substitution $\Gamma_l(\ep)=0$ solves the self-consistency
equation~\rref{eqSCBA1}. However, in accordance with the discussion
of Sec.~\ref{sec:problem} the order of limits is important, so
that we have to investigate the stability of the solution
$\Gamma_l(\ep)=0$ with respect to small but finite coupling to the
bath. The smallness of $\Gamma_l(\ep) \propto b$ enables us to linearize
the spectral density in \req{eqSCBA1} as
\be
{A}_l(\ep) \approx \delta\left(\ep-\xi_l\right)+ \frac{1}{\pi}
\frac{{\Gamma}_l(\ep)} {\left(\ep-\xi_l\right)^2 },
\label{linA} \ee
where the second term is understood as the principal value 
\be
\frac{1}{(\ep-\xi_l)^2} \to \Re \frac{1}{(\ep-\xi_l+i0)^2}.
\label{meaning} \ee
We substitute \req{linA} into the equation for $\Gamma_l(\ep)$, see
\req{eqSCBA1}, notice that the probablity of matching the levels
into exact resonance equals to zero, and keep the terms only
linear in $\Gamma_l(\ep)$. Taking into account that,
according to Sec.~\ref{sec:canonical}, $n_l(\ep)=n_l=\pm
1$, we obtain
\begin{subequations}
\be
\begin{split}
&\Gamma_l(\ep)=\Gamma_l^{(bath)}(\ep) 
+  \sum_{l_1,\vec{a}}
\frac{I^2\delta_{\zeta}^2 \Gamma_l(\ep,\vec\rho+\vec{a})
\theta_\Delta[\ep-\xi_{l_1} (\vec\rho+\vec{a})]}
{\left[\ep-\xi_{l_1} (\vec\rho+\vec{a})\right]^2 }
\\
&\qquad\quad+ \sum_{l_1,l_2,l_3} \frac{\lambda^2\delta_\zeta^2
Y_{l_1,l_2}^{l_3,l}F_{l_1, l_2; l_3}^\Rightarrow
\theta_\Delta(\ep-\xi_{l_1}-\xi_{l_2}+\xi_{l_3})
}
{\left(\ep-\xi_{l_1}-\xi_{l_2}+\xi_{l_3}\right)^2 }
\\
&\qquad \qquad\times \Big[2\Gamma_{l_1}(\ep-\xi_{l_2}+\xi_{l_3}) +
\Gamma_{l_3}(\xi_{l_1}+\xi_{l_2}-\ep) \Big],
\end{split}
\label{linSCBA2} \ee
where the denominators are defined in the sense of \req{meaning},
and
\be
\begin{split}
F^\Rightarrow_{l_1, l_2;l_3}&= \frac{1}{4} \Big\{1+ n_{l_1} n_{l_2}
-n_{l_3} \left[n_{l_1} + n_{l_2 }\right] \Big\}
= \left\{
\begin{matrix}
1; & {\rm if}\quad  n_{l_1}=n_{l_2}= -n_{l_3}=\pm 1;
\\
0; & {\rm otherwise}.
\end{matrix}
\right.
\end{split}
\label{linF} 
\ee 
The decay due to the connection with the bath
\be
\Gamma_l^{(bath)}(\ep)=\frac{(\ep-\xi_{l})b(\ep-\xi_{l})}{2}
\left[\coth \frac{(\ep-\xi_{l})}{2T} + n_l\right] \label{lbath} \ee
is a smooth function of $\ep$. As before, the coordinate
$\vec{\rho}$ is assumed to be the same in all terms in the equations
unless it is specified explicitly otherwise.
\end{subequations}

To deal with certain superfluous logarithmic divergences
in the future analysis of \req{linSCBA2}, 
we introduced an ultraviolet cut-off with the help of the function
\be
\theta_\Delta(x)=\left\{
\begin{matrix}
1; & |x| \leq \Delta\\
0; & |x| > \Delta.
\end{matrix}
\right.
\label{thetaDelta}
\ee
This ultraviolet cut-off is introduced for technical
convenience only and $\delta_\zeta\ll\Delta\lesssim
M\delta_\zeta$, compare with \req{orthog}. The contribution from
the part excluded by cut-off does not contain small
denominators. This part is self-averaging, proportional to
$b(\omega)$, and, thus, it is not relevant for  the question of
the stability of the insulator, see  \req{criterion2}.

The second term in \req{linSCBA2} is the effect of the one
particle tunneling into the neighboring localization cells. Due
to the condition \rref{Ismall} this term alone does not lead to
any significant effects. As this term corresponds to the solution of
the one-particle Schr\"odinger equation, it does not depend on the
occupation numbers $n_l$.
On the other hand, the last  term in \req{linSCBA2} describes the
decay due to the excitation of electron-hole pairs. The
availability of orbitals for such a process is controlled by
the set of  $n_l=\pm 1$.

Three following subsections are devoted to the statistical analysis
of \req{linSCBA2} with respect to the distribution
\rref{averaging}. Namely, we formally solve \req{linSCBA2} as
an expansion in small parameters $\lambda^2,\ I^2$:
\be
\begin{split}
&\Gamma_l(\ep)= \sum_{k,m}\Gamma_l^{(k,m )}(\ep);\quad
\Gamma_l^{(0,0 )}(\ep)\equiv \Gamma_l^{bath}(\ep); \quad
\Gamma_l^{(k,m )}(\ep) \propto \lambda^{2k}I^{2m},
\label{expansion}
\end{split}
\ee 
and calculate the characteristic functions
\be
W^{(k,m )}(s)=\left\langle \exp\left(-s  \Gamma_l^{(k,m )}(\ep)
\right)\right\rangle. 
\label{linearLaplace} 
\ee 
for each term. The
insulator is stable, see condition \rref{Wgeneral},
 if the typical value of each term decreases,
{\em i.e}
\be
\lim_{b\to 0}\lim_{\mathcal{V}\to \infty}
\lim_{k,m\to \infty} W^{(k,m )}(s) = 1, \label{linearcondition}
\ee 
for any fixed $s>0$.

\subsection{Mayer-Mayer cluster expansion for the characteristic functions}
\label{sec:Mayer}

\begin{figure}
\includegraphics[width=0.6\textwidth]{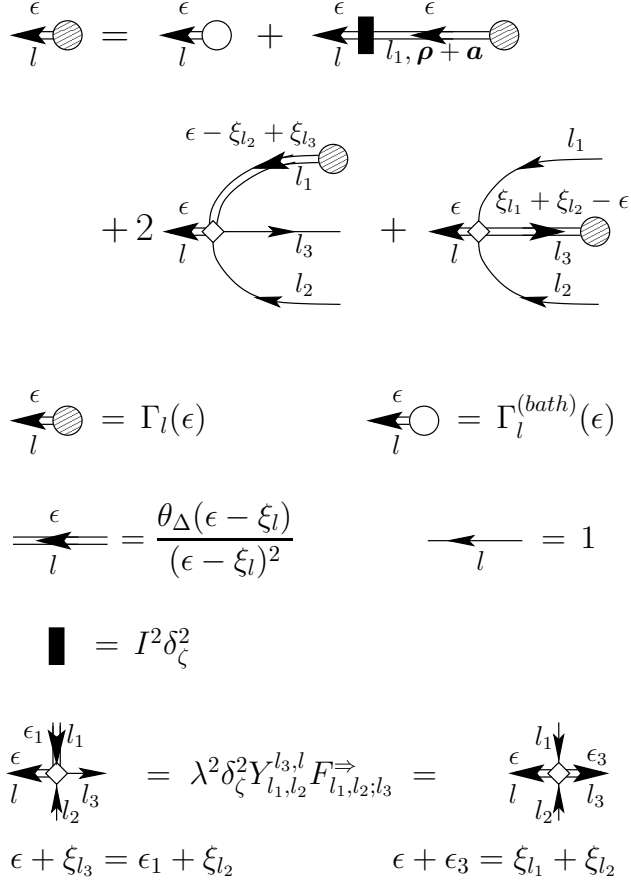}
\caption{ 
Diagrammatic representation of \req{linSCBA2}.
Solid lines bear an orbital energy $\xi_l$ and an occupation
number $n_l=\pm 1$. Double lines bear an orbital energy $\xi_l$,
an occupation number $n_l=\pm 1$, and an energy~$\ep$. The energy
carried by the double line is conserved in each hopping vertex
(bold rectangle) and changes in the interaction vertex (white
square). The algebraic sum of the energies entering the
interaction vertex through the double lines and the orbital
energies entering through the solid lines is conserved (signs are
determined by arrow directions). All the other rules are defined
on the figure, and $\theta_\Delta$ is defined in \req{thetaDelta}.} 
\label{lcbadiag}
\end{figure}

In the actual calculation of the characteristic function
\rref{linearLaplace} we will use the analogy of this function and
averaging procedure \rref{averaging} with the partition function
of the system of classical particles interacting through a certain
many-particle potential. The energy $\xi_l$ plays the role of the
coordinate, occupation number
$n_l=\pm 1$ is equivalent to the particle spin, and
parameter $s$ is analogous to the ``inverse temperature''. In order to
find these interaction potentials we introduce the diagrammatic
representation of \req{linSCBA2}. This representation and all of
the notation are shown on  Fig.~\ref{lcbadiag}.

It is obvious that $k+m$ iterations of the diagrammatic equation
shown on Fig.~\ref{lcbadiag} produce diagrams for each term
$\Gamma^{(k,m)}$ of \req{expansion}. Every term depends on
all the orbital energies and all the occupation numbers.
At the same time, it can be represented as a  sum of terms depending only
on the set $\left\{\xi_{l},n_l\right\}_{l=1}^{3k+m}$, such that
\be
 \Gamma^{(k,m)} =
\sum_{l_1,l_2,...,l_{3k+m}} U(\xi_{l_1},n_{l_1}; \dots;
\xi_{l_{3k+m}},n_{l_{3k+m}}), \label{Gamma=U}
\ee where we
omitted the spatial coordinate $\vec\rho$. Thus, averaging
\rref{averaging} in the expression \rref{linearLaplace} is indeed
equivalent to the calculation of the partition function of $N\to
\infty$ particles interacting via $(3k+m)$-particle potential $U$.

\begin{figure}
\includegraphics[width=0.7\textwidth]{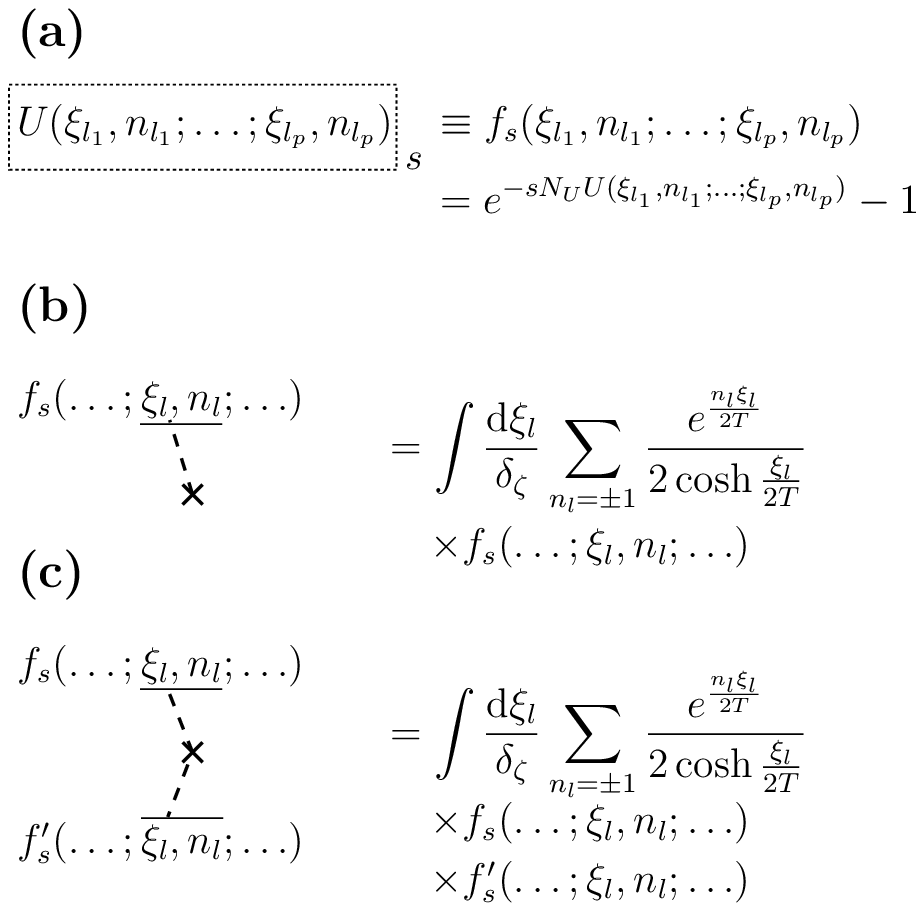}
\caption{(a) Definition of the cluster function~$f$; $N_U$ is the
degeneracy of the diagram -- the number of line permutations
keeping the diagram intact. (b)~Independent averaging over the
position~$\xi_l$ and the occupation~$n_l$ of a level~$l$.
(c)~Connected average of two different cluster functions
$f$~and~$f'$ (an arbitrary number of cluster functions can be
averaged analogously). The contribution of each linked cluster to
$\ln{W}^{(k,m)}_p$ ($p$~being the number of levels averaged over)
should be divided by the number of symmetries of the averaged
diagram. This coefficient will be written explicitly in each
specific case. } \label{clusterexpansion}
\end{figure}

To perform this calculation we employ the procedure known as
Mayer-Mayer cluster expansion~\cite{Mayer}:
\begin{equation}
\ln{W}^{(k,m)}=\sum_{p=3k+m}^\infty\ln{W}^{(k,m)}_p,
\end{equation}
where the summation is performed over linked clusters of
$p$~particles. Analogously to the calculation of the partition
function of a classical gas interacting via a two-particle
potential $U(\vec{r}_1-\vec{r}_2)$, whose leading term of the
cluster expansion is given by
\[
\ln{W}_2(T)=\int\frac{\diff^3\vec{r}}{V}
\left[\eexp^{-\frac{U(\vec{r})}{T}}-1\right],
\]
for each diagram \rref{Gamma=U} we introduce the cluster function
$f=\eexp^{-sU}-1$, as shown in Fig.~\ref{clusterexpansion}. On the
same figure we show the diagrammatic representation of the
averaging procedure and linking the clusters.

This procedure is especially suitable for the present problem as
it systematically takes into account contributions from regions of
the phase space where the interaction potentials~$U$ assume large
values, {\em i.e.}, of small resonant denominators responsible for
delocalization. The next two subsections are dedicated to the
application of this procedure.

\subsection{Statistical analysis for zero-dimensional system.}
\label{0D}

In this section we consider only transitions inside one
localization cell, $m=0$, or, in other words, $I\to 0$. It
suffices for our purposes to limit ourselves with
\be
|\ep-\xi_l|\ll M\delta_\zeta.
\label{almostshell}
\ee
To understand the most crucial features
of the cluster expansion, we will consider a few lowest order
terms explicitly, and then analyze an arbitrary order term.
It will be useful for us to introduce the notation
\be
\Xi_{l_1,\ldots,l_{N_e}}^{m_1,\ldots,m_{N_h}}
=\sum_{i=1}^{N_e}\xi_{l_i}-\sum_{j=1}^{N_h}\xi_{m_j}
\label{Xi}
\ee
for the energy of an excitation
consisting of $N_e$~electrons and $N_h$ holes.

\begin{subequations}
\label{lowest}
The lowest order terms 
are\footnote{We assume that $b(\omega =0)>0$. This assumption
does not correspond to any physical phonons. Actual frequency
dependence $b(\omega)\propto \omega^n$ is important if one tries
to calculate the physical conductivity in the insulating regime.
It is however not important for the determination of the stability
of the inslulating phase.}
\bea
&&\Gamma^{(0,0)}=Tb(0);\\
\label{lowest0}
&&\Gamma^{(1,0)}=\lambda^2\delta_\zeta^2 \Gamma^{(0,0)}\!
\sum_{l_1,l_2,l_3} 
F_{l_1, l_2; l_3}^\Rightarrow\!
\left[\frac{2 Y_{l_1,l_2}^{l_3,l}
\theta_\Delta\!\!\left(\ep-\Xi_{l_1l_2}^{l_3}\right)}
{\left(\ep-\Xi_{l_1l_2}^{l_3}\right)^2 }
+\frac{Y_{l_1,l_2}^{l_3,l}
\theta_\Delta\!\!\left(\ep-\Xi_{l_1l_2}^{l_3}\right)}
{\left(\ep-\Xi_{l_1l_2}^{l_3}\right)^2 }
\right]\!;\nonumber\\
&&\label{lowest1}
\eea
where $\theta_\Delta$ is given by \req{thetaDelta}. In \req{lowest1},
we chose not to join the electron decay and the hole decay
even though they are  equal to each  other. 
This is
done to keep the structure of the subsequent terms
more transparent. Moreover, this form will
be useful for the discussion of the modification of ImSCBA
in Sec.~\ref{val:shift}.
\end{subequations}

Let us use the machinery of Sec.~\ref{sec:Mayer}
to find the characteristic function of the rate \rref{lowest1}.
The corresponding three-particle potentials are given by
\be
U_{12;3}^e = 2 U_{12;3}^h=
\frac{2\Gamma^{(0,0)}\lambda^2\delta_\zeta^2
Y_{1,2}^{3,l}F^\Rightarrow(n_1,n_2,n_3)}
{\left(\ep-\Xi_{12}^{3}\right)^2 },
\label{3body}
\ee
switched on for certain combinations $\{n_1,n_2,n_3\}$, see
\req{linF}. The cutoff $\theta_\Delta(\ep-\Xi_{12}^3)$ will
be incorporated in the definition of cluster function~$f$ below.

\begin{figure}
\includegraphics[width=0.75\textwidth]{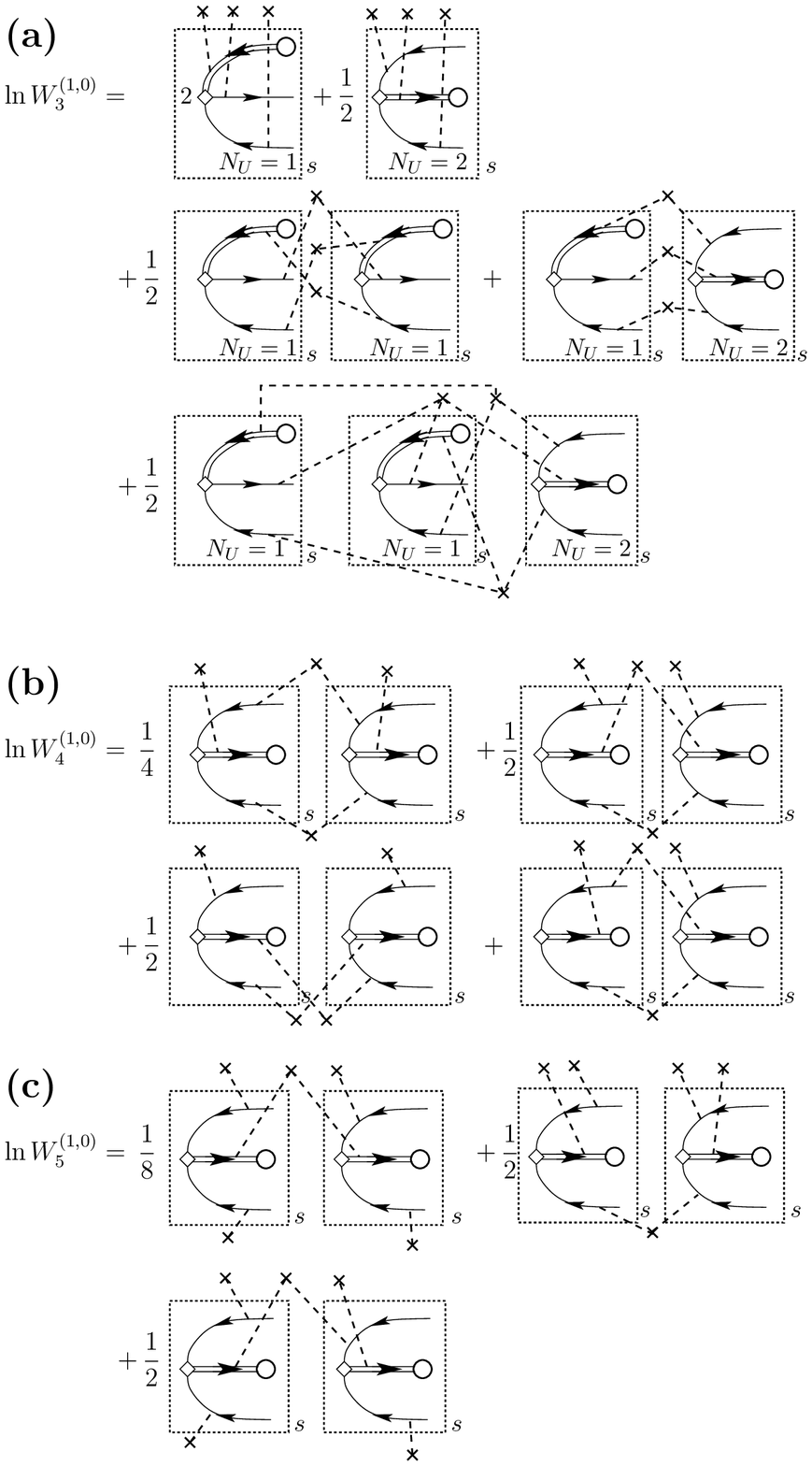}
\caption{
The 3-, 4-, and 5-particle contributions to $\ln{W}^{(1,0)}$.
For three particles (a), the cross-correlations between
the electron and the hole contributions,
the last term in \req{tildef}, are shown in the second and third lines.
For 4 and 5 particles (b,c) only the hole-hole terms are
shown.   
When not displayed, the degeneracy for all diagrams $N_U$ are the same
as on panel (a).}
\label{3pcluster}
\end{figure}

\begin{subequations}
The main conribution comes from clusters with $p=3$ particles,
see Fig.~\ref{3pcluster}a:
{\setlength{\arraycolsep=0pt}
\bea
&&\ln W^{(1,0 )}_3=\prod_{l=1}^{3} \left(
\int\frac{\diff\xi_l}{\delta_{\zeta}} \sum_{n_{l}=\pm 1}
\frac{\exp\frac{n_l\xi_l}{2T} }{2\cosh\frac{\xi_l}{2T}} \right)
\tilde{f}_{12;3};
\\
 &&\tilde{f}_{12;3}={f}_{12;3}^e +\frac{1}{2}{f}_{12;3}^h
\label{tildef}+\frac{1}{2}{f}_{12;3}^e
\left\{{f}_{21;3}^e
+   {f}_{12;3}^h\left[2+ 
{f}_{21;3}^e
\right]\right\}
\\
&&{f}_{12;3}^e=\left(\eexp^{- sU_{12;3}^e
}-1\right)\theta_\Delta\!\!\left(\ep-\Xi_{12}^3\right),
\label{fe}
\\
&&{f}_{12;3}^h=
\left(\eexp^{- 2sU_{12;3}^h}-1\right)
\theta_\Delta\!\!\left(\ep-\Xi_{12}^3\right),
\label{fh}
\eea
where the coefficient $1/2$ and extra
factor of $N_U=2$ in the exponential in ${f}_{12;3}^h$ 
take care of the symmetry
$f_{12;3}^h=f_{21;3}^h$, second term on Fig.~\ref{3pcluster}a.
The last termm in \req{tildef} describes the cross-correlations
between the electron and the hole contributions as 
the potentials $U^e(\xi_1,\xi_2,\xi_3),
\ U^e(\xi_2,\xi_1,\xi_3),$ and $U^h(\xi_1,\xi_2,\xi_3)$ 
diverge for the same sets of 
$\xi_1,\xi_2,\xi_3$.\footnote{
In fact,  using \req{3body}, one can rewrite 
\[
\tilde{f}_{12;3}=
\frac{1}{2}(1+f_{12;3}^e)(1+f_{21;3}^e)(1+f_{12;3}^h)
=\frac{1}{2}\left[\exp(-6U^h)-1\right]
\]
{\em i.e.}, join all the term in one cluster function.
This joining, however, would obscure the structure of the higher
order terms and we choose not to perform this operation.
}
\label{0DW3}
}
\end{subequations}
  
Performing the integration with the help of \req{linF} and definition
of $Y$, see \req{eqSCBA1}, we obtain using the
condition~\rref{almostshell}:
\be
\begin{split}
&\ln W^{(1,0 )}_3= -\frac{1}{\sqrt{3}}
\left({\pi s  \Gamma^{(0,0)}}\right)^{1/2}
\frac{3 \lambda T}{\delta_\zeta}\,
\mathcal{M}\left( \frac{2 T}{\delta_\zeta},\frac{\ep}{2T}\right),
\end{split}
\label{0DW3result}
\ee
where  the numerical coefficient $1/\sqrt{3}$ appears due to
the cross-correlation terms in \req{tildef}. 

Parameter $T\mathcal{M}_d\left(
\frac{2T}{\delta_\zeta},\frac{\ep}{2T}\right)$ roughly corresponds to the
total energy of  all
electron-hole pairs participating in the
process of the decay of a quasiparticle with the energy~$\ep$.
The explicit expression for $\mathcal{M}_d$ is
\be
\begin{split}
&\mathcal{M}\left(y,z\right) = \frac{y}{2} \int 
\frac{\diff{x}_1\diff{x}_2\,
\left|\Upsilon^2(y x_1)- \Upsilon^2(y x_2)\right|\cosh z}
{\cosh(x_1+x_2+z)\prod_{i=1,2}\cosh(x_i+z) }.
\end{split}
\label{calM}
\ee
Performing integration in \req{calM}
with the help of \req{Upsilon},
we obtain for $T \gg M\delta_\zeta$
\be
\mathcal{M}=2M,
\label{calMprime}
\ee
independently of $T$ and $\ep$. This occurs because of the additional
restriction on the phase volume for the decay of one-particle 
excitation into the three-particle excitations by the energy
dependence of the matrix elements.

Although the metal-insulator transition is a property of large
orders of perturbation theory, see \req{linearcondition}, its
precursor can be seen already in the characteristic function of
the lowest order term  \rref{0DW3result}.
Indeed, condition $\ln W^{(1,0 )}(s^*)\simeq 1$ gives the most
probable value $\Gamma^{(1,0)}_{typ} \simeq 1/s^*$ of the
distribution. Requiring $\Gamma^{(1,0)}_{typ} < \Gamma^{(0,0)}$,
we obtain the condition $T < T_*$, where 
\be
\left(\frac{{\lambda} M T_*}{ \delta_\zeta}\right)
\simeq {1},
\label{Tc1}
\ee
{\em i.e.}, $T_*$ is roughly of the order of
the temperature $T_{in}$ limiting the stability of
the metallic phase, see \req{mcondition2}.
Consideration of higher orders of the
perturbation theory to be performed shortly will only slightly
refine expression \rref{Tc1}, see \req{Tc2}.

To complete the analysis of the lowest order term \rref{lowest}, we
verify the validity of the cluster expansion, which amounts to finding
the condition for which
$|\ln W^{(1,0 )}_3(s)|\gg|\ln W^{(1,0 )}_{4,5}(s)|$.
Diagrams for $\ln W^{(1,0 )}_{4,5}(s)$ are shown in
Fig.~\ref{3pcluster}b,c. They give
\be
\begin{split}
\ln W^{(1,0 )}_4(s)=& \prod_{l=1}^{4} \left(
\int\frac{\diff\xi_l}{\delta_{\zeta}} \sum_{n_{l}=\pm 1}
\frac{\exp\frac{n_l\xi_l}{2T} }{2\cosh\frac{\xi_l}{2T}} \right)
\\
& \times \frac{1}{4}\left(\tilde{f}_{12;3}\tilde{f}_{12;4}
+2\tilde{f}_{13;2}\tilde{f}_{14;2}+ 2\tilde{f}_{13;2}\tilde{f}_{24;1} +
4\tilde{f}_{12;3}\tilde{f}_{14;2} \right),\\
\ln W^{(1,0 )}_5(s)=& \prod_{l=1}^{5} \left(
\int\frac{\diff\xi_l}{\delta_{\zeta}} \sum_{n_{l}=\pm 1}
\frac{\exp\frac{n_l\xi_l}{2T} }{2\cosh\frac{\xi_l}{2T}} \right)
\\ & \times
\frac{1}8\left(\tilde{f}_{12;5}\tilde{f}_{34;5}
+4\tilde{f}_{51;2}\tilde{f}_{53;4}+ 4\tilde{f}_{12;5}\tilde{f}_{53;4}\right),
\end{split}
\label{0DW4}
\ee
Using $\tilde{f}_{12,3}$ from \req{tildef}, we can estimate these
expressions as
\be\begin{split}
&\ln W^{(1,0 )}_4(s) \simeq s  \Gamma^{(0,0)}\,
 \frac{\lambda^2 MT}{\delta_\zeta},\quad
\ln W^{(1,0 )}_5(s) \simeq s  \Gamma^{(0,0)}\,
 \frac{\lambda^2 MT^2}{\delta_\zeta^2},
\label{0DW4estimate}
\end{split}\ee
where $M$ is defined in \req{Upsilon}.
Comparing \req{0DW4estimate} with \req{0DW3result}, we find that
for $T \gg \delta_\zeta$ the cluster expansion is justified even
for $\left|\ln W^{(1,0 )}_3\right| \gg 1$. The origin of this
suppression of the larger clusters is the same as the one controlling the
virial expansion for the classical gases -- clusters involving
more particles impose more restrictions on the phase volume.

\begin{figure}
\includegraphics[width=0.7\textwidth]{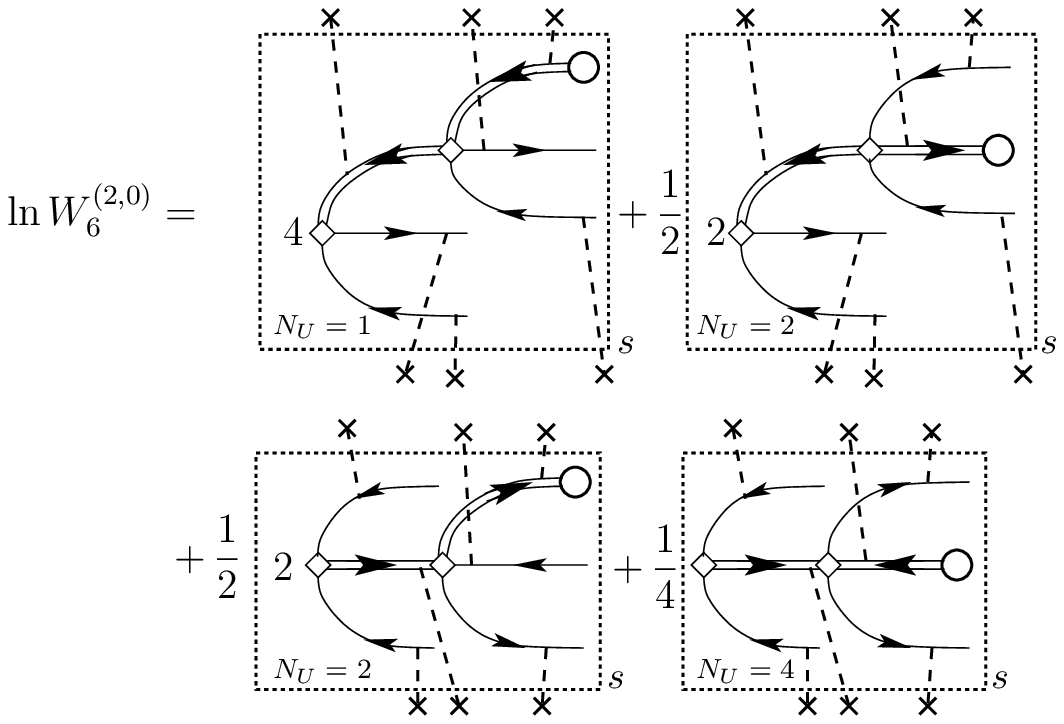}
\caption{The leading contributions to the cluster expansion of
$\ln{W}^{(2,0)}$.
The crosscorrelations between
the electron and the hole contributions,
second and third lines in \req{tildefeh}, are not shown.
}\label{6pcluster}
\end{figure}

Our next task is to consider an arbitrary order term
$\Gamma^{(k,0)}$. To explore how the lowest-order result
\req{0DW3result} is modified, we consider
$\Gamma^{(2,0)}$ first. Performing one more iteration in
\req{linSCBA2} with the help of \req{lowest}, we find
\be
\begin{split}
&\Gamma^{(2,0)}=\Gamma^{(0,0)}\lambda^4\delta_\zeta^4
\sum_{l_1,l_2;l_3} F_{l_1, l_2;l_3}^\Rightarrow{\sum_{l_4,l_5;l_6}}^\prime
 F_{l_4, l_5; l_6}^\Rightarrow
\\
&
\times 
\left\{
\frac{ 2 Y_{l_1,l_2}^{l_3,l}\, 
 \theta_\Delta\!\!\left(\ep-\Xi_{l_1l_2}^{l_3}\right)}
 {\left(\ep-\Xi_{l_1l_2}^{l_3}\right)^2}
\frac{2Y_{l_4,l_5}^{l_1,l_6}\,
\theta_\Delta\!\!\left(\ep-\Xi_{l_2l_4l_5}^{l_3l_6}\right)}
{\left(\ep-\Xi_{l_2l_4l_5}^{l_3l_6}
\right)^2}
\right.
\\ & \qquad 
+
\frac{ 2 Y_{l_1,l_2}^{l_3,l}\, 
 \theta_\Delta\!\!\left(\ep-\Xi_{l_1l_2}^{l_3}\right)}
 {\left(\ep-\Xi_{l_1l_2}^{l_3}\right)^2} 
\frac{Y_{l_4,l_5}^{l_1,l_6}\,
\theta_\Delta\!\!\left(\ep-\Xi^{l_4l_5}_{l_1l_2l_6}\right)}
{\left(\ep-\Xi^{l_3l_6}_{l_2l_4l_5}
 \right)^2}
\\
&
 \qquad +
\frac{  Y_{l_1,l_2}^{l_3,l}\, 
 \theta_\Delta\!\!\left(\ep-\Xi_{l_1l_2}^{l_3}\right)}
 {\left(\ep-\Xi_{l_1l_2}^{l_3}
 \right)^2}
\frac{2Y_{l_3,l_6}^{l_4,l_5}\,
 \theta_\Delta\!\!\left(\ep-\Xi_{l_1l_2l_6}^{l_4l_5}\right)}
 {\left(\ep-\Xi_{l_1l_2l_6}^{l_4l_5}
\right)^2}
\\ & \qquad 
+
\left. \frac{  Y_{l_1,l_2}^{l_3,l}\, 
 \theta_\Delta\!\!\left(\ep-\Xi_{l_1l_2}^{l_3}\right)}
 {\left(\ep-\Xi_{l_1l_2}^{l_3}
 \right)^2}
\frac{Y_{l_3,l_6}^{l_4,l_5}\,
 \theta_\Delta\!\!\left(\ep-\Xi_{l_1l_2l_6}^{l_4l_5}\right)}
 {\left(\ep-\Xi_{l_1l_2l_6}^{l_4l_5}
 \right)^2}
\right\},
\end{split}
\label{Gamma20}
\ee
each term is shown on Fig.~\ref{6pcluster}.
Here prime in the second sum excludes the
terms with $l_4,l_5=l_1,l_2$ and $l_6=l_3$, see
Fig.~\ref{figSCBA}b.2. (Let us note that
the placement of crosses on Fig.~\ref{6pcluster}
automatically takes this exclusion into account.)
Similarly to
the expression for  $\Gamma^{(1,0)}$ in \req{lowest}, we chose not
to join some similar terms together.

The corresponding 6-particle potentials
are given by
\be
\begin{split}
& U^{ee}
= 2U^{eh}
= \frac{4\Gamma^{(0,0)}\lambda^4\delta_\zeta^4 Y_{1,2}^{3,l}
F_{1,2;3}^\Rightarrow Y_{4,5}^{6,1}F_{4,5; 6}^\Rightarrow}
{\left(\ep-\Xi_{12}^{3}\right)^2
\left(\ep-\Xi_{245}^{36}\right)^2};
\\
& U^{he}=2 U^{hh}= \frac{2 \Gamma^{(0,0)}\lambda^4\delta_\zeta^4
Y_{1,2}^{3,l}F_{1,2; 3}^\Rightarrow Y_{4,5}^{6,3}F_{4,5; 6}^\Rightarrow}
{\left(\ep-\Xi_{12}^3\right)^2
\left(\ep-\Xi^{45}_{126}\right)^2}.
\end{split}
\label{6body}
\ee

The leading term of the cluster expansion shown on Fig.~\ref{6pcluster},
is given by [{\em cf.} \req{0DW3}]
\begin{subequations}
{\setlength\arraycolsep{0pt}
\bea
&&\ln W^{(2,0 )}=\prod_{l=1}^{6} \left(
\int\frac{\diff\xi_l}{\delta_{\zeta}} \sum_{n_{l}=\pm 1}
\frac{\exp\frac{n_l\xi_l}{2T} }{2\cosh\frac{\xi_l}{2T}} \right)
\tilde{f}\left({\substack{12;\,3\\45;\,6}}\right);\\
&&\tilde{f}\left({\substack{12;\,3\\54;\,6}}\right)
={f}^{ee}\left({\substack{12;\,3\\45;\,6}}\right)
+ \frac{1}{2}{f}^{eh}\left({\substack{12;\,3\\45;\,6}}\right)
+ \frac{1}{2}{f}^{he}\left({\substack{12;\,3\\45;\,6}}\right)
+  \frac{1}{4}{f}^{hh}\left({\substack{12;\,3\\45;\,6}}\right)
\nonumber\\ &&
\qquad+\frac{1}{2}{f^{ee}\left({\substack{12;\,3\\45;\,6}}\right)}\left\{
f^{ee}\left({\substack{12;\,3\\54;\,6}}\right)
+
f^{eh}\left({\substack{12;\,3\\45;\,6}}\right)
\left[2
+ 
f^{ee}\left({\substack{12;\,3\\54;\,6}}\right)\right]
\right\}
\nonumber\\ &&
\qquad+\frac{1}{4}{f^{he}\left({\substack{12;\,3\\45;\,6}}\right)}\left\{
f^{he}\left({\substack{12;\,3\\54;\,6}}\right)
+
f^{hh}\left({\substack{12;\,3\\45;\,6}}\right)
\left[2
+ 
f^{he}\left({\substack{12;\,3\\54;\,6}}\right)\right]
\right\};
\label{tildefeh}\\
&&
\begin{aligned}
&f^{ee}\left({\substack{12;\,3\\54;\,6}}\right)=
\left[
\eexp^{-sU^{ee}\left({\substack{12;\,3\\54;\,6}}\right)}
-1\right]
\theta_\Delta\!\!\left(\ep-\Xi_{12}^{3}\right)
\theta_\Delta\!\!\left(\ep-\Xi_{245}^{36}\right)
;\\ 
&f^{eh}\left({\substack{12;\,3\\54;\,6}}\right)=
\left[
\eexp^{-2 sU^{eh}\left({\substack{12;\,3\\54;\,6}}\right)}
-1\right]
\theta_\Delta\!\!\left(\ep-\Xi_{12}^{3}\right)
\theta_\Delta\!\!\left(\ep-\Xi_{245}^{36}\right);
\\
&f^{he}\left({\substack{12;\,3\\54;\,6}}\right)=
\left[\eexp^{-2sU^{he}\left({\substack{12;\,3\\54;\,6}}\right)}
-1\right]
\theta_\Delta\!\!\left(\ep-\Xi_{12}^{3}\right)
\theta_\Delta\!\!\left(\ep-\Xi^{45}_{126}\right);
\\
&f^{hh}\left({\substack{12;\,3\\54;\,6}}\right)=
\left[\eexp^{-4 sU^{hh}\left({\substack{12;\,3\\54;\,6}}\right)}
-1\right]
\theta_\Delta\!\!\left(\ep-\Xi_{12}^{3}\right)
\theta_\Delta\!\!\left(\ep-\Xi^{45}_{126}\right).
\end{aligned}
\label{feh}
\eea
\label{0DW6expression}
}
Similarly to \req{tildef}, the second and the third lines 
in \req{tildefeh} express the cross-correlation 
between different terms in the lowest order perturbation theory.
These correlations do not proliferate into higher order terms;
in particular, $U^{ee}$ from \req{6body} is not invariant under
permutation $1\leftrightarrow 2$. 
\end{subequations}

Substituting \req{6body} into
\req{0DW6expression} and using the approximate expression for
the integral
\be
\begin{split}
&\mathcal{Z}_n(y,z) \equiv \int_0^z \diff{x}_1\dots \diff{x}_n
\left[\exp\left(-\frac{y}{x_1^2\dots x_n^2}\right)-1\right]
\\
&=-\sqrt{\pi y}\,\frac{ \left[\ln z^ny^{-1/2} +
\mathcal{O}(1)\right]^{n-1}}{(n-1)!}; \quad y\ll \left[\eexp^{-1}z\right]^{2n},
\end{split}
\label{jn} \ee
we find
\be
\begin{split}
\ln W^{(2,0 )}&= - 
\frac{1}{\sqrt{3}}
\left(\pi s  \Gamma^{(0,0)}\right)^{1/2}
\left[\frac{6\lambda M T}{\delta_\zeta}\right]^2
\times\ln \left(\frac{\Delta^2}
{\left(s\Gamma^{(0,0)}\right)^{1/2}
\left(\lambda\delta_\zeta\right)^2} \right),
\end{split}
\label{0DW6result}
\ee
where $\Delta$ is the scale introduced in \req{thetaDelta},
and the coefficient ${1}/{\sqrt{3}}$ comes from the 
cross-correlations in \req{tildefeh}.

The analysis of the corrections to \req{0DW6result}
 is performed similarly to the estimate of \req{0DW4},
see Appendix~\ref{app2}. It shows that their main
effect is  the more accurate determination
of the cut-off of the logarithm such that
\be
\Delta \to \delta_\zeta,
\label{replacement}
\ee
so that \req{0DW6result} is valid in a leading logarithmic approximation.
This value of the cut-off is in agreement with Ref.~\cite{Mirlin97}.

All the higher order terms are considered analogously, and we find
\be
\begin{split}
&\ln W^{(n,0 )}= - \sqrt{\frac{\pi s  \Gamma^{(0,0)}}{3}}
\left(\frac{6\lambda M T}{\delta_\zeta}\right)^n
\frac{1}{(n-1)!}\ln^{n-1}
\left(\frac{\Delta^n} {\left(s\Gamma^{(0,0)}\right)^{1/2}
(\lambda\delta_\zeta)^n} \right).
\end{split}
\label{0DWnresult} \ee

To investigate metal-insulator transition we have to consider the
behavior at $n\to\infty$. We obtain from \req{0DWnresult} with the
logarthimic accuracy:
\begin{subequations}
\label{0DMIT}
\be
\begin{split}
&\ln W^{(n\gg 1,0 )}\simeq - \left({s
\Gamma^{(0,0)}}\right)^{\gamma(T)/2}
\left(\frac{6\eexp\lambda M{T}}{\delta_\zeta}\,
\ln\frac\Delta{\lambda\delta_\zeta} \right)^n,
\quad 
\gamma(T)=1-\left[\ln\frac{\Delta}{\lambda\delta_\zeta}\right]^{-1}.
\end{split}
\label{0DWinfresult} \ee

\end{subequations}

Equations \rref{0DMIT} constitute the central result of this
section as they describe the complete statistics of each term in
the ImSCBA series. If  many particles are in the excited state ($T$
is finite), formula \rref{0DWinfresult} predicts the phase
transition. Indeed, the stability criterion~\rref{linearcondition}
is violated at $T>T^*$, where
\be
\label{Tc2}
\frac{6\eexp\lambda M {T}_*}{\delta_\zeta}\,
\ln\frac{1}{\lambda}
=1, 
\ee
where we used the replacement \rref{replacement} in the argument
of the logarithm.
Equation \rref{Tc2} refines \req{Tc1}.

This would mean that insulator becomes unstable at $T>T_*$. This
conclusion, however, is an artefact of the averaging procedure
\rref{averaging} which neglected the one-particle level 
repulsion, see \req{levelrepulsion}. To take
this level repulsion into account we use the following qualitative
consideration. Inspection of each term in the perturbaiton
theory, see, {\em e.g.}, \req{0DW3} or \rref{0DW6expression},
shows that $W^{(n,0)}$ is contributed by $2n$ integrations over
$-T<\xi_l < T$. It means that $W^{(n,0)}$ is essentially
determined by the probability to find $2n$ levels within the
interval with the width $T$. According to \req{levelrepulsion},
this probability is suppressed by the level repulsion. As a
result, \req{0DWinfresult} is modified as
\be
\begin{split}
&\ln W^{(n\gg 1,0 )}\simeq - \left({s
\Gamma^{(0,0)}}\right)^{\gamma/2}
\left(\frac{6\eexp\lambda M{T}}{\delta_\zeta}\,
\ln\frac\Delta{\lambda\delta_\zeta} \right)^n
\exp\left[-\mathcal{P}\left(\frac{c n\delta_\zeta}{T}\right)\right],
\end{split}
\label{MITrepulsion}
\ee 
where function $\mathcal{P}$ is defined in
\req{levelrepulsion}, and $c$ is the numerical factor of the order
of unity. Because $\lim_{x\to \infty}x^{-1}\mathcal{P}(x) =\infty$,
we find
\[
\lim_{n\to \infty}\ln W^{(n,0 )}=0 \quad {\rm for\ any}\ T.
\]
Equation \rref{MITrepulsion} describes the impossibility of a
genuine phase tranistion in a finite system where the unphysical
packing of infinite number of levels in a finite enery strip is
forbidden.

Even though for a zero-dimensional system temperature $T_*$ does not
have a physical meaning of the transition temperature, the finite
coupling between the localization cells leads to the phase
transition with $T_c$ very close to $T_*$, as we will show in the
next subsection.


\subsection{Statistical analysis and metal-insulator transition
in finite-dimensional systems.} \label{bulk}

The goal of this subsection is to generalize \reqs{0DMIT} by
including hopping into the neighbouring localization
cells. As a result, the effect of the finite size
\rref{MITrepulsion} will be overcome and the genuine
metal-insulator tranistion will occur.
Similarly to the previous subsection, we will consider explicitly
a few lower orders of the perturbation theory, and then obtain the
result for an arbitrary order.

\begin{figure}
\includegraphics[width=0.7\textwidth]{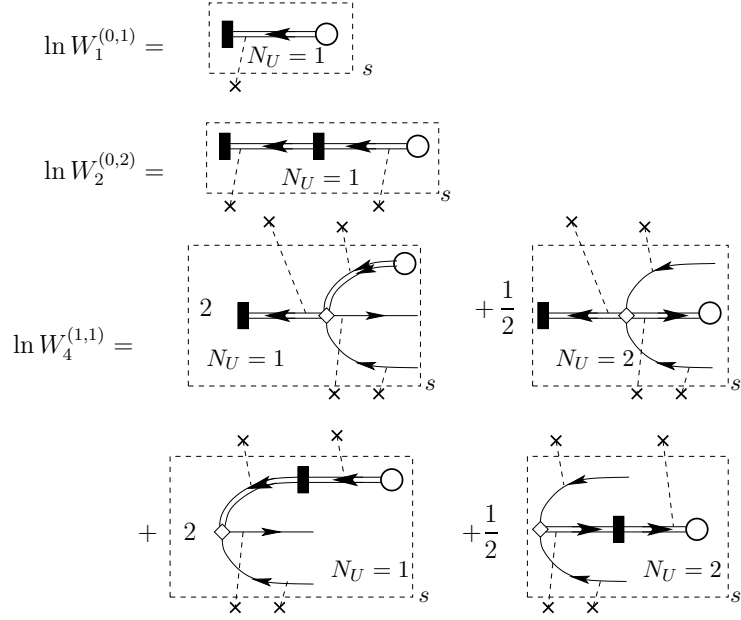}
\caption{
The leading contributions to $\ln{W}^{(0,1)}$,
$\ln{W}^{(0,2)}$, and $\ln{W}^{(1,1)}$.
The cross-correlations between the electron
and hole contributions are not shown.
}
\label{Icluster}
\end{figure}

The contribution which involves hopping only
(its first two terms $\Gamma^{(0,1)}$ and $\Gamma^{(0,2)}$ are
shown in  Fig.~\ref{Icluster}) corresponds to the purely
single-particle problem. Since the exact single-particle
eigenstates are assumed to be localized,
see condition \rref{Ismall}, this contribution does not lead to
delocalization.
Thus, we start with the
term $\Gamma^{(1,1)}$, which involves tunneling of the
particle and creation of an electron-hole pair. By iterating
\req{linSCBA2} twice, we find (see Fig.~\ref{Icluster}):
\be
\begin{split}
\Gamma^{(1,1)}_l(\vec\rho)=&\lambda^2I^2\delta_\zeta^4 \Gamma^{(0,0)}
\sum_{l_1,\dots,l_4;\vec{a}}
\Bigg\{\frac{2
Y_{l_1^+,l_2^+}^{l_3^+,l_4^+}F_{l_1^+, l_2^+; l_3^+}^\Rightarrow
\theta_\Delta \!\! \left(\ep-\xi_{l_4^+}\right)
\theta_\Delta \!\!\left(\ep-\Xi_{l_1^+l_2^+}^{l_3^+}\right)
} 
{
\left(\ep-\xi_{l_4^+}\right)^2 
\left(\ep-\Xi_{l_1^+l_2^+}^{l_3^+}\right)^2 }
\\
& \times 
\Bigg\{\frac{
Y_{l_1^+,l_2^+}^{l_3^+,l_4^+}F_{l_1^+, l_2^+; l_3^+}^\Rightarrow
\theta_\Delta \!\! \left(\ep-\xi_{l_4^+}\right)
\theta_\Delta \!\!\left(\ep-\Xi_{l_1^+l_2^+}^{l_3^+}\right)
} 
{
\left(\ep-\xi_{l_4^+}\right)^2 
\left(\ep-\Xi_{l_1^+l_2^+}^{l_3^+}\right)^2 }
\\
& \quad+\frac{2 Y_{l_1,l_2}^{l_3,l}F_{l_1,
l_2; l_3}^\Rightarrow
\theta_\Delta \!\!\left(\ep-\Xi_{l_1l_2}^{l_3}\right)
\theta_\Delta \!\!\left(\ep-\Xi_{l_2l_4^+}^{l_3}\right)
} {\left(\ep-\Xi_{l_1l_2}^{l_3}\right)^2 
\left(\ep-\Xi_{l_2l_4^+}^{l_3}\right)^2}
\\
& \quad 
+\frac{Y_{l_1,l_2}^{l_3,l}F_{l_1,
l_2; l_3}^\Rightarrow
\theta_\Delta \!\!\left(\ep-\Xi_{l_1l_2}^{l_3}\right)
\theta_\Delta \!\!\left(\Xi_{l_2l_1}^{l_4^+}-\ep\right)
} {\left(\ep-\Xi_{l_1l_2}^{l_3}\right)^2 
\left(\Xi_{l_2l_1}^{l_4^+}-\ep\right)^2} \Bigg\}.
\end{split}
\label{Gamma11} \ee where we used the 
short-hand notation \rref{Xi}, and
$\xi_l\equiv \xi_l(\vec\rho), \
\xi_{l^+}\equiv \xi_l(\vec\rho+\vec{a})$.

From Fig.~\ref{Icluster} we find (omitting overall numerical
coefficient)
\be
\begin{split}
\ln W^{(1,1 )}\simeq &- \left(s
\Gamma^{(0,0)}\right)^{1/2}
\left(\frac{12dI\lambda M
T}{\delta_\zeta}
 \right)
\ln
\left[\frac{\Delta^2}{\left( s
      \Gamma^{(0,0)}\right)^{1/2}\delta_\zeta^2I\lambda}
\right].
\end{split}
\label{W11result} \ee

Comparing \req{W11result} with \req{0DW6result}, we see that by
replacing  the electron-hole pair creation to a single-particle
hopping, we always introduce the additional smallness. Thus, we
can anticipate that the number of hoppings must be as small as
possible to maximize the overall $W^{(m,n)}$ for fixed $m+n$. The
only reason to include hopping at all is to overcome the finite
number of levels restriction which suppressed the transition in
$0$-dimensional case of previous subsection.

The result for the arbitrary order perturbation theory may be
immediately obtained by examining the differences between
\req{W11result} with \req{0DW6result}, and analogy with \req{0DWnresult}:
\be
\begin{split}
&\ln W^{(n,m )}= - \left({  s  \Gamma^{(0,0)}}\right)^{1/2}
\left(
\frac{6\lambda M
T}{\delta_\zeta}
\right)^n I^m
\\ & \times 
\frac{\mathcal{C}(n,m)}{(n+m-1)!}
 \ln^{n+m-1}
 \left[\frac{\left( s
       \Gamma^{(0,0)}\right)^{-1/2}\Delta^{n+m}}
 {\delta_\zeta^{n+m} \lambda^nI^m}
 \right].
\end{split}
\label{Wnmresult} \ee

The most important difference with \req{0DWnresult} is the
presence of the additional factor $\mathcal{C}(n,m)$ which has the
meaning of the number of ways to order the $n$ interactions and
$m$ hoppings with respect to each other. The expression for these
coefficients is
\be
\begin{split}
&\mathcal{C}(n,m)=\prod_{i=1}^{m} \sum_{\vec{a}_i}
\sum\limits_{N_i=0}^\infty \delta_{n,\sum\limits_i N_i}
 \exp\left[-\sum_{\vec\rho} \mathcal{P}
 \left(\frac{c \delta_\zeta\sum\limits_{i=1}^m N_i
\delta_{\vec\rho,\vec\rho_0+\sum_{j=0}^i\vec{a}_j}}{T}\right)\right],
\end{split}
\label{dnm} \ee where $\vec{\rho}_0$ is an arbitrary site on the
lattice, $\vec{a}$ are the vectors connecting each site to its
nearest neighbours. The  meaning of the coefficients,
$\mathcal{C}(n,m)$, is the total number of random lines consisting of
$m$ segments on the $d$-dimensional cubic lattice. Integer $N_i$ is
the number of electron hole pairs emitted by the electron between
the hopping event $i$ and the hopping event $(i+1)$.
We included the effect of the
level repulsion within each localization cell in the same spirit as
in \req{MITrepulsion}. The argument of $\mathcal{P}$ counts the total
number of electron-hole pairs within localization site
$\vec{\rho}$ and takes into account the fact that the random path may
traverse one site more than once.

To investigate the stability of the insulating state, see
\req{linearcondition}, we have to study the limit of $m=m^*(n), \
n \to\infty$, where $m^*(n)$ determines the direction to maximize
$\ln W^{(n,m )}$ for fixed $n$.

As before, we will restrict ourselves by the condition
\be
{I} \simeq \lambda
    \ll \lambda M.
\label{Icondition}
\ee
Simple counting of combinatorial factors in \req{dnm} shows that
under condition \rref{Icondition} maximum is achieved for the
maximal possible number of the electron-hole pairs per a
localization cell, $N_i\simeq T/\delta_\zeta$. It translates
into estimate
\be
m^*(n)\simeq \frac{n\delta_\zeta}{T}. 
\label{mestimate}
 \ee
Substituting \req{mestimate} into \req{dnm}, we find
$\mathcal{C}[n,m^*(n)] \lesssim (2d)^{m^*(n)}$ and obtain from
\req{Wnmresult}:
\be
\begin{split}
&\ln W^{(n \gg 1,m^*(n) )}= - \left({  s
\Gamma^{(0,0)}}\right)^{\gamma(T)/2}
\left(
\frac{6\eexp\lambda M {T}}{\delta_\zeta}\,
I^{c_1\delta_\zeta/T}
\ln\frac\Delta{\lambda\delta_\zeta}
 \right)^n,
\label{Wnomtimal}
\end{split}
\ee where $c_1(d)$
 is the number of the order of unity which we were not
able to calculate, and all the other entries are the same as in
\req{0DWinfresult}.

Applying stability criterion~\rref{linearcondition}, we arrive to
the main conclusion of this section -- {\em insulating state is
stable only for $T<T_c$}, where the critical temperature is given
by
\be
\frac{6\eexp\lambda M {T}_c}{\delta_\zeta}\,
\ln\frac\Delta{\lambda\delta_\zeta}=
 \exp\left[\frac{c_1\delta_\zeta
\left|\ln I\right|}{T_c}\right].
\label{Tc4}
\ee
As $I$ is not
exponentially small, $I \gg
\eexp^{-1/\lambda M}$, \req{Tc4} may be expanded as
\be
T_c =T_*\left[1+\mathcal{O}\left( \lambda M \left|\ln
I\right|\right)\right], 
\label{Tc5} 
\ee
where $T_*$ is given by \req{Tc2}.

At temperatures larger than $T_c$ metallic phase is formed.
Together with the material of Sec.~\ref{sec:metal} proving the
stability of the metallic phase at $T > T_{in} \gg T_c$, see
\req{mcondition2}, this completes the proof of the existence of
the metal insulator transition.

Kinetics of the system near the transition itself is a
complicated problem which we hope to address in a separate
publication. However, some conclusions can be drawn already from
\reqs{Wnmresult}, \rref{dnm}, and the estimate \rref{mestimate}.
Indeed, we concluded that the best paths are those that
maximize the number of electron-hole pairs in a given localization
cell before hopping to a neighbouring one. It means
that the self-intersections  of the random path in the \req{dnm} are
forbidden and the spatial part of the problem becomes equivalent
to the statistics of the self-avoiding random walk.

The latter observation enables us to conjecture the critical
behavior of the spatial localization length $\zeta$ at $T\to
T_c-0$. The latter length is defined from \req{dnm}, as
\[
\zeta= \left|\sum_i\vec{a}_i\right|_{typ}
\]
where the configurations giving the largest statistical weight in
\req{dnm} are meant by typical. 
The correlation length in the Fock space (typical distance between
resonances) diverges in the vicinity of
the transition as~\cite{Efetov87}
\begin{equation}
n_{typ}\simeq\frac{T_c}{|T_c-T|},
\label{treecorrlength=}
\end{equation}
hence, the number of segments
$m\simeq\delta_\zeta/|T_c-T|^{-1}$. We thus conclude
\be
\zeta(T)\simeq \zeta_{loc}
\left(\frac{\delta_\zeta}{T_c-T}\right)^{\nu_d}
\label{zeta}
\ee
where $\nu_d$ is the correlation length index for the
self-avoiding random walk in $d$ dimensions, $\nu_1=1$,
$\nu_2=3/4$, $\nu_3=0.59\dots$, $\nu_{d>4}=1/2$, see
Ref.~\cite{Lawler}.

\section{Validity of ImSCBA scheme}
\label{sec:validity}

This section is devoted to the analysis of the contributions
of the processes not taken into account in our ImSCBA calculational
scheme, see Figs.~\ref{fig:diag2},\ref{fig:diag3}, and \req{ImSCBA}.
The latter approximation corresponds to neglecting the
level shifts due to the electron-electron interaction.
The former will be shown to correspond to the renormalization
of the electron-electron interaction by intermediate
virtual processes, and to certain interference effects.
We will consider the effects of those contributions
separately in the following subsections.

We note here, however, that each remaining diagram,
denoted by $\mathcal{U}$,
is a random quantity. Therefore,
we will have to analyse the statistical
distribution of the  remaining diagrams and check
that their distribution functions have the scale parametrically
smaller than the scale of $\Gamma_l(\ep)$ calculated in the
previous section. The signs of the majority of the diagrams are random,
so it is more convenient to use the characteristic
function, see, {\em e.g.},  \req{linearLaplace},
in a Fourier transform form
\be
W^\mathcal{U}(q)=\left\langle \exp\left(iq\mathcal{U}\right) \right\rangle,
\label{Fourier}
\ee
where the averaging procedure is defined in \req{average}.
To calculate function \rref{Fourier}, all the machinery
of Sec.~\ref{sec:Mayer} is applicable
after the replacement $s\to -iq$, and $\mathcal{U}$
on the Fig.~\ref{clusterexpansion}
should be understood as an analytic expression for the
real or imaginary part of the corresponding diagram.

We will present in detail the analysis for
the insulating phase in Secs.~\ref{sec:renint}
-- \ref{val:shift}; the corresponding consideration
for the developed metallic phase is simple and is
summarized in  Sec.~\ref{sec:valmetal}.

\subsection{Effect of the interaction renormalization.}
\label{sec:renint}

\begin{figure}
\includegraphics[width=0.6\textwidth]{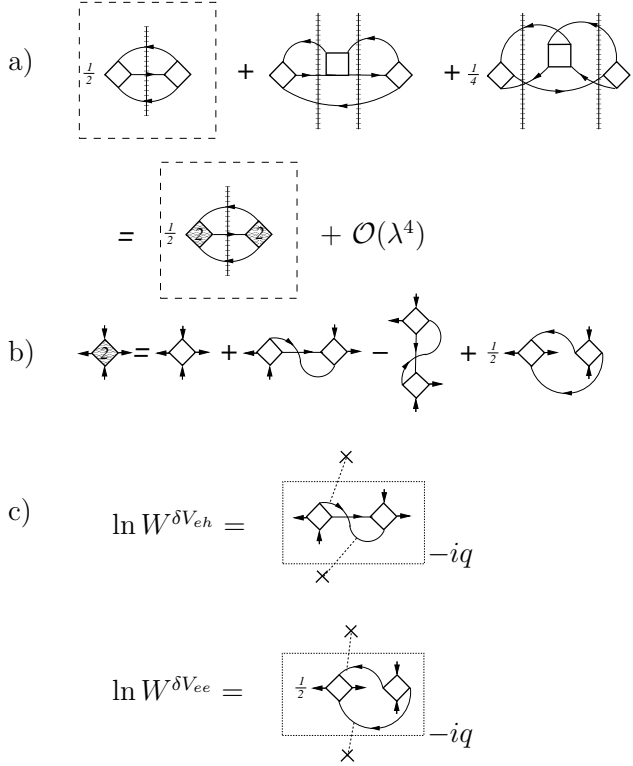}
\caption{Reduction of the third order diagrams (a) to
the renormalization of the interaction constant
in the SCBA scheme (b). Calculation of the characteristic
function (c) using notation of Fig.~\ref{clusterexpansion}.
}
\label{fig:renorm2}
\end{figure}

To understand the origin of the interaction renormalization,
let us consider the contributions c5,c6 from Fig.~\ref{fig:diag2},
see also Fig.~\ref{fig:renorm2}.
These contributions are the only third-order terms that may lead
to the finite decay rate -- all others are either insignificant
corrections to the Hartree-Fock potential (c2,c3),
or (c1,c4) the first order Hartree-Fock shift in the second order diagram
(b1,b3).
Direct calculation of the diagram (c5,c6) yields
\be
\begin{split}
&\Im \Sigma_l^{A({\rm Fig}\ref{fig:diag2}c5)}(\ep)
=\frac{1}{2}
\sum_{l_1,\dots,l_5}\int{\diff\ep_1}\dots{\diff\ep_5}\,
A_{\l_1}(\ep_1)\dots A_{l_5}(\ep_5)
\\
&\quad\times
\Big[V_{ll_3}^{l_1l_2}V_{l_2l_5}^{l_3l_4}V_{l_4l_1}^{l_5l}
\pi \delta\left(\ep-\ep_1-\ep_2+\ep_3\right)
F_{l_1, l_2; l_3}^\Rightarrow(\ep_1,\ep_2;\ep_3)\,
P\frac{n_{l_4}(\ep_4)-n_{l_5}(\ep_5)}{\ep_2-\ep_3-\ep_4+\ep_5}\\
&\qquad+\left(2\leftrightarrow 4;3\leftrightarrow 5 \right)\Big]
;
\\
&\Im \Sigma_l^{A({\rm Fig}\ref{fig:diag2}c6)}(\ep)
=\frac{1}{8}
\sum_{l_1,\dots,l_5}
\int{\diff\ep_1}\dots{\diff\ep_5}\,
A_{\l_1}(\ep_1)\dots A_{l_5}(\ep_5)
\\
&\quad\times
\Big[V_{ll_3}^{l_1l_2}V_{l_1l_2}^{l_4l_5}V_{l_4l_5}^{l_3l}
\pi \delta\left(\ep-\ep_1-\ep_2+\ep_3\right)
F_{l_1, l_2; l_3}^\Rightarrow(\ep_1,\ep_2;\ep_3)\,
P\frac{n_{l_4}(\ep_4)+n_{l_5}(\ep_5)}{\ep_1+\ep_2-\ep_4-\ep_5}\\
&\qquad+\left(1\leftrightarrow 4;2\leftrightarrow 5 \right)\Big],
\end{split}
\label{renorm1}
\ee
where the notation is the same in in \reqs{SCBA},
and $P$ denotes the principal value. It is important to notice
that using ImSCBA Green functions in this expression
rather than the bare Green functions is not an overstepping of the
accuracy of the calculation.

Expression \rref{renorm1} illustrates the well-known
principle of constructing higher order contributions from
the lower ones. Namely, to obtain the imaginary part of any contribution
to the self-energy one can cut the diagram in all possible
ways [cuts are shown by vertical dotted lines  on Fig.~\ref{fig:renorm2}]. The
cross-section produces a $\delta$-function for the energies of
the particles crossing the cut. The two parts of the diagrams on
each side of the cut correspond to transition  amplitudes.
Being real, they can be incorporated into the redefinition of
the constants of the initial Hamiltonian (for the clean Fermi liquid
it was first realized by Eliashberg~\cite{Eliashberg}).

For example, expression \rref{renorm1} can be obtained
from $\Gamma_l^{(in)}(\ep)$ of \req{eqSCBA1} by the replacement
  of the bare interaction matrix
element $V_{l_1l_2}^{j_1j_2}$
with the potential
renormalized by excitation of virtual particle-hole pairs,
$\delta V_{eh}$, or particle-particle pairs, $\delta V_{ee}$,
see  Fig.~\ref{fig:renorm2}b:
\be
\begin{split}
&V_{l_1,l_2}^{l_3l_4}\to V_{l_1,l_2}^{l_3l_4}
+ \left[\delta V_{eh}\right]_{l_1,l_2}^{l_3l_4}
-\left[\delta V_{eh}\right]_{l_2,l_1}^{l_3l_4}
+ \left[\delta V_{ee}\right]_{l_1,l_2}^{l_3l_4}\\
&
\left[\delta V_{eh}\right]_{l_1,l_2}^{l_3l_4}
= \frac{1}{2}
\sum_{l_5,l_6}
\int{\diff\ep_5}{\diff\ep_6}
A_{l_5}(\ep_5)A_{l_6}(\ep_6)
V_{l_1l_5}^{l_3l_6}V_{l_6l_2}^{l_5l_4}
P\frac{n_{l_6}(\ep_6)-n_{l_5}(\ep_5)}{\ep_2-\ep_3-\ep_6+\ep_5};
\\
&
\left[\delta V_{ee}\right]_{l_1,l_2}^{l_3l_4}
= \frac{1}{4}
\sum_{l_5,l_6}
\int{\diff\ep_5}{\diff\ep_6}
A_{l_5}(\ep_5)A_{l_6}(\ep_6)
V_{l_1l_2}^{l_5l_6}V_{l_5l_6}^{l_3l_4}
P\frac{n_{l_6}(\ep_6)+n_{l_5}(\ep_5)}{\ep_1+\ep_2-\ep_6-\ep_5};
\end{split}
\label{eq:renorm1}
\ee
where we are using the short-hand notation $l_i\equiv (l_i,\ep_i)$
in the first line of the expression.

For the insulating regime we
use the leading term in \req{linA},
${A}_l(\ep) \approx \delta\left(\ep-\xi_l\right)$,
and calculate the characteristic functions
$W^{\delta V_{ee}}(q)$ and $W^{\delta V_{eh}}(q)$ as shown
on Fig.~\ref{fig:renorm2}c.
Assuming $|\ep_2-\ep_3|,\ |\ep_2+\ep_1| \gtrsim \delta_\zeta, \
T \gtrsim M\delta_\zeta$, we find
\begin{subequations}
\bea
\ln W^{\delta V_{eh}}(q) &=&
\left\{\begin{matrix} - 2 \pi |q|{\lambda^2 T};
& |\ep_2-\ep_3| \lesssim M\delta_\zeta;\\
\\
-
\displaystyle{\frac{\pi |q|\lambda^2 M\delta_\zeta}{2}};
\ &  |\ep_2-\ep_3| \gtrsim M\delta_\zeta;
\end{matrix}
\right.
\label{WVeh}
\\
\nonumber\\
\ln W^{\delta V_{ee}}(q)
&=&
-  \displaystyle{\frac{\pi |q|\lambda^2 M\delta_\zeta}{2}}.
\label{WVee}
\eea
\label{WV}
\end{subequations}

Equations \rref{WV} describe random quantities with
the characteristic scale  of the distribution $|\delta V|_{typ}$ given
by the coefficient multiplying $|q|$. This width
should be compared with the bare value of the interaction constant
$\lambda \delta_\zeta$. For $T \leq T_*$, [see \req{Tc2}],
we find
\be
\frac{|\delta V|_{typ}}{ \lambda \delta } \simeq
\frac{1}{M}
\left(\frac{T}{T^*}\right),
\label{val:cond1}
\ee
{\em i.e.}, the typical value of the correction to the interaction
constant is parametrically smaller than the bare value.
On the other hand, the distribution of the $\delta V$ has the same
algebraic decay at large values as the distributions of the ImSCBA
quantities $\Gamma$, see, {\em e.g.},    \req{0DWnresult}.
It means that substitution of the renormalized constant
$\delta V$ in, say, second order perturbation theory formula
\rref{lowest} will produce the distribution function
of the form as the fourth order perturbation theory result
\rref{0DW6expression}, but with the coefficient smaller at least by the
factor of $M$. The same is true for any order, and therefore,
the interaction renormalization  \rref{WV} can produce only
perturbative in $1/M$ corrections to the value of the transition
temperature $T_c$.

\begin{figure}[h]
\includegraphics[width=0.7\textwidth]{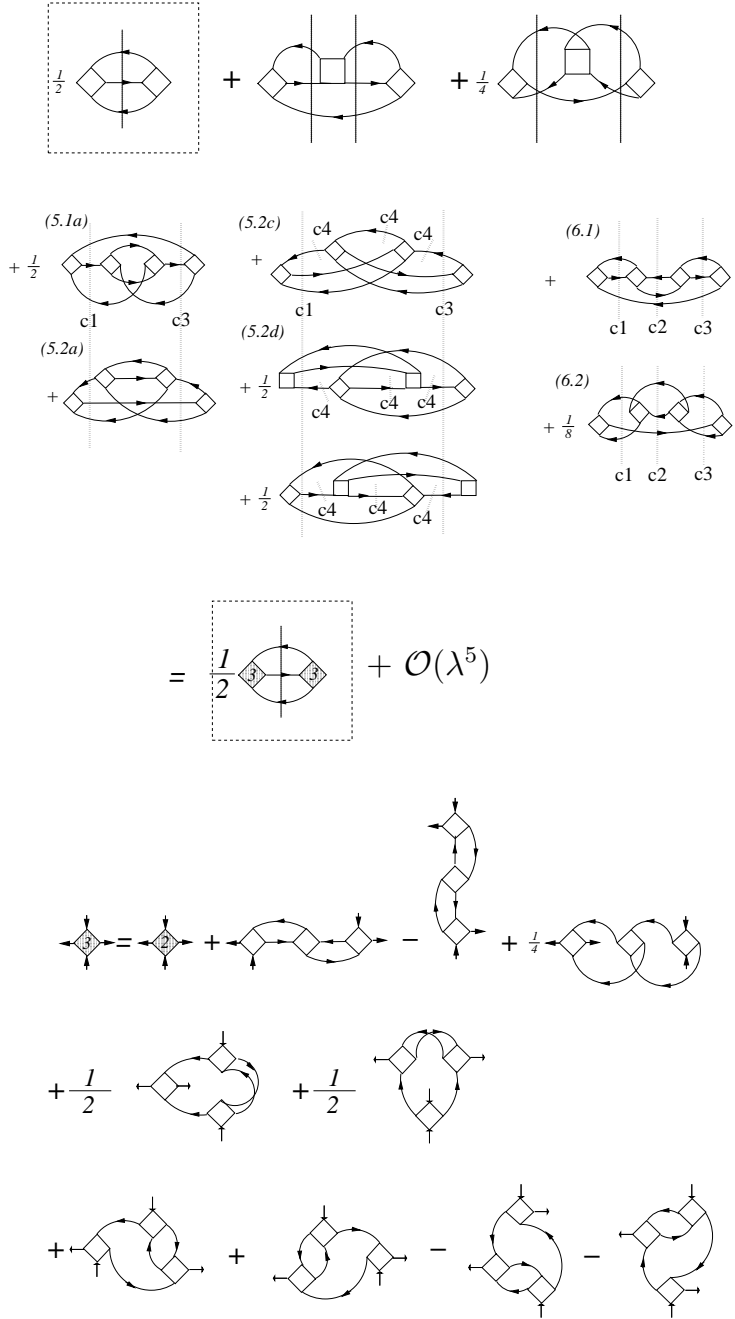}
\caption{Reduction of the certain cross-section
of fourth order diagrams, Fig.~\ref{fig:diag3}, to
the renormalization of the interaction constant
in the SCBA scheme.
}
\label{fig:renorm3}
\end{figure}

\begin{figure}[h]
{\includegraphics[width=0.6\textwidth]{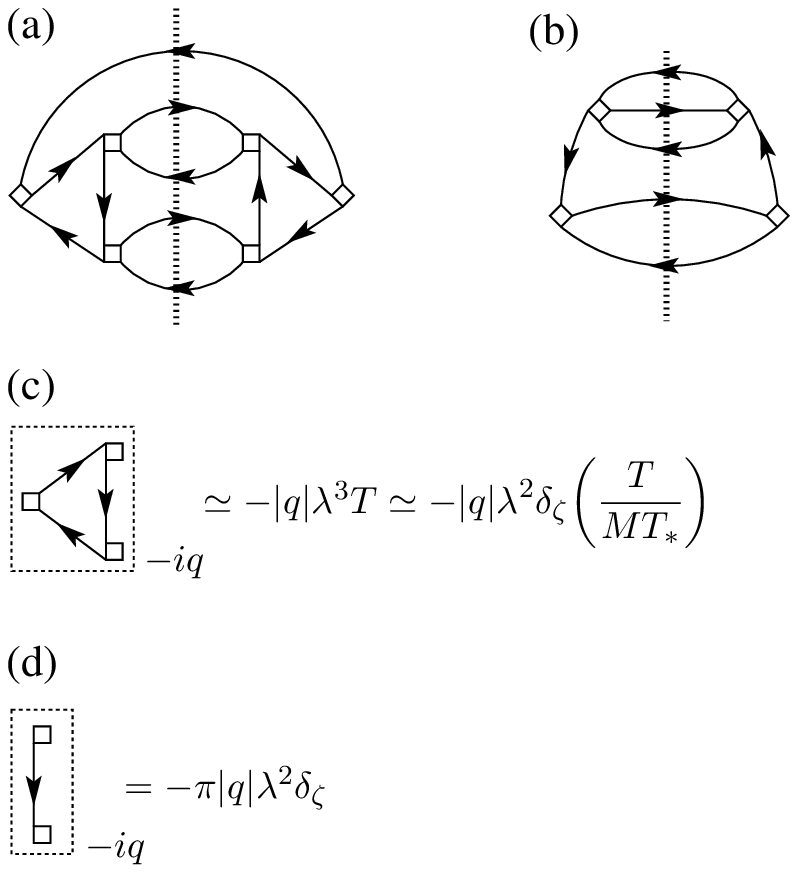}}
\caption{
a) Reduction of a crosssection of the
6-th order diagrams to the three particle interaction.
b) SCBA diagram generating the same state in the crosssection.
c-d) Comparing of the distribution functions of the non-equivalent
blocks in the SCBA diagram (d) and in the 3-particle interaction
diagram (c).
}
\label{fig:6vertex}
\end{figure}

It is not difficult to see that certain cross-sections of
higher-order diagrams may be ascribed to the higher-order
corrections to the interaction vertex, see Fig.~\ref{fig:renorm3}.
However, not all of the cross-sections can be taken into account
in a such a fashion, see, {\em e.g.}, cross-section (c2) of
Fig.~\ref{fig:diag3}, or
Fig.~\ref{fig:6vertex}a. Remaining terms describe the effects of the
particle permutations in the final state which will be discussed in
the following subsection, see Fig.~\ref{fig:exchange},
and generation of the interaction
vertices involving a larger number of the particles, see
Fig.~\ref{fig:6vertex}c.

Statistical analysis of the higher-order corrections to the vertices
is performed in the same fashion and produce distributions similar to
that of \reqs{WV} with the smallness \rref{val:cond1} in higher and higher
powers. We thus conclude that the vertex renormalization does not
lead to any dramatic effects; it only produces perturbative corrections
to the transition temperature $T_c$ calculated within ImSCBA scheme.

\subsection{Effect of the particle permutations in the
final state.}\label{sec:exchange}

\begin{figure*}
\includegraphics[width=0.9\textwidth]{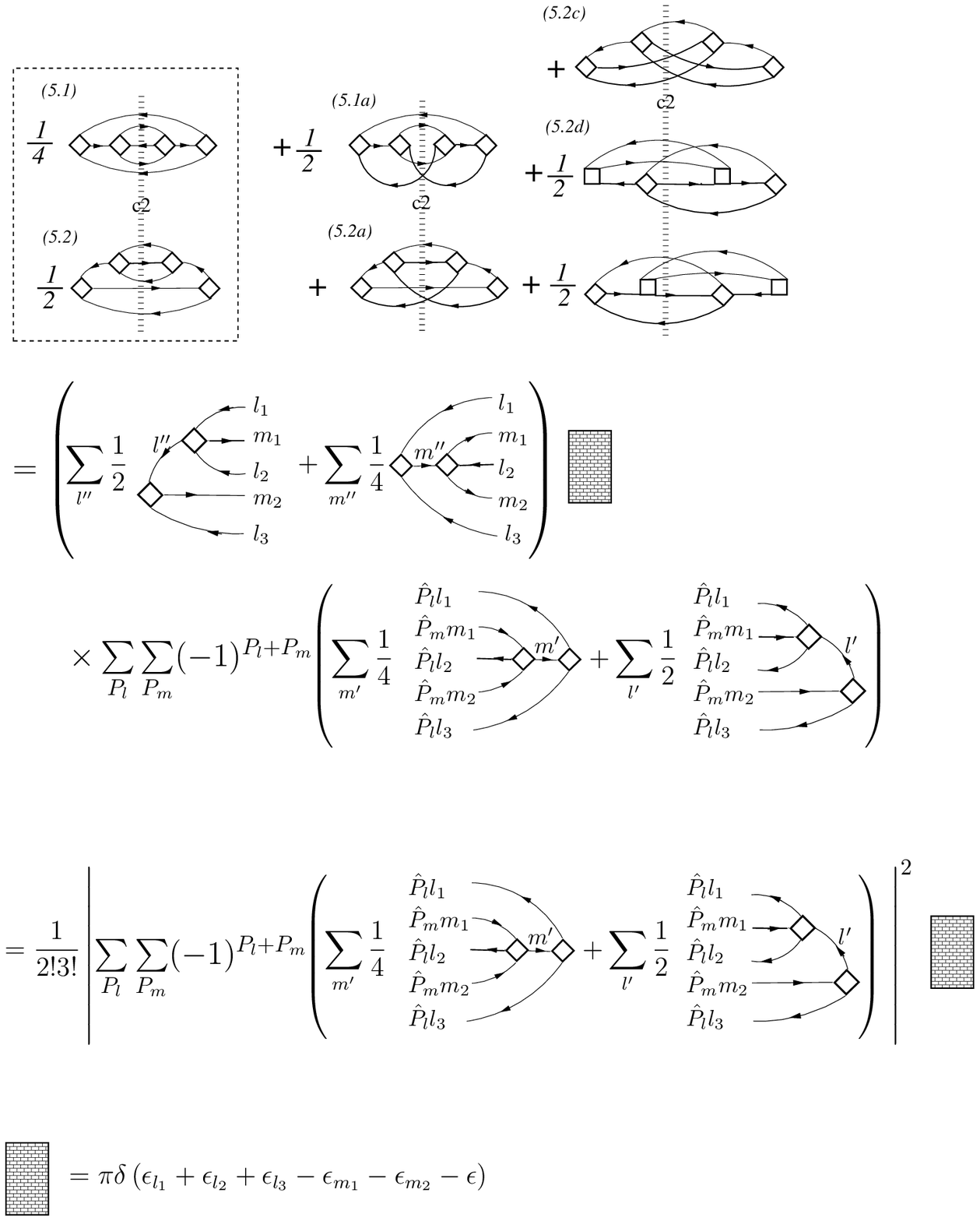}
\caption{
Cross-sections of the fourth-order diagrams for
self-energy, not included in Fig.~\ref{fig:renorm3}. The orbital
indices of the particles crossing the cut ($l_1,l_2,l_3,m_1,m_2$)
are summed over. For fixed values of these indices each of the
two parts of the diagram with the free ends removed corresponds
to the partial transition matrix element. Summation of the latter
over all permutations of the orbital indices of identical particles
in the final state gives the total transition matrix element for
a given final state. The permutations $\hat{P}_l$ and $\hat{P}_m$
are independent; they act on electron ($l_1,l_2,l_3$) and hole
($m_1,m_2$) indices, respectively. The decay rate is then given
by the square of the total matrix element, summed over all final
states. Since different permutations of electron and hole indices
give the same final state, the factor $1/(3!2!)$ in front of the
sum is necessary. Diagrams (2) and (4) of
  Fig.~\ref{figSCBA}c are also generated for $l^{\prime\prime}
\neq l^\prime$ or
$m^{\prime\prime}\neq m^\prime$. }
\label{fig:exchange}
\end{figure*}

Cross-sections of the fourth-order diagrams, not included in
Fig.~\ref{fig:renorm3}, are shown in Fig.~\ref{fig:exchange}
(5.1a--5.2d) together with the two ImSCBA cross-sections (5.1 and
5.2). One can notice that all of these cross-sections contain the
sum over the same final 5-particle states (the same orbital
indices of the particles and holes crossing the cut). Moreover,
the analytic expressions corresponding to the two parts of a cut
ImSCBA diagram for a given final state (transition matrix
elements) are equal. For non-ImSCBA diagrams the two resulting
expressions are different, but they always can be reduced to the
transition matrix elements for the diagrams 5.1 and 5.2 by a
permutation of the orbital indices of the particles in the final
state. This observation enables us to identify the non-ImSCBA
cross-sections 5.1a--5.2d as interference terms in the transition
probability, while ImSCBA corresponds to the replacement of the
square of the sum in Fig.~\ref{fig:exchange} by the sum of squares
of individual terms.

\begin{figure}
\includegraphics[width=0.6\textwidth]{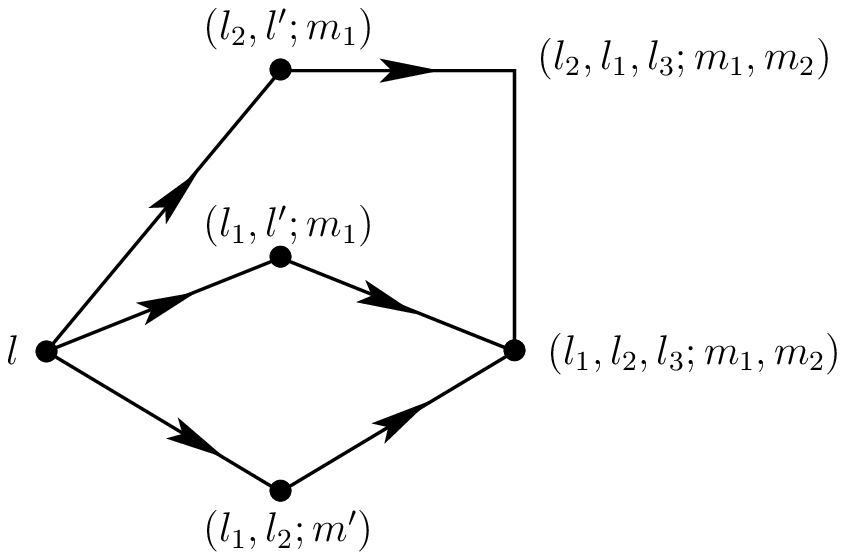}
\caption{Different paths for the transition from a one-particle
state~$l$ to a five-particle state $(l_1,l_2,l_3;m_1,m_2)$,
corresponding to different intermediate three-particle states.
The state  $(l_2,l_1,l_3;m_1,m_2)$ is identical to
$(l_1,l_2,l_3;m_1,m_2)$.}
\label{fig:path}
\end{figure}

This consideration for the fourth-order self-energy illustrates the
general rule of construction of all lowest-order diagrams
containing a cross-section of $N_e$~electrons and $N_h$~holes.
Let us assume for a moment that electrons and holes are distinguishable
particles and introduce the quantum-mechanical amplitude
$\mathcal{A}_\ell(\{l_i\}_{i=1}^{N_e};(\{m_i\}_{i=1}^{N_h})$ of the
transition to the given final state for a given path~$\ell$
(represented by a half-diagram with a definite assignment of orbital
indices to the free electron and hole lines). The decay rate
can then be represented as
\be
\begin{split}
&\Gamma_l\simeq 
\sum_{\{l_i\}, \{m_i\}}\gamma_l 
(\{l_i\},\{m_i\})
\delta\left(\ep-\sum_{i=1}^{N_e}\xi_{l_i}+\sum_{i=1}^{N_h}\xi_{m_i}\right)
;
\\
&\gamma_l 
=\frac{1}{N_e!N_h!}\left|
\sum_\ell\sum_{P_l,P_m}(-1)^{P_l+P_m}
\mathcal{A}_\ell(P_l\{l_i\},P_m\{m_i\})\right|^2.
\label{val:gamma1}
\end{split}
\ee
The factorial
prefactor takes into account the fact that different permutations
of the indices $\{l_i\},\{m_i\}$ correspond to the same final state,
the sum over the permutations represents the usual antisymmetrization
of the quantum-mechanical amplitude\footnote{It is easy to 
see that the diagrams (2) and (4) of
  Fig.~\ref{figSCBA}c
are also included in \req{val:gamma1}}. 

Similarly to the consideration 
of the quantum interference effects for
a single particle in a disorderd potential, 
the resulting double sum in  \req{val:gamma1}
can be separated into diagonal and off-diagonal part 
\be
\Gamma_l\propto\sum_\ell\left|
\mathcal{A}_\ell(\{l_i\},\{m_i\})\right|^2
+(\mbox{off-diag.})
\label{val:gamma2}
\ee
The first term in \req{val:gamma2} is nothing but the ImSCBA series
corresponding to the summation of the probablities of the paths.
Random-sign last term is the quantum interference contribution to
the particle lifetime. 

Even though the number of the interference terms is
much larger than the number of diagonal 
terms one can still argue that they do not affect the value of
the transition temperature. 

Indeed, if the number of
the relevant terms were large, one would be able to apply  the central
limit theorem for the quantum-mechanical amplitudes rather than
for the probabilities. It would result in a distribution
function of the same scale as the one obtained by the 
diagonal approximation but of a different shape
({\em e.g.}, 
for the large number of statiscally independent  amplitudes
Porter-Thomas distribution would replace the Gaussian one).
However, the characteristic functions obtained in Sec.~\ref{0D}
are non-Gaussian (as they are not analytic at $s\to 0$) 
and have  long algebraic  tails at large values of~$\Gamma$.
This indicates that the result is contributed by the largest
term in the sum. 
Therefore, the interference contribution affects
the distribution function in the range of the most probable values
but not the tail of the distribution. 
As     the transition
temperature is controlled by the tail, the interference
term is not important for the position of this 
temperature.\footnote{
Moreover, experience gained
in the study of the critical behavior of
the Anderson transition on the Cayley
tree~\cite{Efetov87}
suggests that the transition itself is associated with
the reconstruction of  the tail of the distribution.
Therefore, we do not see any reason to believe the the interference
terms
can affect the critical behavior in our problem either.}

To  quantify the qualitative consideration above
we evaluated the characteristic function
for the quantity on the right-hand side of \req{val:gamma1},
calculated in the $n$th order of the perturbation theory ($n\gg 1$),
\be
\begin{split}
&\ln \left\langle
\exp\left(-s\delta_\zeta^{-2} \sum_{\{l_i\}, \{m_i\}}\gamma_l 
(\{l_i\},\{m_i\})\right)
\right\rangle
= -\sqrt{s}\left(\frac{T}{T_*}\right)^n
\left(1-\sqrt{s\alpha_{int}}
+\dots\right);
\\
&
\alpha_{int}\simeq \left[\lambda^{n}n!\right]^2.
\end{split}
\label{val:gamma3}
\ee
Derivation of \req{val:gamma3} is relegated to Appendix~\ref{app3},
and we neglected $\ln \lambda$ factors in the expression for $\alpha_{int}$.

Leading at $s\to 0$ term is none but the the diagonal contribution
calculated in SCBA approximation. The subleading term is the
interference term,
and the parameter $\alpha_{int}$ has a 
meaning of the relative contribution
of the off-diagonal terms, affecting the distribution at
small values of $\gamma$.
Values of $\alpha_{int}$ contain the $n!$ factor, so it apparently
becomes large no matter how small the interaction constant $\lambda$
is. However, as we discussed in the derivation of \req{MITrepulsion},
the order of the perturbation theory involving only one localization
cell is limited from above, $n<n^*\simeq T/\delta_\zeta$.
As the result for $I=0$, $n=n^*$ we estimate
\be
\alpha_{int}\simeq 
\left[\frac{\lambda T}{\delta_\zeta}\right]^{2n^*}
\simeq M^{-2n^*} \ll 1,
\label{val:gamma30}
\ee  
which means that the interference processes do not affect
the most important part of the distribution functions even in
the largest possible order of the perturbation theory
without hopping between localization cells.

\begin{figure}
\includegraphics[width=0.6\textwidth]{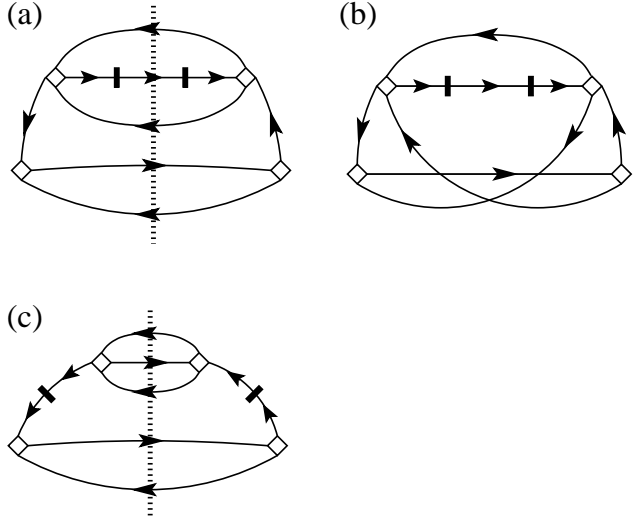}
\caption{
Suppression of the exchange processes shown on 
Fig.~\ref{fig:exchange} due to the inclusion of the
tunneling vertices.
Diagram a) has only one exchange counterpart  b)
whereas the diagram c) does not allow permutations at all
[{\em cf.} diagram (5.2) which produces three exchange counterparts]. 
}
\label{fig:tunexch}
\end{figure}

One could think that the inclusion of the hopping
would allow the growth of the $\alpha_{int}$ beyond the
estimate \rref{val:gamma30} at $n>n^*$.
However, it is not the case. 
Inclusion of tunneling into neighboring
localization cells will lead to further suppression of the
interference effects. The presence of tunneling vertices
on the diagram makes it impossible to interchange particles
residing in different localization cells, compare 
Figs.~\ref{fig:exchange} and~\ref{fig:tunexch}. 
It is possible to check that keeping the combinatorial $n!$
in the perturbation theory involving 
$n$ particles in the final states
spread over $m\ll n$  localization cells would require
the tunneling of all $n$ particles by the distance of
the order of $m$. Each tunneling event
brings additional smallness $I$, and thus we estimate
\be
\alpha_{int}\simeq \left[\lambda^{n}n!
I^{mn}
\right]^2 \lesssim
\left[n\lambda
\exp\left(-\frac{n |\ln I|\delta_\zeta}{T}\right)
\right]^{2n},
\label{val:gamma4}
\ee
where we used $m\simeq m^*$ from \req{mestimate}.
Thus, no accumulation of the factorial terms is possible,
even in the vicinity of the metal-insulator transition,
and the SCBA calculational scheme is valid not only on qualitative
but also on a quantitative level.

\subsection{Effect of the single-particle spectrum renormalization.}
\label{val:shift}

We now turn to the study
of the effects on the interaction and tunneling which
may be viewed as a change in the properties 
of single-particle excitations. This change includes 
the level shifts $\delta \xi_l$ and the possibility
of mixing with other orbitals (which is  present even without
interaction due to
the hopping $I$).
The level shifts are encoded in the statistics of the
real parts of the diagonal part of the 
self-energy ${\rm Re}\Sigma_l^R$ which were
neglected in ImSCBA approximation, whereas the bubbles 
non-diagonal with respect to the initial and final states,
see, {\em e.g.}, diagram (c3) of Fig.~\ref{figSCBA},
are responsible for the mixing.

The lowest order diagram is the Hartree-Fock potential of
Fig.~\ref{fig:diag2}a
\be
U^{HF}_{l_1l_2}=-\frac{1}{2}\sum_{l_3}
\left[n_{l_3}-\mathop{\rm sgn}
 \xi_{l_3}\right] V_{l_1l_3}^{l_3l_2},
\label{valHF}
\ee
where we subtracted the value of the potential at $T=0$.
The latter term is assumed to be incorporated into 
the one-particle Hamiltonian.
The characteristic function \rref{Fourier}
is found using the averaging procedure \rref{average}
and \req{Heffintdist} to be Gaussian
\be
\ln W^{U^{HF}}=-q^2\lambda^2\delta_\zeta T\ln 2
\sim -q^2\delta_\zeta^2\left(\frac{\lambda}{M}\right)
\left(\frac{T}{T_*}\right).
\label{WHF}
\ee
This means that the variation of the Hartree-Fock potential
with the distribution function is much smaller than the level
spacing $\delta_\zeta$ even for $T$ of the order of
transition temperature $T_c$. 

\begin{figure}
\includegraphics[width=0.6\textwidth]{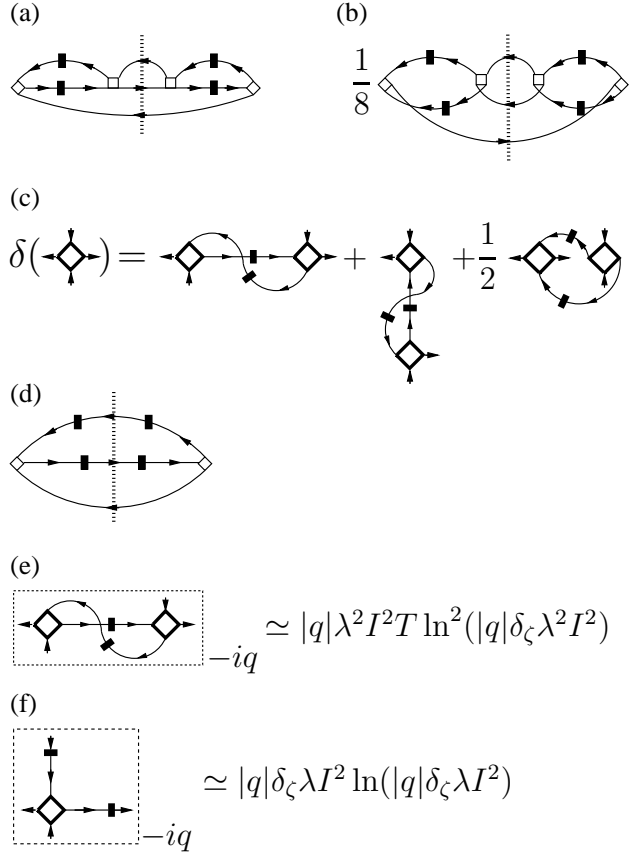}
\caption{
 Reduction of the non-SCBA diagrams (a-b) to
the interaction vertex non-local in space (c).
Characteristic function of the non-local interaction (c)
in comparison with the corresponding block (f) of the
SCBA diagram (d).
}
\label{fig:tunvertex}
\end{figure}

The Hartree-Fock diagram as well as other diagrams, corresponding
to single-particle level shifts and mixing, may be included as
self-energy insertions into single-particle Green's functions
in all diagrams of the previous subsections.
As a result, the Green's functions
become non-diagonal both in the orbital indices~$l$, and
the spatial index~$\vec\rho$, if tunneling is involved. 
Such insertions always introduce an additional smallness, as
discussed in Sec.~\ref{bulk} for the case of tunneling in the ImSCBA.
However, while inclusion of tunneling is necessary to go beyond the
finite state space of a single localization volume, insertions
non-diagonal only in the orbital index do not change the number of
the final states and thus can be ignored. Insertion of tunneling
into the renormalized interaction vertex, see
Fig.~\ref{fig:tunvertex}a, produces interaction, nonlocal in space.
However, the corresponding correction to~$\Gamma_l$ is ``small'' compared
to that of the ImSCBA diagram with Fig.~\ref{fig:tunvertex}b  
with the same final states, see Fig.~\ref{fig:tunvertex}c,d.
Thus we conclude the variations of the Hartree-Fock potential do
not generate any corrections to the transition temperature
or the statistics of $\Gamma_l$ calculated in ImSCBA.

However, there is an important feature in the statistics of
the level shifts $\xi_l$ which changes the numerical factor in
\req{Tc4} and changes the power-law decay in  the tail of
 the distribution of   $\Gamma_l$.
To see this, let us consider now the second-order
contribution shown on Fig.~\ref{fig:diag2}b.2. Real part of this
self-energy $\delta\xi_l(\ep)$ is included in SCBA but neglected in ImSCBA.
To investigate its effect, we apply the Kramers-Kronig relation
to $\Gamma_l^{(in)}(\ep)$ from \req{eqSCBA1}:
\be
\begin{split}
\delta\xi_l(\ep)&=\lambda^2\delta_\zeta^2
\sum_{l_1,l_2,l_3}
Y_{l_1,l_2}^{l_3,l}\int{\diff\ep_1}{\diff\ep_2}{\diff\ep_3}
F_{l_1, l_2; l_3}^\Rightarrow(\ep_1,\ep_2;\ep_3)\,
P
\frac{A_{\l_1}(\ep_1)A_{l_2}(\ep_2)
A_{l_3}(\ep_3)}{\ep-\ep_1-\ep_2+\ep_3}
;
\end{split}
\ee
where all  the notation is introduced in \req{eqSCBA1}.
Substituting $A_{l_i}(\ep_i)=\delta(\ep_i-\xi_{l_i})$, we obtain
\be
\begin{split}
\delta\xi_l(\ep)&=\lambda^2\delta_\zeta^2
\sum_{l_1,l_2,l_3}
\frac{Y_{l_1,l_2}^{l_3,l}F_{l_1, l_2; l_3}^\Rightarrow}
{\ep-\xi_{l_1}-\xi_{l_2}+\xi_{l_3}};
\end{split}
\ee
whose characteristic function is
\begin{equation}
\ln{W}^{\delta\xi}= -2\pi|q|\lambda^2MT\simeq
-|q|\lambda\delta_\zeta(T/T_*).
\label{Wdelta}
\end{equation}
Therefore, the typical value of
$\delta\xi\sim\lambda\delta_\zeta\ll\delta_\zeta$.
One thus might think that the main effect of this contribution is
a weak random shift of the level position, so it can be disregarded
as well.

Nevertheless, despite its smallness, $\delta\xi_l(\ep)$
introduces a new qualitative effect: {\em repulsion between many-body
levels}.
When $\delta\xi_l(\ep)$ is taken
into account, one has to find the position of the shifted
single-particle level from the equation
\be
\ep-\xi_l-\delta\xi_l(\ep)=0.
\label{shiftxi}
\ee
This equation describes repulsion between the single-particle
excitation~$l$ and three-particle excitations $(l_1,l_2;l_3)$:
solutions of Eq.~(\ref{shiftxi}) cannot approach each other
by a distance smaller than $\lambda\delta_\zeta$.

In the considerations of Sec.~\ref{sec:insulator} we
assumed the energies of one-particle states to be independent
of the energies of three-particle states into which these
one-particle states decayed. As a result, resonant denominators
[see, {\em e.g.}, Eq.~(\ref{Gamma20})] could become small independently
of each other. The level repulsion suppresses such effect,
so that Eqs.~(\ref{0DWinfresult}), (\ref{MITrepulsion}), and
(\ref{Wnmresult}) overestimate the strength of the tail
of the distribution.

It is the same effect that was first discussed by
Anderson~\cite{Anderson58} and analyzed rigorously by Abou-Chacra
{\em et al.}~\cite{AbouChacra}. According to their results, this
effect leads to a change in the numerical coefficient
in the equation for the critical disorder strength: $\eexp \to 2$.
Simply adapting this prescription for our \req{Tc4} we
obtain the transition temperature modified by level repulsion
\be
\frac{12\lambda M {T}_c}{\delta_\zeta}\,
\ln\frac{1}{\lambda}=
 \exp\left[\frac{c_1\delta_\zeta
\left|\ln I\right|}{T_c}\right].
\label{Tc4prime}
\ee

\begin{figure}
\includegraphics[width=0.7\textwidth]{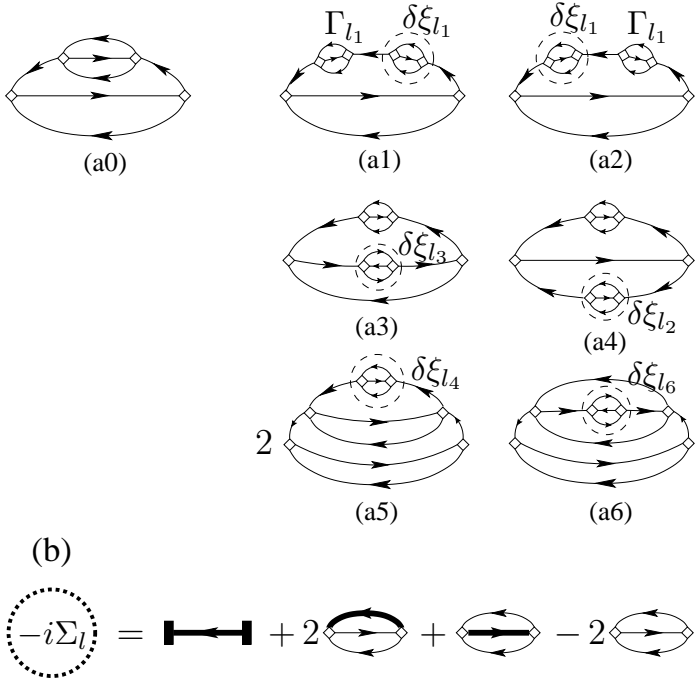}
\caption{(a) Including the level shift into the fourth order 
ImSCBA diagram. Only diagrams a.1 and a.2 are resonant,
{\em i.e.}, divergences in $\delta \xi_l$ and $\Gamma_l$ 
are correlated.
(b) Linearized SCBA approximation which incorporates the
simultaneous divergences in $\delta \xi_l$ and $\Gamma_l$.
}
\label{fig:linSCBA}
\end{figure}

Now we will sketch the derivation of \req{Tc4prime} for our problem.
First, we have to identify the sequence of the diagrams
which may give the level repulsion between the resonant multiparticle
states. To do so, we include the level shift $\delta \xi_l(\ep)$
into the imaginary part of the fourth order SCBA diagram as
shown on Fig.~\ref{fig:linSCBA}a.

Direct inspection of the diagrams  Fig.~\ref{fig:linSCBA}a shows that
for the levels $\xi_{l_1}, \dots, \xi_{l_6}$ maximizing
the skeleton diagram (a0) of Fig.~\ref{fig:linSCBA},
only diagrams (a1) and (a2) give rise to the simultaneous divergence
of $\delta \xi$, whereas in the remaining diagrams (a3)--(a6) 
$\delta \xi$ fluctuates independently of $ \Gamma_{l_1}$.

It means that the role of the level shift $\delta \xi$
in the diagrams (a3)--(a6), is, indeed, just the  broadening
of the distribution of $\xi_l$ by the value of $\lambda \delta_\zeta
(T/T_*)\ll \delta_\zeta$, see \req{Wdelta}. This broadening
does not significantly change the distribution of the resonant
denominators and can be disregarded together with the fluctuations of
the Hartree-Fock potential. On the other hand, the level shifts
on diagrams (a1), (a2) are large whenever  $ \Gamma_{l_1}$ is maximal
and they describe the effect of level repulsion discussed before.

The maximally divergent series incorporating the correlations
in $\delta \xi_l$ and $\Gamma_l$ is generated by the linearized
self-consitent-Born approximation (LSCBA) shown on
Fig.~\ref{fig:linSCBA}b,
as it can be checked by explicit consideration of several iterations.

The analytic expression of the LSCBA self-energy is
[{\em cf.} \reqs{eqSCBA1}, \rref{linSCBA2})]:
\begin{subequations}
\bea
\delta\xi_l (\ep)&=& 
 \sum_{l_1,\vec{a}}
\frac{I^2\delta_{\zeta}^2
\theta_\Delta[\ep-\xi_{l_1}(\vec\rho+\vec{a})]}
{\ep-\xi_{l_1}(\vec\rho+\vec{a})-
\delta \xi_{l_1}(\ep;\vec\rho+\vec{a})
}
\nonumber\\
&+&\lambda^2\delta_\zeta^2
\sum_{l_1,l_2,l_3}
Y_{l_1,l_2}^{l_3,l}F_{l_1, l_2; l_3}^\Rightarrow
\Bigg[
\frac{2\theta_\Delta(\ep-\Xi_{l_1l_2}^{l_3})}
{\ep-\Xi_{l_1l_2}^{l_3}-
\delta\xi_{l_1} (\ep-\Xi_{l_2}^{l_3})}\nonumber\\
&&\qquad+ \frac{\theta_\Delta(\ep-\Xi_{l_1l_2}^{l_3})}
{\ep-\Xi_{l_1l_2}^{l_3}+\delta\xi_{l_3} (\Xi_{l_1l_2}-\ep)}
-\frac{2\theta_\Delta(\ep-\Xi_{l_1l_2}^{l_3})}
{\ep-\Xi_{l_1l_2}^{l_3}}
\Bigg];
\label{eqLSCBAa}
\eea
\bea
\Gamma_l(\ep)&=&\Gamma_l^{(bath)}(\ep) +  \sum_{l_1,\vec{a}}
\frac{I^2\delta_{\zeta}^2 
\theta_\Delta[\ep-\xi_{l_1}(\vec\rho+\vec{a})]\Gamma_l(\ep,\vec\rho+\vec{a})}
{\left[\ep-\xi_{l_1}(\vec\rho+\vec{a})-
\delta \xi_{l_1}(\ep;\vec\rho+\vec{a})\right]^2}
\nonumber\\
&+& \sum_{l_1,l_2,l_3}\lambda^2\delta_\zeta^2
Y_{l_1,l_2}^{l_3,l}F_{l_1, l_2; l_3}^\Rightarrow
 \left\{
\frac{
2\theta_\Delta(\ep-\Xi_{l_1l_2}^{l_3})
\Gamma_{l_1}(\ep-\Xi_{l_2}^{l_3})}
{\left[\ep-\Xi_{l_1l_2}^{l_3}-\delta\xi_{l_1}
(\ep-\Xi_{l_2}^{l_3})
\right]^2}\right. +\nonumber\\
&&\qquad+\left.\frac{\theta_\Delta(\ep-\Xi_{l_1l_2}^{l_3})
\Gamma_{l_3}(\Xi_{l_1l_2}-\ep)} 
{\left[\ep-\Xi_{l_1l_2}^{l_3}+
\delta\xi_{l_3}(\Xi_{l_1l_2}-\ep) 
\right]^2} 
\right\}.\label{eqLSCBAb}
\eea
\label{eqLSCBA}
\end{subequations}
As in \req{eqSCBA1}, the coordinate $\vec{\rho}$ is
assumed to be the same in all terms in the equations unless it is
specified explicitly otherwise. The notation $\Xi_{l_1\dots}^{m_1\dots}$
was introduced in \req{Xi}. The ultraviolet cutoff function 
$\theta_\Delta(x)$ is the same as in \req{meaning}. It is important
to emphasize that it depends on the {\em unshifted} energies of the
levels. This is because the cutoff was introduced in the
non-interacting Green function first; the self-energy appears
in the denominator of the SCBA-dressed Green function as a
result of the summation of a geometric series; such summation
changes the denominator only, keeping the cutoff intact.

Now let us go through the steps of Sec.~\ref{0D} and see how they
are affected by the shifts in the denominators of \reqs{eqLSCBA}.
First, we notice that \req{eqLSCBAb} is still given
by the diagrammatic representation of Fig.~\ref{lcbadiag}
with the change in the rules of reading the double line:
\be
\setlength{\unitlength}{0.05 \textwidth}
\begin{picture}(8,1)
\put(2.2,0.6){\includegraphics[width=1.5\unitlength]{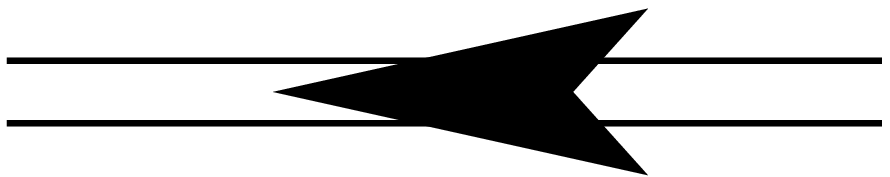}}
\put(2.5,0.2){$l$}
\put(2.5,1.1){$\ep$}
\put(4,0.6){$
\displaystyle{=\frac{
\theta_\Delta 
(\ep-\xi_l )
}
{
\left[
\ep-\xi_l-\delta\xi_l (\ep)
\right]^2
}}
$.}
\end{picture}
\label{change}
\ee
The lowest order term of the cluster expansion, \req{lowest},
see also Fig.~\ref{3pcluster},
becomes
\be
\begin{split}
 \Gamma^{(1,0)}&=2\Gamma^{(0,0)}
\sum_{l_1,l_2,l_3} \frac{\lambda^2\delta_\zeta^2
Y_{l_1,l_2}^{l_3,l}F_{l_1, l_2; l_3}^\Rightarrow\,
\theta_\Delta\!\!\left(\ep-\Xi_{l_1l_2}^{l_3}\right)}
{\left[\ep-\Xi_{l_1l_2}^{l_3}
-\delta\xi_{l_1}(\ep-\Xi_{l_2}^{l_3})\right]^2 }\\
&+\Gamma^{(0,0)}
\sum_{l_1,l_2,l_3} \frac{\lambda^2\delta_\zeta^2
Y_{l_1,l_2}^{l_3,l}F_{l_1, l_2; l_3}^\Rightarrow\,
\theta_\Delta\!\!\left(\ep-\Xi_{l_1l_2}^{l_3}\right)}
{\left[\ep-\Xi_{l_1l_2}^{l_3}
+\delta\xi_{l_3}(\Xi_{l_1l_2}-\ep)\right]^2}.
\end{split}
\label{shiftlowest}
\ee
We see that the corresponding potential is no longer three-particle,
but depends on all the coordinates via~$\delta\xi$, which depend
on the positions of all other levels. However, when
performing the linked cluster expansion, in each term of the sum
over triples $(l_1,l_2,l_3)$ we can shift an
integration variable:
$\xi_{l_1}+\delta\xi_{l_1}\rightarrow\xi_{l_1}$. After this shift
Eq.~(\ref{0DW3}) acquires the form
\bea
&&\ln W^{(1,0 )}_3=\frac{1}{2} \prod_{l=1}^{3} \left(
\int\frac{\diff\xi_l}{\delta_{\zeta}} \sum_{n_{l}=\pm 1}
\frac{\exp\frac{n_l\xi_l}{2T} }{2\cosh\frac{\xi_l}{2T}} \right)
(f^e+f^h);
\label{new0DW3}\\
&& f^e_{12;3}=
\left(\eexp^{-sU_{12;3}^e}-1\right)
\langle\theta_\Delta\!\!\left(\ep-\Xi_{12}^3+\delta\xi_1\right)\rangle',
\nonumber\\
&& f^h_{12;3}=
\left(\eexp^{-2 s U_{12;3}^h}-1\right)
\langle\theta_\Delta\!\!\left(\ep-\Xi_{12}^3-\delta\xi_3\right)\rangle',
\nonumber
\eea
where $\langle\ldots\rangle'$ denotes the average~\rref{average}
with levels $1,2,3$ excluded, and the potentials~$U_{12;3}^{(e,h)}$ 
were defined in \req{3body}. 
Using the fact that $\delta\xi_1(\ep-\Xi_{l_1l_2}^{l_3})$ 
from \req{eqLSCBAa} is not singular at
$\Xi_{12}^3 \rightarrow \ep$ and $|\delta\xi|\ll\Delta$, we can perform
the integration in the same way as in Sec~\ref{0D} to obtain
\be
\begin{split}
&\ln W^{(1,0 )}_3= - \left({\pi s  \Gamma^{(0,0)}}\right)^{1/2}
\frac{6\lambda MT}{\delta_\zeta}.
\end{split}
\label{new0DW3result}
\ee
Here the only effect of the shifts was to make the electron
contribution 
statistically
independent from the hole one. As a result,
the cross-term in \req{tildef} vanishes and
 the
numerical coefficient is changed in comparison with 
\req{0DW3result}.

Let us now consider the modification of the fourth
order result \rref{Gamma20}. After two iterations of 
\req{eqLSCBAb}, see also Fig.~\ref{6pcluster} and \req{change},
 we find
\be
\begin{split}
\Gamma^{(2,0)}&=\Gamma^{(0,0)}\lambda^4\delta_\zeta^4
\sum_{l_1,l_2;l_3} F_{l_1, l_2;l_3}^\Rightarrow{\sum_{l_4,l_5;l_6}}^\prime
 F_{l_4, l_5; l_6}^\Rightarrow
\\
&
\times 
\left\{
\frac{ 2 Y_{l_1,l_2}^{l_3,l}\, 
 \theta_\Delta\!\!\left(\ep-\Xi_{l_1l_2}^{l_3}\right)}
 {\left[\ep-\Xi_{l_1l_2}^{l_3}-\delta\xi_{l_1}\!\!\left(\ep-\Xi_{l_2}^{l_3}
 \right)\right]^2}
\frac{2Y_{l_4,l_5}^{l_1,l_6}\,
\theta_\Delta\!\!\left(\ep-\Xi_{l_2l_4l_5}^{l_3l_6}\right)}
{\left[\ep-\Xi_{l_2l_4l_5}^{l_3l_6}
-\delta\xi_{l_5}\!\!\left(\ep- \Xi_{l_2l_4}^{l_3l_6}
\right)\right]^2}
\right.
\\
& +
\frac{ 2 Y_{l_1,l_2}^{l_3,l}\, 
 \theta_\Delta\!\!\left(\ep-\Xi_{l_1l_2}^{l_3}\right)}
 {\left[\ep-\Xi_{l_1l_2}^{l_3}-\delta\xi_{l_1}\!\!\left(\ep-\Xi_{l_2}^{l_3}
 \right)\right]^2} 
\frac{Y_{l_4,l_5}^{l_1,l_6}\,
\theta_\Delta\!\!\left(\ep-\Xi^{l_4l_5}_{l_1l_2l_6}\right)}
{\left[\ep-\Xi^{l_3l_6}_{l_2l_4l_5} + \delta\xi_{l_6}
\left(\Xi_{l_2l_4l_5}^{l_3}-\ep
 \right)
 \right]^2}
\\
& +
\frac{  Y_{l_1,l_2}^{l_3,l}\, 
 \theta_\Delta\!\!\left(\ep-\Xi_{l_1l_2}^{l_3}\right)}
 {\left[\ep-\Xi_{l_1l_2}^{l_3}+\delta\xi_{l_3}\!\!\left(\Xi_{l_1l_2}-\ep
 \right)\right]^2}
\frac{2Y_{l_3,l_6}^{l_4,l_5}\,
 \theta_\Delta\!\!\left(\ep-\Xi_{l_1l_2l_6}^{l_4l_5}\right)}
 {\left[\ep-\Xi_{l_1l_2l_6}^{l_4l_5}
 + \delta\xi_{l_4}\!\!\left(\Xi_{l_1l_2l_6}^{l_5} -\ep  \right)
\right]^2}
\\
& +
\left. \frac{  Y_{l_1,l_2}^{l_3,l}\, 
 \theta_\Delta\!\!\left(\ep-\Xi_{l_1l_2}^{l_3}\right)}
 {\left[\ep-\Xi_{l_1l_2}^{l_3}+\delta\xi_{l_3}\!\!\left(\Xi_{l_1l_2}-\ep
 \right)\right]^2}
\frac{Y_{l_3,l_6}^{l_4,l_5}\,
 \theta_\Delta\!\!\left(\ep-\Xi_{l_1l_2l_6}^{l_4l_5}\right)}
 {\left[\ep-\Xi_{l_1l_2l_6}^{l_4l_5}
 - \delta\xi_{l_6}\!\!\left(\ep-\Xi_{l_1l_2}^{l_4l_5}  \right)
\right]^2}
\right\},
\end{split}
\label{val:Gamma20}
\ee
where the prime in the second sum 
has the same meaning as in \req{Gamma20}.
Each term in \req{val:Gamma20} produces its own cluster
function, see \req{0DW6expression}. 
Similarly to \req{new0DW3}, the cross-correlation terms vanish
and we obtain after obvious shifts of the variables
\begin{subequations}
{\setlength\arraycolsep{0pt}
\bea
&&\ln W^{(2,0 )} =\prod_{l=1}^{6} \left(
\int\frac{\diff\xi_l}{\delta_{\zeta}} \sum_{n_{l}=\pm 1}
\frac{\exp\frac{n_l\xi_l}{2T} }{2\cosh\frac{\xi_l}{2T}} \right)
\tilde{f}\left({\substack{12;\,3\\45;\,6}}\right);
\label{newW2}
\\
&&
\tilde{f}\left({\substack{12;\,3\\45;\,6}}\right)
= {f}^{ee}\left({\substack{12;\,3\\45;\,6}}\right)
+ \frac{1}{2}\,{f}^{eh}\left({\substack{12;\,3\\45;\,6}}\right)
+ \frac{1}{2}\,{f}^{he}\left({\substack{12;\,3\\45;\,6}}\right)
+  \frac{1}{4}\,{f}^{he}\left({\substack{12;\,3\\45;\,6}}\right);
\label{newtildefeh}
\eea
\be
\begin{split}
f^{ee}&=
(\eexp^{-sU^{ee}}-1)
\Big\langle\theta_\Delta\!\!\left[\ep-\Xi_{12}^{3}
+ \delta\xi_{1}\!\!\left(\ep-\Xi_{2}^{3}
 \right)\right]
\theta_\Delta\!\!\left[\ep-\Xi_{245}^{36}
+\delta\xi_{5}\!\!\left(\ep- \Xi_{24}^{36}\right)
\right]
\Big\rangle^\prime
;\\ 
f^{eh}&=
(\eexp^{-2 sU^{eh}}-1)
\Big\langle\theta_\Delta\!\!\left[\ep-\Xi_{12}^{3}
+ \delta\xi_{1}\!\!\left(\ep-\Xi_{2}^{3}
 \right)\right]
\theta_\Delta\!\!\left[\ep-\Xi_{245}^{36}
-\delta\xi_{6}\!\!\left(\Xi_{245}^{3}-\ep\right)
\right]\Big\rangle^\prime;
\\
f^{he}&=
(\eexp^{-2sU^{he}}-1)
\Big\langle\theta_\Delta\!\!\left[\ep-\Xi_{12}^{3}
- \delta\xi_{1}\!\!\left(\Xi_{12}-\ep
 \right)\right]
\theta_\Delta\!\!\left[\ep-\Xi_{45}^{126}
-\delta\xi_{4}\!\!\left(\Xi_{126}^{5}-\ep\right)
\right]\Big\rangle^\prime;
\\
f^{hh}&=
(\eexp^{-4 sU^{hh}}-1)
\Big\langle\theta_\Delta\!\!\left[\ep-\Xi_{12}^{3}
- \delta\xi_{1}\!\!\left(\Xi_{12}-\ep
 \right)\right]
\theta_\Delta\!\!\left[\ep-\Xi_{126}^{45}
+\delta\xi_{6}\!\!\left(\ep-\Xi_{12}^{45}\right)
\right]\Big\rangle^\prime,
\end{split}
\label{newfeh}
\ee
where $\langle\dots\rangle^\prime$ denotes averaging
\rref{average} with the levels $1,2,3,4,5,6$ excluded.
The potentials $U$ here are defined in \req{6body}.
}
\label{new0DW6expression}
\end{subequations}

Apparently, the role of the level shifts in
\reqs{newfeh} is similar to that in \req{new0DW3} --
perturbative modification of the cut-off. However, unlike
the lowest order perturbative corrections, the
shifts in \req{newfeh} contain a resonant term which depends 
on variables $\xi_1,\dots, \xi_6$ 
only and therefore can not be treated as a 
non-correlated random number. To see the origin of such resonant term,
consider the argument of the first $\theta$-function in $f^{ee}$.
From \req{eqLSCBAa} with $I=0$, we find
after the same variable shifts as in $f^{ee}$
\be
\delta\xi_{1}\!\!\left(\ep-\Xi_{2}^{3}\right)
=
\frac{2\lambda^2\delta_\zeta^2
Y_{4,5}^{6,1}F_{4, 5; 6}^\Rightarrow
\theta_\Delta\left[\ep-\Xi_{452}^{63}+ \dots
\right]
}
{\ep-\Xi_{452}^{63}}
+\dots,
\label{resonantfee}
\ee
where $\dots$ denote the terms which contain extra levels.
Those terms are random and can be disregarded. 
One can check by the same method that the shift
$\delta\xi_{5}\!\!\left(\ep- \Xi_{24}^{36}\right)$ always
depends on extra levels and does not produce resonance.

Having in mind that the $\theta$-functions in \reqs{newfeh}
cut off the logartihmic divergence, we neglect the non-resonant 
terms,\footnote{Keeping them will be beyond the accuracy of LSCBA
  approximations
where the terms of the same order were neglected in the very
beginning.} and simplify \reqs{newfeh} as
\be
\begin{split}
f^{ee}\left({\substack{12;\,3\\45;\,6}}\right)&=
\left[
\eexp^{- sU^{ee}\left({\substack{12;\,3\\45;\,6}}\right)}
-1\right]
\theta_\Delta\!\!\left(\ep-\Xi_{12}^{3}
+ 
\frac{2\lambda^2\delta_\zeta^2
Y_{4,5}^{6,1}
}
{\ep-\Xi_{452}^{63}}
\right)
\theta_\Delta\!\!\left(\ep-\Xi_{245}^{36}
\right);\\ 
 f^{eh}\left({\substack{12;\,3\\45;\,6}}\right)&=
 \left[
 \eexp^{-2 sU^{eh}\left({\substack{12;\,3\\45;\,6}}\right)}
 -1\right]
 \theta_\Delta\!\!\left(\ep-\Xi_{12}^{3}
 + 
 \frac{2\lambda^2\delta_\zeta^2
 Y_{4,5}^{6,1}
 }
 {\ep-\Xi_{452}^{63}}
 \right)
 \theta_\Delta\!\!\left(\ep-\Xi_{245}^{36}
 \right);
\\
 f^{he}\left({\substack{12;\,3\\45;\,6}}\right)&=
 \left[\eexp^{-2sU^{he}\left({\substack{12;\,3\\45;\,6}}\right)}
 -1\right]
  \theta_\Delta\!\!\left(\ep-\Xi_{12}^{3}
  + 
  \frac{
\lambda^2\delta_\zeta^2
  Y^{4,5}_{3,6}
  }
  {\ep-\Xi_{45}^{126}}
  \right)
 \theta_\Delta\!\!\left(\ep-\Xi_{45}^{126}
  \right);
 \\
 f^{hh}\left({\substack{12;\,3\\45;\,6}}\right)&=
 \left[\eexp^{-4 sU^{hh}\left({\substack{12;\,3\\45;\,6}}\right)}
 -1\right]
  \theta_\Delta\!\!\left(\ep-\Xi_{12}^{3}
  + 
  \frac{\lambda^2\delta_\zeta^2
  Y^{4,5}_{3,6}
  }
  {\ep-\Xi_{45}^{126}}
  \right)
  \theta_\Delta\!\!\left(\ep-\Xi_{45}^{126}
  \right).
\end{split}
\label{leadingfeh}
\ee

Substituting \req{leadingfeh} into \reqs{newW2}--\rref{newtildefeh}
and performing integration we obtain analogously to \req{0DW6result}
\be
\ln W^{(2,0 )}= - 
\left(\pi s  \Gamma^{(0,0)}\right)^{1/2}
\left[\frac{6\lambda M T}{\delta_\zeta}\right]^2
\ln \left(\frac{\Delta^2}
{\lambda^2\delta_\zeta^2} \right).
\label{new0DW6result}
\ee
In addition to the change in the overall numerical factor similar
to \req{new0DW3result}, \req{new0DW6result} shows new important
feature. Namely, the argument of the logartihm is no longer dependent
on the parameter $s$, or, in other words, the algebraic tail
in the distribution function is suppressed in comparison with 
\req{0DW6result}. This is the manifestation of the level repulsion
in complete analogy with the arguments of Anderson~\cite{Anderson58}.

\begin{figure}
\includegraphics[width=0.33\textwidth]{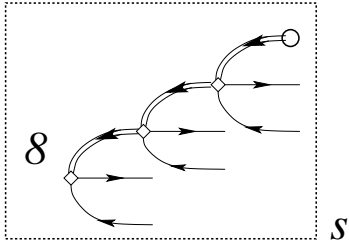}
\caption{The cluster function $f^{eee}$
for the six order perturbation theory expansion. 
The notation is defined on Figs.~\ref{lcbadiag},
~\ref{clusterexpansion} and in \req{change}.
}
\label{9pcluster}
\end{figure}

The procedure outlined above can be continued to all orders
of perturbation theory. In particular, for the sixth order
cluster function, Fig.~\ref{9pcluster}, one finds
\be
\begin{split}
f^{eee}=&
\left[
\eexp^{- sU^{eee}\left({\substack{12;\,3\\45;\,6\\78; 9}}\right)}
-1\right]
\theta_\Delta\!\!\left(\ep-\Xi_{12}^{3}
+ 
\frac{2\lambda^2\delta_\zeta^2
Y_{4,5}^{6,1}
}
{\ep-\Xi_{245}^{36}}
\right)
\\
&\times
\theta_\Delta\!\!\left(\ep-\Xi_{245}^{36}
+ 
\frac{2\lambda^2\delta_\zeta^2
Y_{7,8}^{9,4}
}
{\ep-\Xi_{2578}^{369}}
\right)
\theta_\Delta\!\!\left(\ep-\Xi_{2578}^{369}
\right);
\\
U^{eee}=&
\frac{
8\Gamma^{(0,0)}\lambda^6\delta_\zeta^6 Y_{1,2}^{3,l}
F_{1,2;3}^\Rightarrow Y_{4,5}^{6,1}F_{4,5; 6}^\Rightarrow
Y_{7,8}^{9,4}F_{7,8; 9}^\Rightarrow
}
{\left(\ep-\Xi_{12}^{3}\right)^2
\left(\ep-\Xi_{245}^{36}\right)^2\left(\ep-\Xi_{2578}^{369}
\right)^2},
\end{split}
\label{feee}
\ee
which means that the scale of the integration is
determined by the energy of the previous generation only.
This transfer matrix structure of the cluster functions repeats
itself in all orders and  makes
it possible to perform the integration in any order.
Instead of \req{Wnmresult} we find
\be
\begin{split}
&\ln W^{(n,m )}= - \left({  s  \Gamma^{(0,0)}}\right)^{1/2}
 {\mathcal{C}(n,m)}
\\
&\quad
\times
\left[
\frac{6\lambda M
T}{\delta_\zeta}
\ln
 \left(\frac{\Delta^{2}}
 {\delta_\zeta^{2} \lambda^{2}}
 \right)
\right]^n \left[I
\ln
 \left(\frac{\Delta^{2}}
 {\delta_\zeta^{2} I^{2}}
\right)
\right]
^m,
\end{split}
\ee
{\em i.e.}, once again the algebraic tail becomes ``more short-range'' due to
the suppression of the large denominators.
Coefficients ${\mathcal{C}(n,m)}$ are insensitive to the
resonant denominators and they are still given by
\req{dnm}.
We, therefore, can repeat all of the
arguments after \req{dnm} and obtain \req{Tc4prime}
after replacement \rref{replacement}.

To conclude, the level repulsion present in SCBA but neglected
in SCBA leads to the change in the numerical prefactor in the
expression for the transition temperature [cf. \req{Tc4prime}
with \req{Tc4}] but does not affect the statement about the
existence of the transition itself. All other corrections
lead to the perturbative corrections to~$T_c$.

\subsection{Validity of ImSCBA in the metallic phase}
\label{sec:valmetal}

In the metallic phase the classification of non-SCBA diagrams into
interaction vertex corrections and interference terms is the same
as presented in Secs.~\ref{sec:renint} and~\ref{sec:exchange} for the
insulating phase. What changes is that the resonant terms
in all of the expressions now acquire the finite width $\Gamma^{(in)}$
\be
\begin{split}
&\pi \delta(x) + P \frac{\Gamma}{ x^2} \to \frac{\Gamma^{(in)}}{x^2 +
\left[\Gamma^{(in)}\right]^2};\qquad
P \frac{1}{x} \to \frac{x}{x^2 +\left[\Gamma^{(in)}\right]^2},
\end{split}
\label{valreplacement1}
\ee
where $\Gamma^{(in)}$ is given by \req{gammam1}. 
This finite width prohibits vanishing of the denominators, thereby
cutting off the power-law tails of the distribution functions for
all the quantities considered  in
Secs.~\ref{sec:renint}--\ref{val:shift}.
As a result, all the cumulants of the distribution function become
finite. Moreover, under the condition \rref{mcondition2} the
distribution functions may be considered Gaussian.

Another feature is that the level occupation \rref{n} is no longer
binary, $n_l(\ep)=\pm 1$, as in insulating phase, but
it is kept close to its equlibrium value $n_l(\ep)=\tanh\frac{\ep}{2T}$
by the energy relaxation. This further suppresses the fluctuations
by the factor of $\delta_\zeta/T$. 

On the other
hand, in Sec.~\ref{sec:renint}, we established that
the role of the higher-order correction to the vertices is
the perturbative renormalization of the SCBA results.
As the scale of the fluctuation of those vertices is suppressed
even further in the metallic phase, neglecting this
renormalization is justified even more in this phase.

The role  of the particle permutations and interference in the
final state, see Sec.~\ref{sec:exchange} is investigated by direct
evaluation of the mesoscopic fluctuations and averages of the
diagrams of Fig.~\ref{fig:exchange} in a fashion of Fig.~\ref{figSCBAm}.
Those contribution are smaller than SCBA values by a factor of $M$
at least.
Therefore, deep in the metallic phase the kinetic equation 
consideration of transport in Sec.~\ref{sec:metalB} is well justified.

Finally, in the insulating phase the most important effect, not
included in ImSCBA, was  many-mody level repulsion,
see discussion around \req{shiftxi}. The energy scale of this
repulsion is $\lambda\delta_\zeta$. On the other hand, at
$T > T_{in}$, where  $T_{in}$ is defined in \req{mcondition2},
we find $\Gamma^{(in)} > \lambda \delta_\zeta$. This means
that the spectrum structure produced by
this repulsion is smeared by inelastic processes and
need not be taken into account.

Thus, we conclude that the ImSCBA is justified parametrically in
deep in the metallic phase.

\section{Conclusions}\label{sec:conclusions}

This paper is devoted to the analysis of the low-temperature transport
in disordered conductors with localized one-particle
states. The main question is whether the electron-electron interaction
alone is sufficient to establish the
thermal equilibrium in the system. The same question can be formulated
even more boldly -- {\em whether there is a many-body mobility
threshold}, {\em i.e.}, energy separating the many-body states
localized in the Fock space
of the system from the states which are delocalized.

One can apply weak-localization arguments to show that conductivity at
high enough temperatures is non-zero. It is not disputable as long as
the inelastic dephasing rate $1/\tau_\phi$ exceeds
the level spacing in one-particle localization volume (cell)
$\delta_\zeta$. Extension of this approach to lower temperatures is
problematic as the quantum corrections to conductivity diverge.
For this reason, in order to describe the low-temperature behavior,
we adopt a different strategy.
We formulate the effective
Hamiltonian description for the processes with
the energy transfer of the order of~$\delta_\zeta$.
Reduction  of the original Hamiltonian to the effective one is not
performed systematically. Nevertheless, we believe that it is an
apropriate low-energy limit of the theory of electrons 
in disorder potential.

Statisitical analysis based on the effective Hamiltonian enables us to
demonstrate the stability of two qualitatively different phases
-- {\em metallic}, for $T>T_c$ and {\em insulating} for $T<T_c$,
where $T_c$ is given by \req{Tc4prime}. This corresponds to
the existence of the many-body mobility threshold $\mathcal{E}_c$
related to $T_c$ by the thermodynamic formula \rref{Temp}.

We show that deep in the metallic phase, $T\gg T_c$ (see
Sec.~\ref{sec:metal}) the transport
coefficients in the system are self-averaging. Using this fact, we
derived the quantum Boltzman equation. The temperature
dependence of the electrical conductivity~$\sigma(T)$, following from this
equation (see Sec.~\ref{sec:metalB}), is quite non-trivial even for
$T\gg{T_c}$. Namely, $\sigma(T)$ increases with~$T$ as $T^2$ at low
temperatures, while at high temperatures it saturates, approaching the
Drude limit. Thermal conductivity deviates from the Wiedemann-Franz
law with the decreasing temperature. However,
this deviation is never strong, see Fig.~\ref{Plot1d}.

In the insulating phase, $T<T_c$, we use Feynmann diagram technique to
determine the probability distribution function for quantum-mechanical
transition rates. The probability of an escape rate from a given
quantum state to be finite turns out to vanish in every order of the
perturbation theory in electron-electron interaction.
Thus, in the absence of coupling to any
external bath (phonons) electron-electron interaction alone is
unable to cause the relaxation and establish the thermal equilibrium.
In other words, the insulating phase is stable.

Although $\sigma (T)=0$ exactly as long as $T<T_c$, the stability of the 
insulator decreases as $T$ approaches $T_c$. 
It means that effects of interaction 
of the electrons with the external bath (phonons) become more and more 
pronounced. More precisely, if the 
electron-phonon coupling is weak and $T \ll T_c$, one phonon can cause at 
most one electron to hop because the phase volume of the accessible final 
states is quite small. The closer is $T$ to $T_c$, the bigger is this 
phase volume, and in the vicinity of the transition point one phonon can 
initiate a whole cascade of the electronic hops.
The size of the cascades fluctuates strongly, depending on the
realization of disorder in the system.
The typical size of the cascade grows at $T\rightarrow{T}_c$ [see
Eq.~(\ref{treecorrlength=})].
It means, that even the infinitesimal 
electron-phonon interaction would produce a finite conductivity.
This is the onset of the metallic phase.

We also conclude that the phonon-induced hopping conductivity in the 
insulating phase close to the transition is strongly enhanced by the 
electron-electron interaction. This conclusion can be relevant for the 
numerous
experiments~\cite{1999Pepper,2000Berge,2000Gershenson,2002Pepper,2003Yakimov},
where the observed conductivity
in the strongly localized phase of disordered conductors was
too large to be explained by  conventional theory of phonon-assisted
hopping conductivity.

It should be emphasized that the many-body localization, which we discuss 
in this paper, is qualitatively different from conventional finite 
temperature Metal to Insulator transitions, such as formation of a 
band insulator due to the structural phase transition or Mott-Hubbard 
transition~\cite{Motttr,Hubbard}. In these two cases, at a certain
temperature $T^*$ a gap appears in the spectrum of charge excitation
(Mott insulator) or all 
excitations (band insulator). However the conductivity remains finite 
although exponentially small as long as $T>0$. This is not the case for 
many-body localization, which causes exactly zero conductivity in the 
low-temperature phase.

Is the many-body localization a true thermodynamical phase transition with 
corresponding singularities in all equilibrium properties?
This question definitely requires
additional studies, however, some speculations can be put forward.
The physics described in the present paper is associated with
the change of the characteristics of the many-body wavefunctions.
It is well known that for non-interacting systems localization-delocalization
transition does not influence the average density of states,
{\em i.e.}, it does not affect
any macroscopic thermodynamic properties. Application of the same
logic to the exact many-body eigenvalues would indicate that the
many-body localization tranisition
is not followed by any singularities in the static specific heat, etc.
On the other hand, at this point we can not rule out the possibility
that this conclusion is an artefact of treating the real parts of the
electron self-energies
with an insufficient accuracy. Most likely scenario, to our opinion, 
is that the insulating phase 
behaves like a glass (spin or structural) and demonstrates all the glassy 
properties~\cite{Glass}, like absence of ergodicity (even when some
coupling with  phonons is included), effects of aging, etc. 
Question of the equilibrium susceptibilities in the latter case
becomes quite meaningless. 

The quantitative theory built in this paper assumes that the
interaction is weak. On the other hand, qualitative consideration
of the localization of many-body excitations does not rely on this
assumption. The important ingredients are (i)~localization of
single-particle excitations, and (ii)~Fermi statistics. 
Consider, as an example, Wigner crystal~\cite{Wigner}.
It is well known that strong enough interaction leads to a spontaneous
breaking of the translational symmetry in $d$-dimensional clean
systems at $d\geq{2}$. In a clean system Wigner crystallization is either
a first-order phase transition ($d=3$), or a Kosterlitz-Thouless
transition ($d=2$). Even weak disorder destroys both translational and
orientational order~\cite{Larkin} and pins the crystal. The symmetry
of this state is thus not different from the symmetry of a liquid, and
the thermodynamic phase transition is commonly believed to be reduced
to a crossover.

We argue that the many-body localization provides the correct
scenario for the finite-temperature ``melting'' transition between the
insulating phase, which may be called ``solid'', and the metallic
phase, which may be called ``liquid''. Indeed, the conductivity of the
pinned Wigner crystal is provided by the motion of defects. At low
temperatures and in the absence of the external bath, all defects are
localized by the one-particle Anderson mechanism. Phonon modes of the
Wigner crystal are localized as well, so the system should behave as a
many-body insulator. As the temperature is increased, the many-body
metal-insulator transition occurs, though it is not clear at present,
whether it occurs before or after the crystalline order is destroyed
at distances smaller than Larkin's scale. Construction of effective
theory of such a transition is a problem which deserves further
investigation.

\section{Acknowledgements}
We are grateful to M.~E.~Gershenson and V.~E.~Kravtsov for valuable
remarks. We also want to thank A.~D.~Mirlin for
clarifying statements made in their preprint~\cite{Gornyiv1} and
revealing the content of Eqs.~(19)--(20) of Ref.~\cite{Gornyiv1} to us.

I.L.A. was supported by Packard foundation,
B.L.A. acknowledges financial support from
US DOE Office of Science under contract No. W-31-109-ENG-38, ARO/ARDA 
(DAAD19-02-1-0039) DARPA under QuIST program and NSF grant DMR 0210575.

{\em Note added}-- When the
current manuscript was almost completed, a new version of 
preprint~\cite{Gornyiv1} appeared~\cite{Gornyiv2}, which deals with some of the
problems  constituting the content of the present paper.

\appendix

\section{Slow energy relaxation?}
\label{subslow}

Let us now come back to the assumption of the strong inelastic relaxation
which was used to validate expansion \rref{sol1}.
One could think that
in some cases this assumption is not valid, which would lead to the
deviation of the temperature dependences from those of \reqs{observables}.
To investigate the limits of validity of
Eqs.~(\ref{observables})--(\ref{WF}),
we employ the following qualitative arguments. These arguments
only slightly modify the discussion after Eq.~(\ref{results}).

We notice that the transport still occurs through rare pin-holes.
However, the insufficient rate of the
inelastic processes allows the electron to explore an energy strip of the
width $\bar\ep\ll{T}$ before leaving the site via tunneling to the neighboring
sites. Correspondingly, instead of using Eq.~(\ref{sol1})
one may look for the solution in the form
\begin{equation}
\Phi(\ep,\vec\rho;t)=\frac{\delta\mu(\vec\rho,t)}{\bar\ep}\,
\tilde\beta(\ep/\bar\ep)\,,\quad\int \diff{x}\tilde\beta(x)\simeq{1}\,,
\end{equation}
where $\bar\ep$~is a scale to be found self-consistently. The
conductivities are still given by Eqs.~(\ref{observables}), but
the random quantities $\Beta_\sigma,\Beta_\kappa$ replaced by
different ones $\tilde\Beta_\sigma,\tilde\Beta_\kappa$, which
are defined through $\tilde\beta(xT/\bar\ep)$ rather than through
$\beta_{\sigma,\kappa}(x)$. Thus, either a straightforward calculation, or a
qualitative argument, similar to that after Eq.~(\ref{results}),
give for the typical~$\tilde\Beta$ an estimate
\begin{equation}\label{tildeBeta=}
\tilde\Beta\simeq\frac{\Gamma^{(in)}\bar\ep}{\delta_\zeta^2}\,,
\end{equation}
valid if $\tilde\Beta\ll 1$. To find~$\bar\ep$, we balance the
elastic and inelastic terms in Eq.~(\ref{finalsmooth}). The estimate
for the elastic term is
\begin{equation}\begin{split}
I^2\delta^3\sum_{\vec{a}}
\langle{A}(\ep,\vec\rho)\,{A}(\ep,\vec\rho+\vec{a})\rangle
\Phi
\simeq{I}^2\delta_\zeta\tilde\Beta\Phi \simeq
\Gamma^{(in)}\,\frac{I^2\bar\ep}{\delta_\zeta}\,\Phi,
\end{split}\end{equation}
while the inelastic term is estimated from Eq.~(\ref{hatStPhi=}) as:
\begin{equation}\begin{split}
\widehat{\mathrm{St}}_\Phi\Phi
&\simeq
D^{(\ep)}\partial_\ep^2\Phi
\simeq\Gamma^{(in)}\,\frac{\delta_\zeta^2 M^2}{\bar\ep^2}\Phi
\end{split}
\end{equation}
Requiring these two rates to be of the same order, we estimate the
scale~$\bar\ep$ as
\begin{equation}\label{barep=}
\bar\ep=\delta_\zeta\left(\frac{M}{I}\right)^{2/3}.
\end{equation}

Conditions $\bar\ep\ll{T}$ and $\tilde\Beta\ll{1}$ would produce a
parametric temperature region
\begin{equation}
\label{slowrel}
\delta_\zeta\left(\frac{M}{I}\right)^{2/3}
\lesssim T \lesssim
\frac{\delta_\zeta}{\lambda^2{M}}\left(\frac{I}{M}\right)^{2/3}
,
\end{equation}
in which the non-equilibrium function $\Phi(\ep)$ does not have
the quasi-equilibrium shape~(\ref{sol1}).
However, existence of such a regime requires the condition
\be
\left(\frac{I}{\lambda}\right)^{4/3}
\left(\frac{1}{\lambda M^2}\right)^{7/6}\lambda^{1/2} \gg 1,
\label{alas}
\ee
which is not consistent with the additional assumptions
$M^2\lambda \simeq 1,\ I\simeq \lambda \ll 1$.
Therefore, the regime of the slow energy relaxation is not
feasible for this model and \reqs{observables}, \rref{fromLaplace}
describe the
entire temperature dependence for $T \gtrsim T_{in}$.

\section{Probability distributions for $\Beta_{\sigma,\kappa}$.}
\label{app1}

Let us represent the definitions~(\ref{smallg}) in the form
\be
\begin{split}
&\Beta=\sum_{l,l'}B(\xi_l,\xi_{l'})=\sum_{l,l'}\frac{\delta_\zeta}{2T}\,
\beta\!\left(\frac{\xi_{l}}{2T}\right)
\frac{2\delta_\zeta\Gamma/\pi}{(\xi_l-\xi_{l'}')^2+4\Gamma^2}\,,
\end{split}
\label{largeB}
\ee
where the functions $\beta_\sigma(x)$ and $\beta_\kappa(x)$ are
given by \req{smallbeta}.
(We omit the superscript ``$(in)$'' of $\Gamma^{(in)}$ everywhere
in this appendix). The positions of the levels $\xi_l,\xi_{l'}'$
are assumed to be completely uncorrelated,
so the sought characteristic function
$\tilde{P}(s)=\left\langle{\eexp^{-s\Beta}}\right\rangle$ can be represented
as
\be
\begin{split}
\tilde{P}(s)=\lim_{N\rightarrow\infty}
\left(\int\limits_{-N\delta_\zeta/2}^{N\delta_\zeta/2}
\prod_{l=1}^N
\frac{\diff\xi_l^{+}\diff\xi_{l}^{-}}{(N\delta_\zeta)^2}
\right)
\eexp^{-s\sum\limits_{l,l'=1}^NB(\xi_l^+,\xi_{l'}^-)}.
\end{split}
\label{A2}
\ee
Equation is equivalent to the partition function of the
classical gas of two species, $(\pm)$, and interacting
to each other via pair potential  $B(\xi_l^+,\xi_{l'}^-)$.
It can be immediately
evaluated using Mayer-Mayer cluster expansion~\cite{Mayer}.
We will keep
the contributions upto four-particle clusters
to justify the further approximations:
{\setlength\arraycolsep{0pt}
\begin{subequations}
\label{P234}
\bea
&&\ln\tilde{P}(s)= \ln\tilde{P}^{(2)}(s)+  \ln\tilde{P}^{(3)}(s) +
\ln\tilde{P}^{(4)}(s)
\dots\\
&&
\ln\tilde{P}^{(2)}(s)=
\int\limits_{-\infty}^{\infty}
\frac{\diff\xi_1\diff\xi_2}{\delta_\zeta^2}f_{12};
\label{P2}\\
&&\ln\tilde{P}^{(3)}(s)
=\frac{1}{2}
\int\limits_{-\infty}^{\infty}\frac{\diff\xi_1\diff\xi_2\diff\xi_3}{\delta_\zeta^3}
\left[
f_{12}f_{32} + f_{21}f_{23}
\right],
\\
&&\ln\tilde{P}^{(4)}(s)
=
\int\limits_{-\infty}^{\infty}
\frac{\diff\xi_1\dots \diff\xi_4}{\delta_\zeta^3}
\left[
f_{12}f_{32}f_{34}
+\frac{f_{12}f_{32}
f_{34}f_{14}}{2}
\right],
\nonumber \\
\label{P4}
\eea
where
\be
f_{12}=\exp\left[-s B(\xi_1,\xi_{2})\right]-1,
\label{f}
\ee
is the Mayer's $f$-function which vanishes rapidly
with the distance between the levels $|\xi_1-\xi_2|$.
\end{subequations}
}
Performing integrations in \reqs{P2}--\rref{P4} with
the help of definition \rref{largeB} and using $\Gamma \gg T$,
we find
\be
\begin{split}
&\ln\tilde{P}^{(2)}(s)=-
\int\limits_{-\infty}^{\infty}\diff{x}
\left[r S_2\left(\frac{s\beta(x)}{r}\right)\right];
\\
&
\ln\tilde{P}^{(3)}(s)=\frac{\delta_\zeta}{2T}
\int\limits_{-\infty}^{\infty}\diff{x}
\left[r S_2\left(\frac{s\beta(x)}{r}\right)\right]^2;
\\
&\ln\tilde{P}^{(4)}(s)=\left(\frac{\delta_\zeta}{2T}\right)^2
r^3 \int\limits_{-\infty}^{\infty}\diff{x}
\left[-S_2^3\left(\frac{s\beta(x)}{r}\right)
+ S_4\left(\frac{s\beta(x)}{r}\right)\right];
\\
&S_2(y)=y
\eexp^{-y}\left[\mathcal{I}_0(y)+\mathcal{I}_1(y)\right].
\\
&S_4(y)=
\int_{-\infty}^\infty
\frac{\diff{x}_1\diff{x}_2\diff{x}_3}{(2\pi )^3}
\\
&\quad\times
R\left(y,x_2\right)R\left(y,x_2-x_3\right)
R\left(y,x_3-x_4\right)R\left(y,x_4\right)\\
&
R(y;x)=\exp\left(-\frac{2y}{1+x^2}\right)-1.
\end{split}
\ee
Here $\mathcal{I}_0$~and~$\mathcal{I}_1$
are the modified Bessel functions. We also have
introduced the parameter $r\equiv{8}\pi\Gamma{T}/\delta_\zeta^2$,
see \req{rT}.

Next step is to notice that $T \gg \delta_\zeta$ and the
observable quantities are contributed by $s$ such as
$r S_2\left({s}/{r}\right) \lesssim 1$. Therefore,
the contribution from the higher cluster are suppressed
and can be neglected. All the further calculation
is performed using $\tilde P^{(2)}$ only.

Two limiting cases can be considered. First, if $r\gg{1}$,
$S(x)$ can be expanded
$S(x)=x(1-x/2+\dots)$, which gives
\begin{equation}\begin{split}
r\gg{1}:\quad\left\{\begin{matrix}\ln\tilde{P}_\sigma(s) \\
\ln\tilde{P}_\kappa(s)\end{matrix}\right\}
&\approx-s+\frac{s^2}{2r}
\int\limits_{-\infty}^{\infty}\diff{x}
\left\{\begin{matrix}
\beta_\sigma^2(x) \\ \beta_\kappa^2(x)
\end{matrix}\right\}\\
&= -s+\frac{s^2}{2r}
\left\{\begin{matrix}
1/3 \\ 7/5-12/\pi^2
\end{matrix}\right\}.
\end{split}\end{equation}
In the opposite case,
$r\ll{1}$, we use the asymptotic expansions of the Bessel functions
to approximate
$S(x)\approx \sqrt{2/(\pi x )}$, to obtain
\begin{equation}\begin{split}
r\ll{1}:\quad\left\{\begin{matrix}\ln\tilde{P}_\sigma(s) \\
\ln\tilde{P}_\kappa(s)\end{matrix}\right\}
\approx-\sqrt{\frac{2}{\pi}\,rs}
\int\limits_{-\infty}^{\infty}\diff{x}
\left\{\begin{matrix}
\sqrt{\beta_\sigma(x)} \\ \sqrt{\beta_\kappa(x)}
\end{matrix}\right\}\\
= -\sqrt{\frac{2}{\pi}\,rs}
\left\{\begin{matrix}
\pi/\sqrt{2} \\ 4\sqrt{6}\mathrm{G}/\pi
\end{matrix}\right\}\equiv
\left\{\begin{matrix}
-\sqrt{c_\sigma{rs}} \\ -\sqrt{c_\kappa{rs}} \end{matrix}\right\},
\end{split}\end{equation}
where $\mathrm{G}\approx{0}.916$ is the Catalan's constant.

It is instructive to calculate
the inverse Laplace transform of $\tilde{P}_{\sigma,\kappa}(s)$
and find the distributions functions $P_{\sigma,\kappa}(\Beta)$. In
both limiting cases of large and small $r$
one can use the saddle point approximation.
For $r\gg{1}$ the distributions are approximately Gaussian with
$\langle\Beta_{\sigma,\kappa}\rangle=1$ and
\be
\langle\Beta_\sigma^2\rangle-\langle\Beta_\sigma\rangle^2=
\frac{1}{3r}\,,\quad
\langle\Beta_\kappa^2\rangle-\langle\Beta_\kappa\rangle^2=
\frac{7/5-12/\pi^2}{r}\,,
\ee
corresponding to self-averaging of~$\Beta_{\sigma,\kappa}$.
For $r\ll{1}$ have:
\begin{equation}\begin{split}
&P(\Beta)=\int\limits_{-i\infty}^{i\infty}\frac{\diff{s}}{2\pi{i}}\,
\eexp^{s\Beta-\sqrt{crs}}
\approx
\frac{\sqrt{cr/\pi}}{2\Beta^{3/2}}\,\eexp^{-cr/(4\Beta)}.
\end{split}\end{equation}
It is noteworthy that,
for $\Beta\gg{r}$ this ditribution coincides with the distribution
of the largest of $n=\sqrt{8\pi{c}}\,(T/\delta_\zeta)$ independent
random variables
\begin{equation}
\frac{\delta_\zeta}{4T}\frac{2\delta_\zeta\Gamma/\pi}{\xi_1^2}
\,,\;\ldots\,,\;
\frac{\delta_\zeta}{4T}\frac{2\delta_\zeta\Gamma/\pi}{\xi_n^2}
\end{equation}
[see Eq.~(\ref{smallg})], with $\xi_i$ uniformly distributed in
the range $0<\xi_i<\delta_\zeta/2$.

\section{Cancellation of the cutoff~$\Delta$}
\label{app2}

The formal reason to introduce the cutoff~$\Delta$ in
Sec.~\ref{sec:insulator} was the
insufficient ability of the $3n$-particle potentials
to confine particles, which resulted in logarithmic divergences in
the leading terms of the cluster expansion for $n>1$. On the other
hand, since the transition is associated with anomalously small
values of the energy denominators, rather than anomalously large
ones, it is natural to expect the transition condition not to
contain~$\Delta$ at all.
This means that~$\Delta$ should cancel
out when higher-order terms of the cluster expansion are taken
into account. Moreover, this cancellation should occur for each
component of the potential separately ({\em e.g.}, for each of the
6-particle potentials $U^{ee}$, $U^{eh}$, $U^{he}$, and $U^{hh}$).
In this appendix we first show how the cancellation occurs for the
6-particle potential $U^{hh}$,
and then discuss $3n$-particle potential $U^{hh\ldots{h}}$ for
an arbitrary~$n$. The terms corresponding to an arbitrary
sequence of electron and hole decays can be analyzed analogously.

\begin{figure}
\includegraphics[width=0.6\textwidth]{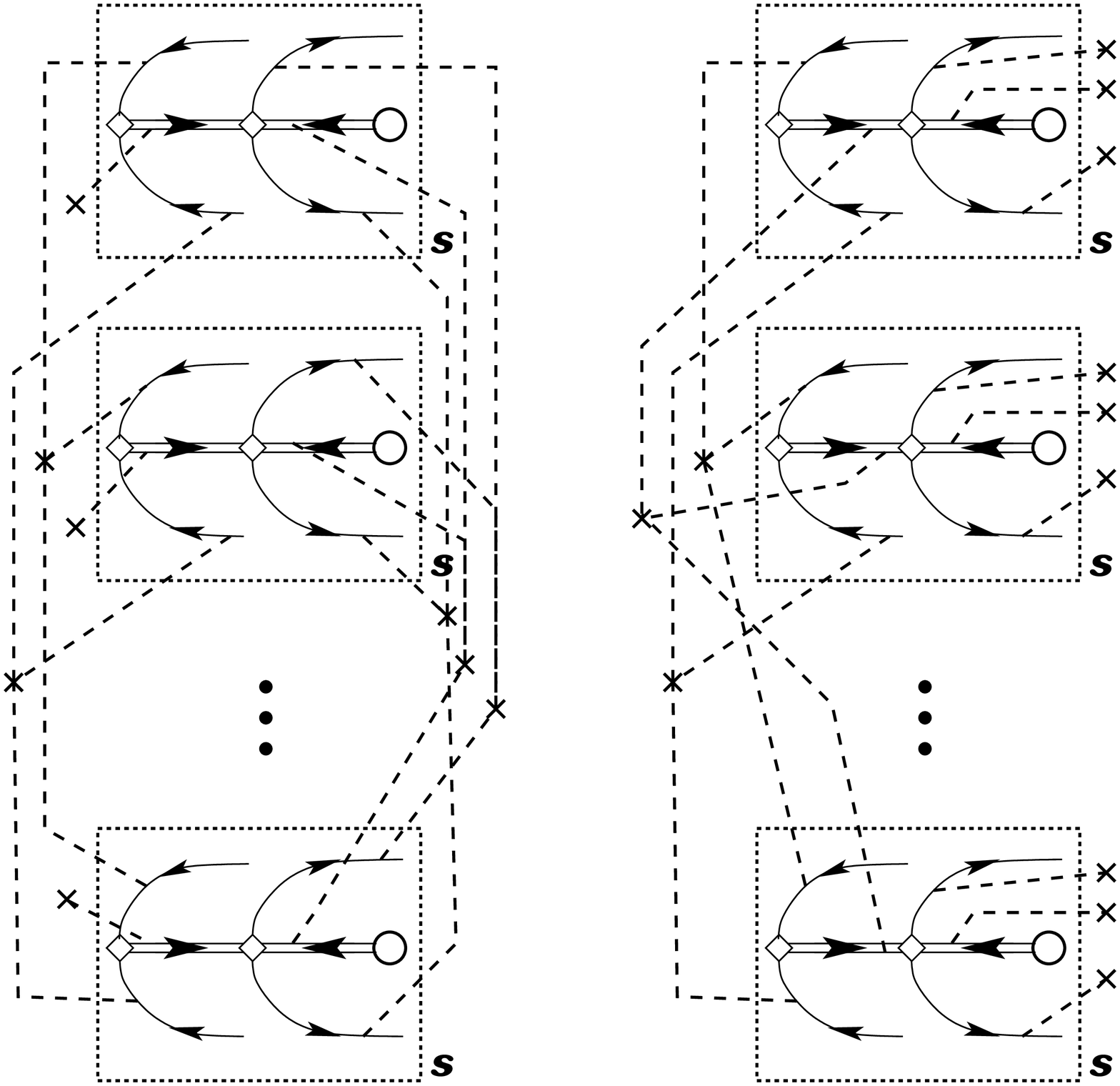}
\caption{The diagrams canceling the cutoff~$\Delta$ in the
fourth diagram of Fig.~\ref{6pcluster}.}\label{fig:cancelcutoff}
\end{figure}

Consider connected terms of the cluster expansion containing
$k$~functions $f^{hh}(\xi_1^{(1)},\ldots,\xi_6^{(1)})$,\ldots,
$f^{hh}(\xi_1^{(k)},\ldots,\xi_6^{(k)})$. To maximize the
corresponding contribution one should connect them in such
a way that one of the energy denominators coincides for all
terms (two denominators cannot coincide, as it would require
the coincidence of $\xi_i^{(j)}$ for all~$j$ and each~$i$,
which is prohibited by the rules of the cluster expansion).
Having fixed one of the denominators, one should keep the
remaining free level positions independent of each other in
order not to obtain smallness in $M$ or $T/\delta_\zeta$.
As a result, two diagrams should be evaluated for each $k$,
shown in Fig.~\ref{fig:cancelcutoff}. Summing over~$k$ and
adding the leading term, we obtain
\begin{equation}\begin{split}
&\ln{W}^{(2,0)}_{hh}=-\sqrt{\pi{s}\Gamma^{(0,0)}}
\left(\frac{2\lambda{M}T}{\delta_\zeta}\right)^2\left[
\ln\frac{(\Delta/\lambda\delta_\zeta)^2}{\sqrt{{s}\Gamma^{(0,0)}}}
\right.\\
&\quad+\sum_{k=2}^\infty\frac{(-\Delta/\delta_\zeta)^{k-1}}{k!\,(k-1)}
\int\limits_{-\infty}^\infty\frac{\diff{x}}{4}
\frac{\cosh\frac{(k-1)x}{2}}{\cosh^{k+1}\frac{x}{2}}
\left.+\sum_{k=2}^\infty\frac{(-\Delta/\delta_\zeta)^{k-1}}{k!\,(k-1)}
\left(\frac{2\lambda{M}T}{\delta_\zeta}\right)^{k-1}\right],
\label{lnWhhcanc=}
\end{split}\end{equation}
where the logarithmic term in the square brackets is the leading
one, corresponding to the fourth diagram of Fig.~\ref{6pcluster},
the second and third terms correspond to the diagrams of
Fig.~\ref{fig:cancelcutoff}. The summation over~$k$ is performed
as
\begin{equation}\begin{split}
&\sum_{k=2}^\infty\frac{(-A)^{k-1}}{k!(k-1)}=
-\int\limits_0^A \diff{x}\sum_{k=2}^\infty\frac{(-x)^{k-2}}{k!}
=-\int\limits_0^A \frac{\diff{x}}{x^2}\left(\eexp^{-x}-1+x\right)\\
&=-\int\limits_{1/A}^\infty \diff{x}\left(\eexp^{-1/x}-1+\frac{1}x\right)
\approx-\ln{A}\,,\quad{A}\gg{1}\,.
\end{split}\end{equation}
As a result, we obtain
\begin{equation}
\ln{W}^{(2,0)}_{hh}=-\sqrt{\pi{s}\Gamma^{(0,0)}}
\left(\frac{2\lambda{M}T}{\delta_\zeta}\right)^2
\ln\frac{MT/(\lambda^2\delta_\zeta)}{\sqrt{{s}\Gamma^{(0,0)}}}\,.
\end{equation}
This seems to correspond to the replacement
\begin{equation}
\Delta\to\sqrt{MT\delta_\zeta}\,,
\end{equation}
rather than to the promised Eq.~(\ref{replacement}).

To show that the transition point is nevertheless determined by
Eq.~(\ref{replacement}), we analyze the hole channel for an
arbitrary~$n$. In this case the subleading logarithmic correction
is given by diagrams where $n-1$ energy denominators are fixed
by connections, and only one is independent of others. The
independent denominator can be chosen in $n$~ways, of which
$n-1$ produce diagrams analogous to the first diagram of
Fig.~\ref{fig:cancelcutoff} (each $f$ has only one
argument independent of other $f$'s) and the last denominator
produces a diagram similar to the second diagram of
Fig.~\ref{fig:cancelcutoff} (each $f$ has three last arguments
independent of other $f$'s). The result is
\begin{equation}\begin{split}
&\ln{W}^{(n,0)}_{h\dots{h}}=-\sqrt{\pi{s}\Gamma^{(0,0)}}
\left(\frac{2\lambda{M}T}{\delta_\zeta}\right)^n\\
&\times\left[\frac{1}{(n-1)!}
\ln^{n-1}\frac{(\Delta/\lambda\delta_\zeta)^n}{\sqrt{{s}\Gamma^{(0,0)}}}
-\frac{1}{(n-2)!}
\ln^{n-2}\frac{(\Delta/\lambda\delta_\zeta)^n}{\sqrt{{s}\Gamma^{(0,0)}}}
\ln\left(\frac{MT}{\delta_\zeta}\frac{\Delta^n}{\delta_\zeta^n}\right)
\right],
\end{split}\end{equation}
The expression in the square brackets is nothing else but the first
two terms of the binomial
\begin{equation}\begin{split}
&\frac{1}{(n-1)!}\left[
\ln\frac{(\Delta/\lambda\delta_\zeta)^n}
{\sqrt{{s}\Gamma^{(0,0)}}}-
\ln\left(\frac{MT}{\delta_\zeta}\frac{\Delta^n}{\delta_\zeta^n}\right)
\right]^{n-1}\\
&=\frac{1}{(n-1)!}\ln^{n-1}
\left(\frac{\delta_\zeta}
{MT\lambda^n}\frac{1}{\sqrt{{s}\Gamma^{(0,0)}}}\right).
\label{logbinomial}
\end{split}\end{equation}
To obtain the next terms of the binomial one would have to connect
the cluster functions in a way that fixes $n-2$, $n-3$, {\it etc.}
denominators. As seen from Eq.~(\ref{logbinomial}), the factor
$MT/\delta_\zeta$ under the logarithm becomes unimportant at large~$n$.
This justifies the rule~(\ref{replacement}).

\section{Derivation of \req{val:gamma3}}
\label{app3}

\begin{figure}
\includegraphics[width=0.4\textwidth]{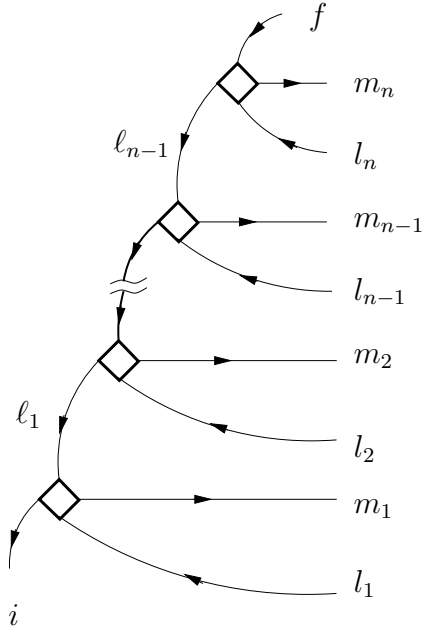}
\caption{Amplitude of $n$th order of the
peturbation theory used for the estimate of the distribution
function \rref{app3:eq1}. Being squared, these amplitudes generate
the electron-like decay processes only (the first diagrams on
Figs.~\ref{3pcluster},~\ref{6pcluster}).
External lines are assumed to be amputated and shown for
the notation of orbitals and energy. Path $\ell$ in \req{val:gamma1}
corresponds to the sequence $\{\ell_1,\ell_2,\dots \ell_{n-1}\}$
on this figure.
}
\label{fig:amplitude1}
\end{figure}

Our goal  in this Appendix is to evaluate the 
most dangerous contribtion from the  interference corrections
to the SCBA result. For this purpose, it suffices to demonstrate the
calculation of $\alpha_{int}$ from  \req{val:gamma3} without numerical
coeficient. Therefore, we will take into account only electron-like
processes shown on Fig.~\ref{fig:amplitude1} and disregard other
processes, cf. Figs.~\ref{3pcluster},~\ref{6pcluster}.
This leads to the underestimate of the overall numerical coefficient
but does not affect the factorials.

The expression for such $n$th order amplitude in the insulating region
reads
\be
\begin{split}
\mathcal{A}_i^f\left(\{\ell_k\};\{l_k\};\{m_k\}\right) &\simeq
\frac{ V_{il_1}^{\ell_1 m_1}F^{\Rightarrow}_{\ell_1,l_1; m_l}}
{\ep_i-
\Xi_{l_1;\ell_1}^{m_1}}\,
\frac{ V_{\ell_1l_2}^{\ell_2 m_2}F^{\Rightarrow}_{\ell_2,l_2; m_2}}
{\ep_i-
\Xi_{l_1,l_2;\ell_2}^{m_1,m_2}}\,
\dots
\,\frac{ V_{\ell_{n-1}l_n}^{f m_n}F^{\Rightarrow}_{f,l_n; m_n}}
{\ep_i-
\Xi_{l_1,l_2\dots,l_n;\ell_{n-1}}^{m_1,m_2,\dots,m_n}}\,,
\end{split}
\label{app3:eq1}
\ee
where the notation was introduced in \reqs{linF}  and \rref{Xi}.

Expression \rref{app3:eq1} has the same form as those considered in
Secs.~\ref{0D} and~\ref{sec:renint}.
so that the machinery of Sec.~\ref{sec:Mayer} can be applied with
small modification.
First we fix the electron and hole levels $\xi_{l_1},\dots, \xi_{l_n}; 
\ \xi_{m_1},\dots, \xi_{m_n}$ in \req{app3:eq1}, and sum over
all the intermediate states ${\ell}_1,{\ell}_2; \dots, \ell_n$.
Calculating the characteristic
function, we perform the {\em partial averaging} -- {\em i.e.}, procedure
of \req{averaging} with $2n$ levels ${l_1},\dots, {l_n}; 
\ {m_1},\dots, {m_n}$ excluded.

Using the linked cluster expansion and \req{Heffintdist}, we find 
\be
\begin{split}
&
\ln \tilde{W}\left(q; \{l_i\};\{m_i\}\right)\equiv
\ln \left\langle 
\exp\left[iq\sum_{\ell_k}\mathcal{A}_i^f\left(\{\ell_k\};\{l_k\};\{m_k\}\right)\right]
\right\rangle_{ \{k_i\};\{m_k\}}
\\
& \quad\simeq
-\pi |q|\delta_\zeta \lambda \left[ 2\lambda \ln \frac{1}{\lambda}
  \right]^{n-1}
\prod_{k=1}^n\frac{1-n_{l_k}n_{m_k}}{2}
\left[
\Upsilon^2\left(
\Xi_{l_1}^{m_1}\right) 
- \Upsilon^2\left(
\Xi_{l_1}^{i}\right) 
\right]
\\
& \quad\times
\left[
\Upsilon^2\left(
\Xi_{l_2}^{m_2}\right) 
- \Upsilon^2\left(
\Xi_{l_1l_2}^{i m_1}\right) 
\right]
\dots \left[
\Upsilon^2\left(
\Xi_{l_n}^{m_n}\right) 
- \Upsilon^2\left(
\Xi_{l_1l_2\dots l_n}^{i m_1m_2\dots m_{n-1}}\right) 
\right]
.
\end{split}
\label{app3:eq2}
\ee
where we were dealing with the logarithmic integrals as in
Sec.~\ref{val:shift}. For the sake of brievity 
we replaced $\Xi_{l_1\dots}^{m_1\dots}/\delta_\zeta \to
\Xi_{l_1\dots}^{m_1\dots}$ in the last two lines of the equation.

Equation \rref{app3:eq2} enables us to estimate the
distribution function of the amplitudes summed over the permutations among
$n$ electrons and $n$ holes, see \req{val:gamma1}.
Approximating the amplitudes obtained by the permutations of the final
states to be independent of each other, {\em cf.}
Fig.~\ref{fig:path}\footnote{This
corresponds to the leading logarithmic approximation in the linked
cluster expansion. Note, however, that  
it does not affect the terms divergent as $n!$ at all.}
we find
\be
\begin{split}
\mathcal{W}\left(q; \{l_i\};\{m_i\}\right)
&\equiv
\left\langle 
\exp\left[iq\sum_{\substack{\ell_k\\P_lP_m}}
(-1)^{P_l+P_m}
\mathcal{A}_i^f\left(P_l\{l_k\};P_m\{m_k\}\right)\right]
\right\rangle
\\
&=\prod_{ P_lP_m}{\tilde W}\left(q; P_l\{l_i\};P_m\{m_i\}\right), 
\end{split}
\label{app3:eq3}
\ee
where we used the standard expression for the characteristic function
of sum, we suppressed the $\ell_k$ arguments of the amplitude
for brevity, and the partial averaging as in \req{app3:eq2} is implied.

\be
\begin{split}
&\ln \left\langle
\exp\left(-s\delta_{\zeta}^{-2} \sum_{\{l_k\}, \{m_k\}}\gamma_l 
(\{l_k\},\{m_k\})\right)
\right\rangle
\\ &
=
\int \frac{\delta_\zeta\, \diff q}{\left(2\pi s\right)^{1/2}}
\eexp^{-\frac{q^2\delta_{\zeta}^2}{2s}}
\prod_{k=1}^{2 n}\frac{\diff\xi_k}{\delta_\zeta}
\sum_{n_k=\pm 1}\frac{\exp\frac{\xi_kn_k}{2}}
{2\cosh \frac{\xi_kn_k}{2T}}
\\
&
\qquad\qquad\times \frac{\mathcal{W}\left(q; \{1,2,\dots n\};\{n+1, \dots,
  2n\}\right)
-1}{\left(n!\right)^2}.
\end{split}
\label{app3:eq4}
\ee
Finally, expanding the last in 
\req{app3:eq4} 
in powers of $|q|$, and performing remaining
integrations, we obtain the structure of \req{val:gamma3}.
Indeed, exponent contains $(n!)^2$ terms itself, so that the factorial
factors cancel at all in the leading term. In the next term
product $\mathcal{W}\left(q; \{l_k\};\{m_k\}\right)
\mathcal{W}\left(q; P_l\{l_k\};P_m\{m_k\}\right)$ 
gives the same volume for the $\xi$ integration only for $P_l=P_m$
and impose the additional restricitions otherwise. The factor
of $n!$ is, thus, just a number of permutations electron-hole pairs
without their destruction.


\end{document}